\documentclass[11pt] {article}

\usepackage{epsfig}
\catcode`@=11

\def\@citex[#1]#2{\if@filesw\immediate\write\@auxout{\string\citation{#2}}\fi
  \def\@citea{}\@cite{\@for\@citeb:=#2\do
    {\@citea\def\@citea{,\penalty\@m}\@ifundefined
      {b@\@citeb}{{\bf ?}\@warning
       {Citation `\@citeb' on page \thepage \space undefined}}%
\hbox{\csname b@\@citeb\endcsname}}}{#1}}

\def\citer{\@ifnextchar [{\@tempswatrue\@citexr}{\@tempswafalse\@citexr[]}}

\def\@citexr[#1]#2{\if@filesw\immediate\write\@auxout{\string\citation{#2}}\fi
  \def\@citea{}\@cite{\@for\@citeb:=#2\do
    {\@citea\def\@citea{--\penalty\@m}\@ifundefined
       {b@\@citeb}{{\bf ?}\@warning
       {Citation `\@citeb' on page \thepage \space undefined}}%
\hbox{\csname b@\@citeb\endcsname}}}{#1}}
\catcode`@=12
\newcommand{\dis}{\displaystyle}

\def\s{\hat{s}}
\def\u{\hat{u}}
\def\z{v \cdot \hat{q}}

\def\ml{\hat{m}_l}
\def\ms{\hat{m}_s}

\def\mc{\hat{m}_c}

\def\lo{\hat{\lambda}_1}
\def\lt{\hat{\lambda}_2}

\def\q{\hat{q}}
\def\bxsll{$B \rightarrow X_s \ell^+ \ell^- $}
\def\absvcb{\left| V_{cb} \right|}

\title{  { \bf
Perturbative QCD- and Power-Corrected Hadron Spectra 
and Spectral Moments in the Decay
 \bxsll }}
\author{\vspace{1cm}\\
   {\bf A.~Ali\thanks{E-mail address: ali@x4u2.desy.de} ~and 
G.~Hiller\thanks{E-mail address: ghiller@x4u2.desy.de}}\\
         Deutsches Elektronen-Synchrotron DESY, Hamburg \\
        \vspace{5mm}\\}
\date{}

\begin{document}
\setlength{\baselineskip}{24pt}

\maketitle
\begin{picture}(0,0)
       \put(325,260){DESY 98-030}
       \put(325,245){March 1998}
\end{picture}
\vspace{-24pt}
\setlength{\baselineskip}{7mm}

\begin{abstract}
We compute the leading order (in $\alpha_s$) perturbative QCD and power 
($1/m_b^2)$ corrections 
to the hadronic invariant mass and hadron energy spectra in the decay $B 
\to X_s \ell^+ \ell^-$ in standard model. The computations are carried out 
using the heavy quark expansion technique (HQET) and a perturbative-QCD
improved Fermi motion (FM) model which takes into account $B$-meson
wave-function effects. The corrections in the 
hadron energy ($E_H$) spectrum are found to be small over a good part of this 
spectrum in both methods. However, the expansion in  $1/m_b$ in HQET 
fails near the lower kinematic end-point and at the $c\bar{c}$ threshold.
The hadronic 
invariant mass ($S_H$) spectrum is calculable only over a limited range
$S_H > \bar{\Lambda}m_B$ in the heavy quark expansion, where
$\bar{\Lambda} \simeq m_B-m_b$. We also present results for the
first two hadronic moments $\langle S_H^n\rangle$ and $\langle 
E_H^n\rangle$, $n=1,2$, working out their sensitivity on the 
HQET and FM model parameters. For equivalent values of these parameters,
the moments in these methods are remarkably close to each other.
The constraints following from assumed values of $\langle S_H^n\rangle$ on
the HQET parameters $\lambda_1$ and $\bar{\Lambda}$ are worked out.
Data from the forthcoming B facilities could be
used to measure the short-distance contribution in \bxsll and 
constrain the HQET parameters $\lambda_1$ and $\bar{\Lambda}$.
This could be combined with complementary constraints from the decay $B 
\to X \ell \nu_\ell$ to determine these parameters precisely.
We also study the effect of the experimental cuts,
used recently by the CLEO collaboration in searching for the decay 
\bxsll, on the branching ratios, hadron spectra and hadronic invariant mass 
moments using the FM model.

\vspace*{1.5cm}
\centerline{(Submitted to Physical Review D)}

\end{abstract}

\thispagestyle{empty}
\newpage
\setcounter{page}{1}

\section{Introduction}

  The semileptonic inclusive decays $B \to X_s \ell^+ \ell^-$ , where
$\ell^\pm = e^\pm,\mu^\pm,\tau^\pm$, offer, together with the
radiative electromagnetic penguin decay $B \to X_s + \gamma$, presently the
most popular testing grounds for the standard model (SM) in the flavor
sector. This is reflected by the impressive experimental and theoretical
activity in this field, reviewed recently in \cite{skwarnicki97} and 
\cite{aliapctp97}, respectively. We shall concentrate here 
on the decay $B \to X_s \ell^+ \ell^-$ for which 
the first theoretical calculations were reported a decade ago  
\citer{HWS87,JW90}, emphasizing the
sensitivity of the dilepton mass spectrum and decay rate to the top quark
mass in the short-distance contribution. With the discovery of the top
quark and a fairly accurate measurement of its mass \cite{PDG}, 
theoretical emphasis has changed from predicting the top quark mass
using this decay to using its measured value  as  
input and making theoretically accurate predictions for the decay rates and
spectra. This will help confront the predictions in the SM with experiment
more precisely and will allow to search for new phenomena, such as 
supersymmetry \citer{BBMR91,Hewett96}.

Since these early papers, considerable theoretical work
has been done on the decay \bxsll~in the context of the standard model.
This includes, among other aspects, the calculation of the complete leading 
order perturbative corrections in the QCD coupling constant $\alpha_s$ 
to the dilepton invariant mass spectrum \cite{burasmuenz,misiakE},
forward-backward (FB) asymmetry of the 
leptons \cite{amm91,AHHM97}, and, additionally, leading order power 
corrections in  
$1/m_b^2$ to the decay rate, dilepton invariant mass spectrum and  
FB asymmetry \cite{AHHM97}, using the heavy quark expansion technique 
(HQET) \citer{georgi,FLSold}. We recall that the $1/m_b^2$ corrections to 
the dilepton spectrum and decay 
rate in \bxsll were calculated in ref.~\cite{FLSold} but their  
results were at variance with the ones derived later in 
ref.~\cite{AHHM97}. The power corrected
dilepton mass spectrum and FB asymmetry have been rederived for the 
massless  $s$-quark case recently \cite{BI98}, confirming the  
results in ref.~\cite{AHHM97}. Corrections of order $1/m_c^2$
to the dilepton mass spectrum  away from the 
$(J/\psi,\psi^\prime,...)$-resonant regions have also 
been worked out \cite{buchallaisidorirey,chenrupaksavage}, making use of 
earlier work on similar power corrections in the decay rate for 
$B \to X_s + \gamma$ \cite{Voloshinbsg,powermc}.
 The $1/m_b^2$ power corrections to the left-right
asymmetry \cite{LD96,BP97} have been presented
in \cite{BI98} correcting an earlier calculation of the same
\cite{BP97}. Likewise, the longitudinal
polarization of the lepton, $P_L$, in $B \to X_s \tau^+ \tau^-$ at the
partonic level has been worked out \cite{Hewettpol}; the other two
orthogonal polarization components $P_T$ (the component in the decay plane)
and $P_L$ (the component normal to the decay plane) were subsequently worked 
out in ref.~\cite{KSpol96}. As an alternative to HQET, $B$-meson 
wave-function effects in the
decay \bxsll ~have also been studied for the dilepton invariant mass
spectrum and FB asymmetry \cite{AHHM97}, using the Fermi motion (FM) 
model \cite{aliqcd}. Some of the cited works have also addressed 
the long-distance aspect of the decay \bxsll having to do with the 
resonant structure of the dilepton invariant mass spectrum. We shall 
leave out the $J/\psi,\psi^\prime,...$-resonant contributions 
in this paper and will present a detailed  phenomenological study 
including them elsewhere \cite{AH98-3}.

  This theoretical work, despite some uncertainties associated with the 
LD-part, will undoubtedly  
contribute significantly to a meaningful comparison of the SM and experiment
in the decay \bxsll. Still, concerning the SD-contribution,
some aspects of this decay remain to be studied theoretically. 
In the context of experimental searches for \bxsll,
it has been emphasized (see, for 
example, the CLEO paper \cite{cleobsll97}) that theoretical
estimates of the hadronic invariant  mass and hadron energy
spectra in this decay will greatly help in providing
improved control of the signal and they will also be needed to correct 
for the experimental acceptance. In addition to their
experimental utility, hadron spectra in heavy hadron decays are also of 
considerable 
theoretical interest in their own right, as reflected by similar studies 
done for the charged current induced semileptonic
decays $B \to X_c \ell \nu_\ell$ and $B \to X_u \ell \nu_\ell$
\citer{bargerkim90,DU97}, where the main 
emphasis has been on testing HQET and/or in determining the
Cabibbo-Kobayashi-Maskawa (CKM) matrix elements $V_{cb}$ and $V_{ub}$. The 
hadronic invariant mass spectra in $b \to s\ell^+ \ell^-$ and $b \to u \ell^- 
\nu_\ell$ decays have 
striking similarities and differences. For example, both of these processes
have at the parton level a delta function behavior  
$d\Gamma/ds_0 \propto \delta(s_0-m_q^2)$, $q=u,s$, where $s_0$ is
the hadronic invariant mass at the parton level. Thus, the 
entire invariant mass spectrum away from $s_0=m_q^2$ is 
generated perturbatively (by gluon bremsstrahlung) and through the 
$B$-hadron non-perturbative effects. Hence,  measurements of these
spectra would lead to direct information on the QCD dynamics
and to a better determination of the non-perturbative parameters. There
are also obvious differences in these decays, namely the decay $B \to X_u 
\ell \nu_\ell$ is intrinsically a lot simpler due to the absence of the
resonating $c\bar{c}$ contributions, which one must 
include to get the inclusive spectra in \bxsll, or else use data in 
restricted phase space where the $c\bar{c}$-resonant contributions are
subleading.

 Having stated the motivations, we study hadron spectra in 
the  decay \bxsll ~in this paper. 
We first compute the leading order (in $\alpha_s$) perturbative QCD and 
power ($1/m_b^2)$ corrections to the hadronic invariant mass and 
hadron energy spectra at the parton level.
 In addition to the bremsstrahlung contribution $b \to (s+ g) \ell^+ 
\ell^-$, there are important
non-perturbative effects even in ${\cal O}(\alpha_s^0)$ that come from the 
relations between the 
$b$ quark mass and the $B$ meson mass. In HQET, this takes the form   
$m_B=m_b+\bar{\Lambda} -(\lambda_1+ 3\lambda_2)/2m_b +...$, where
$\bar{\Lambda}$, $\lambda_1$ and $\lambda_2$ are the HQET parameters 
\citer{georgi,FLSold}. 
Keeping, for the sake of simplicity just the $\bar{\Lambda}$ term, the
hadronic invariant mass $S_H$ is related to $s_0$ and the partonic energy 
$E_0$ by  $S_H=s_0 + 2\bar{\Lambda}E_0 + \bar{\Lambda}^2$.
This gives rise to a non-trivial spectrum in the entire region
$\bar{\Lambda}^2 < S_H < M_B^2$. Including both the ${\cal O} (1/m_b^2)$
and ${\cal O}(\alpha_s)$ terms generates hadron energy and hadronic
invariant mas spectrum with terms of ${\cal O}(\bar{\Lambda}/m_B)$,
${\cal O}(\alpha_s\bar{\Lambda}/m_B)$, ${\cal O}(\lambda_1/m_B^2)$ 
and ${\cal O}(\lambda_2/m_B^2)$. The power- and perturbatively
corrected hadron spectra
up to and including these terms are presented here.
 The $1/m_b^2$ corrections in the hadron energy spectrum
are found to be small over a good part of this spectrum. However, the     
expansion in $1/m_b$ fails near the lower end-point  and near the
$c\bar{c}$ threshold.
The hadronic invariant mass spectrum
is reliably calculable over a limited region only, namely for $S_H > 
\bar{\Lambda}m_B$. Hadronic moments  
$\langle S_H^n\rangle$ and $\langle E_H^n\rangle$, on the other hand,
are calculable in HQET and we have summarized the results for the first two
moments $n=1,2$ in a letter \cite{AH98-1}, based on this study. The
hadronic invariant mass moments are sensitive to the HQET parameters
$\bar{\Lambda}$ and $\lambda_1$.
This provides potentially an independent determination of these quantities.
We think that the hadron spectra in \bxsll and $B \to X_u \ell \nu_\ell$
can be related to each other over limited phase space and this could help
in vastly improving the present precision on  $V_{ub}$
\cite{PDG} and the parameters
$\lambda_1$ and $\bar{\Lambda}$ \cite{gremm,neubert}.

 In view of the continued phenomenological interest in the FM model 
\cite{aliqcd}, motivated in part by its close resemblance to the HQET 
framework \cite{Bigietal94,MW},
we also compute the hadron spectra in \bxsll~in this model,
taking into account the ${\cal O}(\alpha_s)$ perturbative QCD corrections.
The FM model is characterized by two parameters which are usually taken as 
$p_F$, the Gaussian width of the $b$-quark momentum, 
and $m_q$, the spectator quark mass in the $B$ hadron; the $b$-quark mass is 
a momentum-dependent quantity (see section 6 for details).
The matrix element of the kinetic energy operator, $\lambda_1$ and the
binding energy $\bar{\Lambda}$ can be calculated in terms of the FM model
parameters. The difference between the effective 
$b$-quark mass, which is a derived quantity in the FM model, and the 
$B$-meson mass can also be expressed  via an HQET-type relation, 
$m_B=m_b^{\mbox{eff}} + \bar{\Lambda} - \lambda_1/2 m_b^{\mbox{eff}}$.
However, there is no analog of $\lambda_2$ in the FM model. Having 
defined the equivalence between the FM model and HQET parameters, we shall 
use $\bar{\Lambda}$ and $\lambda_1$ to also characterize the FM model 
parameters. The dependence of the hadron spectra in the FM model in
the decay \bxsll~on the parameters $\bar{\Lambda}$ and 
$\lambda_1$ is studied in this paper. We find that
the hadron energy spectrum in \bxsll in the FM model is
stable against variations of the model parameters.
The hadron energy spectra in the FM model and HQET are also found to be  
close to each other in regions where HQET holds.
This feature was also noticed  in the context of the
decay $B \to X_u \ell \nu_\ell$ in ref.~\cite{greubrey}. The hadronic 
invariant mass spectrum depends sensitively on
the parameters of the FM model -  a behavior which has again its parallel
in studies related to the decay $B \to X_u \ell \nu_\ell$ \cite{FLW} as
well as in HQET.
Hadronic moments $\langle S_H^n \rangle$ and  $\langle E_H^n \rangle$
are computed in the FM model and are found to be
remarkably close to their counterparts calculated in HQET for equivalent 
values of the parameters $\bar{\Lambda}$ and $\lambda_1$. The picture
that emerges from these comparisons is that the spectra and moments in the
two approaches are rather similar, though not identical.  
We also study the effects of the CLEO experimental cuts on the branching 
ratios, hadron spectra and hadronic moments in \bxsll~in the FM model
and the results are presented here. These
can be compared with data when they become available.

This paper is organized as follows. In section 2, we define the kinematics 
of the process $B \to X_s \ell^+ \ell^-$ and introduce the quantities of
dynamical interest in the framework of an effective Hamiltonian. Leading
order (in $\alpha_s$) perturbative corrections to the hadron energy
and hadronic invariant mass spectra at the parton level are derived in 
section 3, where we also present the Sudakov-improved spectrum $d{\cal B}/
ds_0$. Using the HQET relation between $m_B$ and $m_b$, we calculate the
corrected hadronic invariant mass spectrum $d{\cal B}/dS_H$.
 In section 4, we present the leading power corrections
(in $1/m_b^2)$ for the Dalitz distribution $d^2{\cal B}/dx_0 d\hat{s}_0$
(here $x_0$ and $\hat{s}_0$ are the scaled partonic energy and 
hadronic invariant mass, respectively) and 
derive analytic expressions for the hadron energy spectrum
$d{\cal B}/dx_0$ and the resulting spectrum is compared with the one in 
the parton model.
 In section 5, we calculate the 
moments in the hadron energy and hadronic invariant mass in HQET and give
the results for $\langle S_H \rangle$, $\langle S_H^2 \rangle$,
$\langle E_H \rangle$ and  $\langle E_H^2 \rangle$  in terms of 
the corresponding moments in the partonic variables. 
Section 6 describes the wave-function effects in the FM model
\cite{aliqcd} in the hadron energy and hadronic invariant mass 
spectra.
 We also give here numerical estimates of the hadronic
moments in HQET and the FM model.
In section 7, we study the effects of the 
experimental cuts used in the CLEO analysis of \bxsll on the hadron 
spectra and hadronic moments using the FM model.
 Estimates of the
branching ratios ${\cal B}(B \to X_s \ell^+ \ell^-)$ for $\ell =\mu,e$
are also presented here, together with estimates of the survival
probability for the CLEO cuts, using the FM model.
 Section 8 contains a summary of our work and some 
concluding remarks. 
 Definitions of various auxiliary
functions and lengthy expressions appearing in the derivation of our  
results, including the partonic moments $\langle x_0^n\rangle$,
$\langle (\hat{s}_0-\hat{m}_s)^n\rangle$ and $\langle x_0 
(\hat{s}_0-\hat{m}_s)\rangle$ for $n=1,2$  are relegated to the Appendices 
A - D.      

\section{\bf The Decay \bxsll~in the Effective Hamiltonian Approach}

\subsection{Kinematics}

We start with the definition of the kinematics of the decay at the parton 
level,
\begin{equation}
b (p_b) \to s (p_s) (+g (p_g))+\ell^{+} (p_{+})+\ell^{-}(p_{-}) ~,
\end{equation} 
where $g$ denotes a gluon from the $O(\alpha_s)$ correction
(see Fig.~1). The 
corresponding kinematics at the hadron level can be written as:
\begin{equation}
B (p_B) \to X_s (p_H)+\ell^+ (p_{+})+\ell^{-} (p_{-})~.
\end{equation}
We define the momentum transfer to the lepton pair and the invariant 
mass of the dilepton system, respectively, as
\begin{eqnarray}
q &\equiv & p_{+}+p_{-} \; , \\
s &\equiv & q^2 \; .
\end{eqnarray}
The dimensionless variables with a
hat are related to the dimensionful variables by the scale $m_b$, 
the $b$-quark mass, e.g., $\s= \frac{s}{m_b^2}$, $\ms=\frac{m_s}{m_b}$ etc..
Further, we define a 4-vector $v$, which denotes the velocity of both the
$b$-quark and the $B$-meson, $p_b=m_b v$ and $p_B=m_B v$. We shall also
need the variable $u$ and the scaled variable $\u=\frac{u}{m_b^2}$, 
defined as:
\begin{eqnarray}
u  & \equiv& -(p_b-p_+)^2+(p_b-p_{-})^2 \; , \\
\u & =& 2 v \cdot (\hat{p}_{+}-\hat{p}_{-}) \; .
\end{eqnarray}
The hadronic invariant mass is denoted by  $S_H \equiv p_H^2$ and 
$E_H$ denotes the hadron energy in the final state. 
The corresponding quantities at parton level are the 
invariant mass $s_0$ and the scaled parton energy $x_0\equiv \frac{E_0}{m_b}$.
In parton model without gluon bremsstrahlung, this simplifies to
$s_0=m_s^2$ and $x_0$ becomes directly related to the dilepton invariant mass 
$x_0=1/2(1-\s +\ms^2)$.
{}From momentum conservation the following equalities hold in the $b$-quark,
equivalently $B$-meson, rest frame ($v=(1,0,0,0)$):
\begin{eqnarray}
x_0  &=& 1- v \cdot \q \, \, ,
~~~\s_0 = 1 -2 v \cdot \q + \s \, \, ,\label{eq:kin} \\
E_H &=& m_B-v \cdot q \, \, ,
~~~S_H = m_B^2 -2 m_B v \cdot q  + s \, \, .
\end{eqnarray}
The relations between the kinematic variables of the parton model and the 
hadronic states , using the
HQET mass relation, can be written as
\begin{eqnarray}
  E_H&=&\bar\Lambda-{\lambda_1+3\lambda_2\over2m_B}+\left(m_B-\bar
  \Lambda+
  {\lambda_1+3\lambda_2\over2m_B}\right) x_0+\dots\,, \nonumber\\
  S_H&=&m_s^2+\bar\Lambda^2+(m_B^2-2\bar\Lambda m_B+\bar\Lambda^2
  +\lambda_1+3\lambda_2)\,(\hat s_0-\hat m_s^2) \nonumber \\
  &&\qquad\qquad\mbox{}+(2\bar\Lambda
m_B-2 \bar\Lambda^2-\lambda_1-3\lambda_2) x_0
  +\dots\,,
\label{ehe0shs0}
\end{eqnarray}
where the ellipses denote terms higher order in
$1/m_b$.

\subsection{Matrix element for the decay \bxsll}

The effective 
Hamiltonian obtained by integrating out the top quark and the $W^\pm$ bosons
is given as
\begin{eqnarray}\label{heffbsll}
{\cal H}_{eff}(b \to s + X, \, X=\gamma, \, \ell^{+} \ell^{-}) 
= - \frac{4 G_F}{\sqrt{2}} V_{ts}^* V_{tb}
 \left[ \sum_{i=1}^{6} C_i (\mu)  O_i 
+ C_7 (\mu) \frac{e}{16 \pi^2}
          \bar{s}_{\alpha} \sigma_{\mu \nu} (m_b R + m_s L) b_{\alpha}
                F^{\mu \nu} 
\right. \nonumber \\
 \left.
+C_8 (\mu) O_8 
+ C_9 (\mu) \frac{e^2}{16 \pi^2}\bar{s}_\alpha \gamma^{\mu} L b_\alpha
\bar{\ell} \gamma_{\mu} \ell 
+ C_{10}  \frac{e^2}{16 \pi^2} \bar{s}_\alpha \gamma^{\mu} L
b_\alpha \bar{\ell} \gamma_{\mu}\gamma_5 \ell \right] \, ,
\end{eqnarray}
where $L$ and $R$ denote chiral projections, $L(R)=1/2(1\mp \gamma_5)$,
 $V_{ij}$ are the CKM matrix elements and the
CKM unitarity has been used in factoring out the product $V_{ts}^\ast
V_{tb}$. The operator basis is taken from \cite{AHHM97}, where also the 
Four-Fermi operators $O_{1},\dots ,O_{6}$ and the chromo-magnetic 
operator $O_8$ can be seen.
Note that $O_8$ does not contribute to the decay \bxsll in the 
approximation which we use here. The $C_i(\mu)$ are the Wilson coefficients,
which depend, in general, on the renormalization scale $\mu$,
except for $C_{10}$. 

The matrix element for the decay \bxsll can be factorized 
into a leptonic and a hadronic part as
\begin{equation}
        {\cal M (\mbox{\bxsll})} =
        \frac{G_F \alpha}{\sqrt{2} \pi} \, V_{ts}^\ast V_{tb} \, 
        \left( {\Gamma^L}_\mu \, {L^L}^\mu 
        +  {\Gamma^R}_\mu \, {L^R}^\mu \right) \, ,
\end{equation}
with
\begin{eqnarray}
        {L^{L/R}}_\mu & \equiv & 
                \bar{l} \, \gamma_\mu \, L(R) \, l \, , \\
        {\Gamma^{L/R}}_\mu & \equiv & 
                \bar{s} \left[ 
                R \, \gamma_\mu 
                        \left( C_9^{\mbox{eff}}(\s) \mp C_{10} 
                          + 2 C_7^{\mbox{eff}} \, 
                        \frac{\hat{\not{q}}}{\s} \right)
                + 2 \hat{m}_s \, C_7^{\mbox{eff}} \, \gamma_\mu \, 
                        \frac{\hat{\not{q}}}{\s} L 
                \right] b \, .  
        \label{eqn:gammai}
\end{eqnarray}
The effective Wilson coefficient $C_9^{\mbox{eff}}(\s)$ receives 
contributions from various pieces. The
resonant $c\bar{c}$ states also contribute to $C_9^{\mbox{eff}}(\s)$; hence
the contribution given below is just the perturbative part:
\begin{eqnarray}
C_9^{\mbox{eff}}(\s)=C_9 \eta(\s) + Y(\s) \, .
\end{eqnarray}
Here $\eta(\s)$ and $Y(\s)$ represent the ${\cal{O}}(\alpha_s)$ correction 
\cite{jezkuhn}
and the one loop matrix element of the 
Four-Fermi operators \cite{burasmuenz,misiakE}, respectively. 
While $C_9$ is a renormalization scheme-dependent quantity, this
dependence cancels out with the corresponding one in the function $Y(\s)$
(the value of $\xi$, see below).
To be self-contained, we list  
the two functions in $C_9^{\mbox{eff}}(\s)$:
\begin{eqnarray}
\label{Ypert}
        Y(\s) & = & g(\mc,\s)
                \left(3 \, C_1 + C_2 + 3 \, C_3
                + C_4 + 3 \, C_5 + C_6 \right)
\nonumber \\
        & & - \frac{1}{2} g(1,\s)
                \left( 4 \, C_3 + 4 \, C_4 + 3 \,
                C_5 + C_6 \right) 
         - \frac{1}{2} g(0,\s) \left( C_3 +   
                3 \, C_4 \right) \nonumber \\
        & &     + \frac{2}{9} \left( 3 \, C_3 + C_4 +
                3 \, C_5 + C_6 \right) 
             - \xi \, \frac{4}{9} \left( 3 \, C_1 +
                C_2 - C_3 - 3 \, C_4 \right),
                \label{eqn:y} \\
        \eta(\s) & = & 1 + \frac{\alpha_s(\mu)}{\pi}
                \omega(\s) ~,
\end{eqnarray}
 \begin{equation}
        \xi = \left\{
                \begin{array}{ll}
                        0       & \mbox{(NDR)}, \\
                        -1      & \mbox{(HV)},
                \end{array}
                \right.
\end{equation}

\begin{eqnarray}
\label{gpert}
g(z,\hat{s}) &=& -\frac{8}{9}\ln (\frac{m_b}{\mu})
 -\frac{8}{9} \ln z + \frac{8}{27} +\frac{4}{9}y
-\frac{2}{9}(2 + y) \sqrt{\vert 1-y \vert}\nonumber\\
&\times & \left[\Theta(1-y)(\ln\frac{1+\sqrt{1-y}}{1-\sqrt{1-y}} -i\pi )
+\Theta(y-1) 2 \arctan \frac{1}{\sqrt{y-1}} \right] ~, \\
g(0,\hat{s})& =& \frac{8}{27}-\frac{8}{9}\ln (\frac{m_b}{\mu})
              -\frac{4}{9}\ln \hat{s} + \frac{4}{9}i\pi ~,
\end{eqnarray}
where $y=4z^2/\hat{s}$, and
\begin{eqnarray}
\omega(\hat{s}) &=& -\frac{2}{9}\pi^2 -\frac{4}{3}{\mbox Li}_2(\hat{s})
-\frac{2}{3}
\ln \hat{s} \ln(1-\hat{s}) -
\frac{5+4\hat{s}}{3(1+2\hat{s})}\ln(1-\hat{s})\nonumber\\
&-& \frac{2\hat{s}(1+\hat{s})(1-2\hat{s})}{3(1-\hat{s})^2(1+2\hat{s})}
\ln \hat{s} + \frac{5 + 9\hat{s} -6\hat{s}^2}{6(1-\hat{s})(1+2 \hat{s})}~.
\label{omegahats}
\end{eqnarray}
Above, (NDR) and (HV) correspond to the naive dimensional regularization
and the 't Hooft-Veltman schemes, respectively.
The one gluon correction to $O_9$ with respect to
$x_0$ will be presented below in eq.~(\ref{c9eff}).
The Wilson coefficients in leading logarithmic approximation 
can be seen in \cite{burasmuenz}.

With the help of the above expressions, the differential
decay width becomes on using $p_{\pm}=(E_{\pm}, \mbox{\boldmath $p_{\pm}$})$,
\begin{equation}
        {\rm d} \Gamma = \frac{1}{2 m_B} 
                \frac{{G_F}^2 \, \alpha^2}{2 \pi^2} 
                \left| V_{ts}^\ast V_{tb} \right|^2 
                \frac{{\rm d}^3 \mbox{\boldmath $p_+$}}{(2 \pi)^3 2 E_+} 
                \frac{{\rm d}^3 \mbox{\boldmath $p_-$}}{(2 \pi)^3 2 E_-} 
                \left( {W^L}_{\mu \nu} \, {L^L}^{\mu \nu} 
                +  {W^R}_{\mu \nu} \, {L^R}^{\mu \nu} \right) \, ,
\end{equation}
where $W_{\mu \nu}^{L,R}$ and $L_{\mu \nu}^{L,R}$ are the
hadronic and leptonic tensors, respectively.
The hadronic tensor $W_{\mu\nu}^{L/R}$
is related to the discontinuity in the forward scattering amplitude, 
denoted by
$T_{\mu \nu}^{L/R}$, through the relation $W_{\mu \nu} = 2 \, {\rm Im} \, 
T_{\mu \nu}$.  Transforming the integration variables 
 to $\hat{s}$, $\hat{u}$ and $v \cdot \hat{q}$, one can express the 
Dalitz distribution in \bxsll as:  
\begin{equation}
        \frac{{\rm d} \Gamma}{{\rm d}\u \, {\rm d}\s \, {\rm d}(\z)} = 
                \frac{1}{2 \, m_B}
                \frac{{G_F}^2 \, \alpha^2}{2 \, \pi^2} 
                \frac{{m_b}^4}{256 \, \pi^4}
                \left| V_{ts}^\ast V_{tb} \right|^2 
                \, 2 \, {\rm Im} 
                \left( {T^L}_{\mu \nu} \, {L^L}^{\mu \nu}
                +  {T^R}_{\mu \nu} \, {L^R}^{\mu \nu} \right) \, ,
        \label{eqn:dgds}
\end{equation}
with
\begin{eqnarray}
\label{eq:hadtensor}
        {T^{L/R}}_{\mu \nu} & = & 
        i \, \int {\rm d}^4 y \, e^{-i \, \hat{q} \cdot y}      
        \left< B \left| {\rm T} \left\{ 
                {{\Gamma_1}^{L/R}_\mu} (y), 
                {\Gamma_2}^{L/R}_\nu (0) \right\} \right| B \right>\, , \\
        {L^{L/R}}^{\mu \nu} & =&
         2 \left[ {p_+}^\mu \, {p_-}^\nu + {p_-}^\mu \, {p_+}^\nu 
                - g^{\mu \nu} (p_+ \cdot p_-) 
                \mp i \epsilon^{\mu \nu \alpha \beta} \, 
                        {p_+}_\alpha \, {p_-}_\beta \right] \, , 
\end{eqnarray}
where ${{\Gamma_1}^{L/R}_\mu}^\dagger = {\Gamma_2}^{L/R}_\mu = 
\Gamma^{L/R}_\mu $, and is given in eq. (\ref{eqn:gammai}).
Using Lorentz decomposition, the tensor $T_{\mu \nu}$ can be expanded in 
terms of three structure functions $T_i$,
\begin{equation}
        T_{\mu \nu} = -T_1 \, g_{\mu \nu} + T_2 \, v_\mu \, v_\nu 
                + T_3 \, i \epsilon_{\mu \nu \alpha \beta} \, 
                        v^\alpha \, \hat{q}^\beta \, ,
\label{eq:hadrontensor}
\end{equation}
where the structure functions which do not contribute to the
amplitude in the limit of massless leptons have been neglected.
The problem remaining is now to determine the $T_i$, to which 
we shall return in section \ref{sectionhqet}.
\begin{table}[h]
        \begin{center}
        \begin{tabular}{|l|l|}
        \hline
        \multicolumn{1}{|c|}{Parameter} & 
                \multicolumn{1}{|c|}{Value}     \\
        \hline \hline
        $m_W$                   & $80.26$ (GeV) \\
        $m_Z$                   & $91.19$ (GeV) \\
        $\sin^2 \theta_W $      & $0.2325$ \\
        $m_s$                   & $0.2$ (GeV)   \\
        $m_c$                   & $1.4$ (GeV) \\
        $m_b$                   & $4.8$ (GeV) \\
        $m_t$                   & $175 \pm 5$ (GeV)     \\
        $\mu$                   & ${m_{b}}^{+m_{b}}_{-m_{b}/2}$        \\
        $\alpha^{-1}$     & 129           \\
        $\alpha_s (m_Z) $       & $0.117 \pm 0.005$ \\
        ${\cal B}_{sl}$         & $(10.4 \pm 0.4)$ \%   \\
        \hline
        \end{tabular}
        \end{center}
\caption{\it Default values of the input parameters and errors used in the 
numerical calculations.}
\label{parameters}
\end{table}
\begin{table}[h]
        \begin{center}
        \begin{tabular}{|c|c|c|c|c|c|c|c|c|c|}
        \hline
        \multicolumn{1}{|c|}{ $C_1$}       & 
        \multicolumn{1}{|c|}{ $C_2$}       &
        \multicolumn{1}{|c|}{ $C_3$}       & 
        \multicolumn{1}{|c|}{ $C_4$}       &
        \multicolumn{1}{|c|}{ $C_5$}       & 
        \multicolumn{1}{|c|}{ $C_6$}       &
        \multicolumn{1}{|c|}{ $C_7^{\mbox{eff}}$}       & 
        \multicolumn{1}{|c|}{ $C_9$}       &
                \multicolumn{1}{|c|}{$C_{10}$} &
 \multicolumn{1}{|c|}{ $C^{(0)}$ }     \\
        \hline 
        $-0.240$ & $+1.103$ & $+0.011$ & $-0.025$ & $+0.007$ & $-0.030$ &
   $-0.311$ &   $+4.153$ &    $-4.546$    & $+0.381$     \\
        \hline
        \end{tabular}
        \end{center}
\caption{ \it Values of the Wilson coefficients used in the numerical
          calculations corresponding to the central values 
          of the parameters given in Table \protect\ref{parameters}.
Here, $C_7^{\mbox{eff}} \equiv C_7 -C_5/3 -C_6$, and for $C_9$ we use the 
NDR scheme.} \label{wilson}
\end{table}

\section{Perturbative QCD Corrections in $O(\alpha_s)$ in the Decay \bxsll}

\begin{figure}[htb]
\vskip -0.4truein
\centerline{\epsfysize=7in
{\epsffile{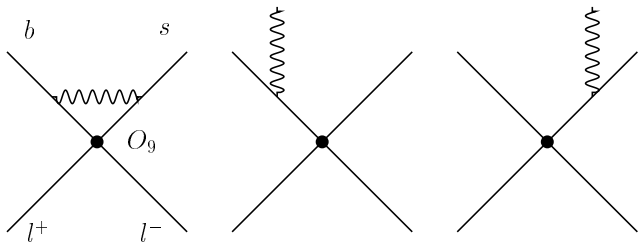}}}
\vskip -5.0truein
\caption[]{ \it Feynman diagrams contributing to the explicit
order $\alpha_s$ corrections of the operator $O_9$. 
Curly lines denote a gluon. Wave function corrections are not shown.}
\label{fig:o9}
\end{figure}

In this section the $O(\alpha_s)$ corrections to the hadron spectra 
are investigated. 
Only $O_9$ is subject to $\alpha_s$ corrections and the renormalization 
group improved perturbation series for $C_9$ is 
${\cal{O}}(1/\alpha_s)+{\cal{O}}(1)+{\cal{O}}(\alpha_s)+ \dots$,
 due to the 
large logarithm in $C_9$ represented by ${\cal{O}}(1/\alpha_s)$
\cite{burasmuenz}.
The Feynman diagrams, which contribute to the matrix element of $O_9$ in 
$O(\alpha_s)$, corresponding to the virtual one-gluon and 
bremsstrahlung corrections, are shown in Fig.~\ref{fig:o9}.
The effect of a finite $s$-quark mass on the 
${\cal O}(\alpha_s)$ correction function is found to be very 
small. After showing this, we 
have neglected the $s$-quark mass in the numerical calculations of the
${\cal{O}}(\alpha_s)$ terms.

\subsection{Hadron energy spectrum}

\begin{figure}[htb]
\vskip -0.2truein
\centerline{\epsfysize=3.5in
{\epsffile{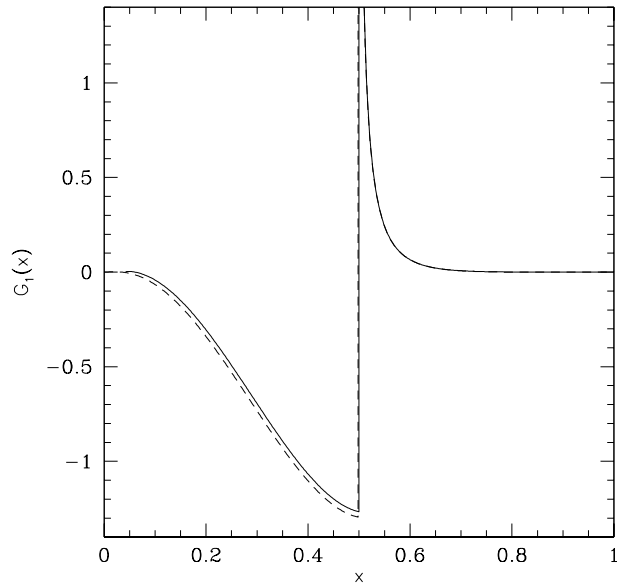}}}
\vskip -0.4truein
\caption[]{ \it The function $G_1(x)$ is shown for 
$m_s=0.2 \, {\mbox{\rm GeV}}$ (solid line)
and for the massless case corresponding to eq.~(\ref{eq:g1}) (dashed line).}
\label{fig:g1}
\end{figure}

The explicit order $\alpha_s$ correction to $O_9$ can be obtained
by using the existing results in the literature as follows:
The vector current $O_9$ can be decomposed as
$V=(V-A)/2 + (V+A)/2$.
We recall that the $(V-A)$ and $(V+A)$ currents yield the same  
hadron energy spectrum \cite{aliold}
and there is no interference term  present in this spectrum for massless 
leptons. So, the correction for the vector current case in \bxsll  
can be taken from the corresponding result for the charged $(V-A)$ case
\cite{aliqcd,jezkuhn}, yielding
\begin{eqnarray}
C_9^{{\mbox{eff}}}(x_0)=C_9 \rho(x_0) + Y(x_0)
\label{c9eff}
\end{eqnarray}
with
\begin{eqnarray}
\rho(x)&=& 1+ \frac{\alpha_s}{\pi} \sigma(x) , \\
\sigma(x)&=&
\frac{1}{(3 x -4 x^2-2 \ms^2+3 \ms^2 x)} \frac{G_1(x)}{3 \sqrt{x^2-\ms^2}} \; ,
\end{eqnarray}
where
$Y(x_0)\equiv Y(\hat{s})$ with $\hat{s}=1-2 x_0 +\ms^2$.
The expression for $G_1(x)$ with  $m_s\neq 0$ has been calculated in 
\cite{jezkuhn}.
The effect of a finite $m_s$ is negligible in $G_1(x)$,
as can be seen in Fig.~\ref{fig:g1}, where this function
is plotted both with a 
finite $s$-quark mass, $m_s=0.2$ GeV, and for the  massless case, $m_s=0$.
A numerical difference occurs at the lowest order
end point $x_0^{max}=1/2 (1+\ms^2)$ (for $m_l=0$), where the function 
develops a 
singularity from above ($x_0 >x_0^{max}$) and the position of which 
depends on the value of $m_s$.        
The function $G_1(x)$ for a massless $s$-quark  is given and 
discussed below \cite{jezkuhn}.
 \begin{eqnarray}
G_1(x)&=& x^2 \{ \frac{1}{90} (16 x^4 -84 x^3 +585 x^2-1860 x+1215) +
(8 x-9) \ln(2 x) \nonumber \\
&+& 2 (4 x-3) \left[ \frac{\pi^2}{2} +Li_2(1-2 x) \right] \} \, \, \, \, 
{\mbox{for}} \, \,  0 \leq x \leq 1/2 \nonumber \, \, , \\
G_1(x)&=&\frac{1}{180} (1-x)(32 x^5-136 x^4+1034 x^3-2946 x^2+1899 x+312) 
\nonumber \\
&-&\frac{1}{24} \ln(2 x-1) ( 64 x^3-48 x^2-24 x-5) \nonumber \\
&+& x^2 (3 -4 x) \left[ \frac{\pi^2}{3}-4 Li_2(\frac{1}{2 x}) +\ln^2(2 x-1)  -2
\ln^2(2 x) \right] \, \, \, {\mbox{for}} \, \, 1/2 < x \leq 1 \, \, ,
\label{eq:g1}
\end{eqnarray}
where $Li_2(z)$ is the dilogarithmic function. 

The ${\cal{O}}(\alpha_s)$ correction has a double logarithmic 
(integrable) singularity for $x_0 \to 1/2$ from above ($x_0 >1/2$).
Further, the value of the order $\alpha_s$ corrected Wilson coefficient
$C_9^{{\mbox{eff}}}(x_0)$ is reduced compared to its value 
with  $\alpha_s=0$,
therefore also the hadron energy spectrum is 
reduced after including the explicit order $\alpha_s$ QCD correction
for $0 < x_0 <1/2$.
Note that the hadron energy spectrum for \bxsll  
receives contributions for $1 \geq x > 1/2 $ only from the order 
$\alpha_s$ bremsstrahlung corrections.

\subsection{Hadronic invariant mass spectrum\label{sec:qcdmass}}

\begin{figure}[htb]
\vskip -0.0truein
\centerline{\epsfysize=3.5in
{\epsffile{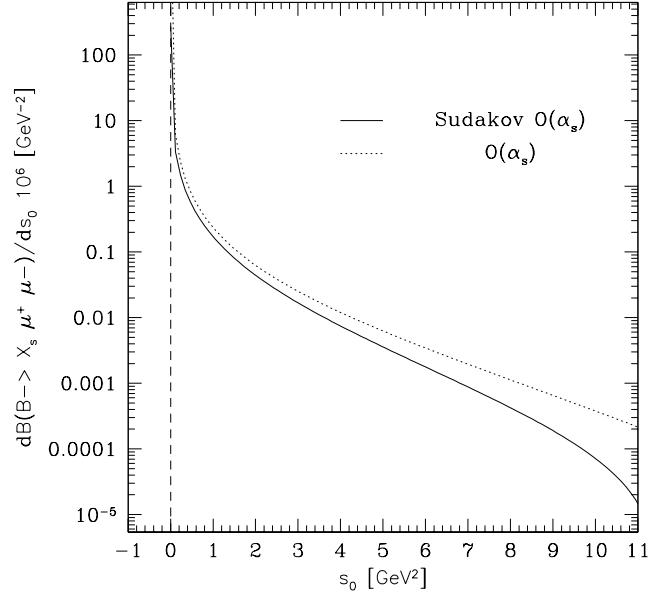}}}
\vskip -0.0truein
\caption[]{ \it The differential branching ratio 
$ \frac{{\rm d}{\cal B}(B \to X_s \ell^+ \ell^-)}{{\rm d} s_0}$ 
in the parton model 
is shown in the ${\cal{O}}(\alpha_s)$ bremsstrahlung region.
The dotted (solid) line corresponds to eq.~(\ref{dbds0}), (eq.~(\ref{eq:sud})).
The vertical line denotes the one particle pole from $b \to s \ell^+ 
\ell^-$. 
We do not show the full spectra in the range $0 \leq s_0 \leq m_b^2$ as 
they tend to zero for larger values of $s_0$.}
\label{fig:sh0}
\end{figure}
\begin{figure}[htb]
\vskip -0.0truein
\centerline{\epsfysize=3.5in
{\epsffile{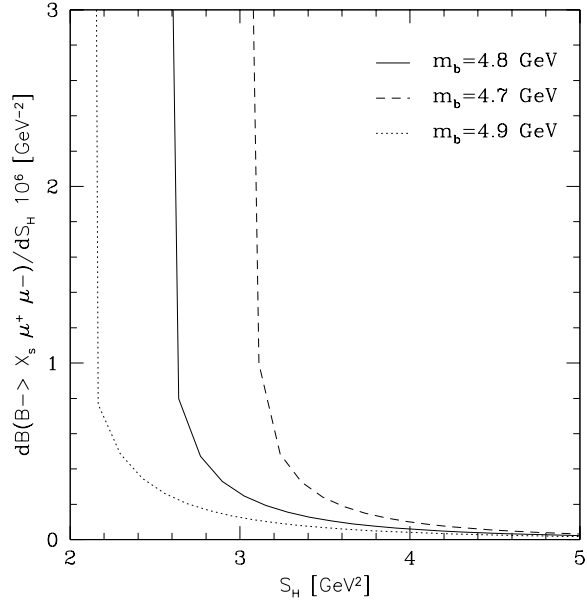}}}
\vskip -0.0truein
\caption[]{ \it The differential branching ratio 
$ \frac{{\rm d}{\cal B}(B \to X_s \ell^+ \ell^-)}{{\rm d} S_H}$  
in the hadronic invariant mass, $S_H$, shown for different values of 
$m_b$ in the range where only bremsstrahlung diagrams 
contribute. We do not show the result in the full kinematic range 
as the spectra tend monotonically to zero for larger values of $S_H 
\leq m_B^2$.} \label{fig:sH}
\end{figure}
We have calculated the  order $\alpha_s$  perturbative QCD correction
for the hadronic invariant mass in the range
$\ms^2 < \s_0 \leq 1 $. Since the decay $b \to s + \ell^+ + \ell^-$
contributes in the parton model only at $\s_0 =\ms^2$,  only 
the bremsstrahlung graphs $b \to s + g + \ell^+ + \ell^-$ contribute in this 
range.
This makes the calculation much simpler than in the full $\s_0$ range 
including virtual gluon diagrams. 
We find 
\begin{eqnarray}
\frac{{\rm d}{\cal{B}}}{{\rm d } \s_0}=\frac{2}{3} {\cal{B}}_0
{\alpha_s\over \pi}{1\over\s_0}
  \{ \frac{(\s_0-1)}{27} (93-41\hat s_0-95 \hat s_0^2+55 \s_0^3)
+ {4\over 9}\ln\hat s_0 (-3-5\hat s_0+9 \hat s_0^2-2 \hat s_0^4) 
\} C_9^2 \, \, .
\label{dbds0}
\end{eqnarray}
Our result for the spectrum in \bxsll is in agreement with the corresponding 
result for the $(V-A)$ 
current obtained for the decay $B \to X_q \ell \nu_{\ell}$ in the $m_q=0$ 
limit in ref.~\cite{FLS} (their eq.~(3.8)), once one takes into account 
the difference in the  normalizations.
We display the hadronic invariant mass distribution  in Fig.~\ref{fig:sh0} 
as a function of $s_0$ (with 
$s_0=m_b^2 \s_0$), where we also show the Sudakov improved spectrum,
obtained from the ${\cal{O}}(\alpha_s)$ spectrum 
in which the double logarithms have been resummed. For the decay 
$B \to X_u \ell \nu_{\ell}$, this  has been derived in 
ref.~\cite{greubrey}, where all further details can be seen.
We confirm eq.~(17) of ref.~\cite{greubrey} for the Sudakov exponentiated 
double differential 
decay rate $\frac{{\rm d^2}{\Gamma}}{{\rm d } x {\rm d } y}$  
and use it after changing the normalization 
$\Gamma_0 \to {\cal{B}}_0 \frac{2}{3} C_9^2$ for the decay
\bxsll . The constant ${\cal{B}}_0$ is given later. Defining the
 kinematic variables $(x,y)$ as
\begin{eqnarray}
q^2 &=& x^2 m_b^2 \, , \nonumber \\
v \cdot q &=& (x+\frac{1}{2} (1-x)^2 y) m_b \, ,
\end{eqnarray}
the Sudakov-improved Dalitz distribution  is given by
\begin{eqnarray}
\label{doubleexpon}
{d^2 {\cal{B}} \over d x d y}(B \to X_s \ell^+ \ell^-) &=& - {\cal{B}}_0 
 \frac{8}{3} x (1 - x^2)^2 (1 + 2 x^2) \,
\exp\Big( - {2 \alpha_s \over 3 \pi} \ln^2 (1 - y) \Big)
  \\
&\times &  \left\{
{4 \alpha_s \over 3 \pi} {\ln(1-y) \over (1-y)}
\Big[ 1 - {2 \alpha_s \over 3 \pi} \big( G(x) + H(y) \big) \Big]
-{2 \alpha_s \over 3 \pi} {d H \over d y}(y) \right\}  C_9^2
 \, ,\nonumber
\end{eqnarray}
where \cite{greubrey}
\begin{eqnarray}
G(x) &=&
\frac{[8x^2(1-x^2 -2x^4) \ln x
+ 2(1-x^2)^2 (5 + 4 x^2) \ln(1-x^2) -
(1-x^2)(5 + 9x^2 - 6 x^4) ]}{2 (1-x^2)^2 (1 + 2x^2)}
\nonumber \\
&& + \pi^2 + 2 Li_2 (x^2) - 2 Li_2 (1-x^2) \quad \, , \\
H(y)& = & \int_0^{y} dz \Big( {4 \over 1 - z}
\ln{2 - z(1-x) + \kappa
\over 2}          \nonumber \\
&&- {(1-x)(3 + x + xz - z) \over (1+x)^2}
\Big[\ln(1-z) - 2 \ln{2 - z(1-x) + \kappa \over 2} \Big]
\nonumber \\
&&- { \kappa \over 2 (1 + x)^2 (1 + 2x^2) }
\Big[{7 (1+x) (1 + 2 x^2) \over 1 - z} + (1-x)(3 - 2 x^2) \Big]\Big) \, .
\label{h0}
\end{eqnarray}
The quantity $\kappa$ in eq. (\ref{h0}) is defined as
$\kappa \equiv \sqrt {z^2 (1-x)^2 + 4 xz}$.

To get the hadronic invariant mass spectrum for a 
$b$ quark decaying at rest we change variables from $(x,y)$
to $(q^2,s_0)$ 
followed by an integration over $q^2$,
\begin{eqnarray}
\frac{{\rm d}{\cal{B}}}{{\rm d } s_0}=\int_{4 m_l^2}^{(m_b-\sqrt{s_0})^2}
{\rm d} q^2 \frac{{\rm d^2}{\cal{B}}}{{\rm d } x {\rm d } y} 
\frac{1}{2 m_b^4 x (1-x)^2} \, \, .
\label{eq:sud}
\end{eqnarray}

The most significant effect of the bound state is the difference between
$m_B$ and $m_b$,  
which is dominated by $\bar{\Lambda}$. 
Neglecting $\lambda_1, \, \lambda_2$, i.e., using
$\bar{\Lambda}= m_B-m_b$, 
the spectrum $\frac{{\rm d}{\cal{B}}}{{\rm d } S_H}$
is obtained along the lines as given above for $\frac{d{\cal{B}}}{ds_0}$, 
after changing variables
from $(x,y)$ to $(q^2,S_H)$ and performing an integration over $q^2$.
It is valid in the region
$m_B \frac{m_B\bar{\Lambda}-\bar{\Lambda}^2+m_s^2}{m_B-\bar{\Lambda}}
< S_H \leq m_B^2 $ (or $m_B \bar{\Lambda} \leq S_H \leq m_B^2$, neglecting
$m_s$) which excludes the zeroth order and virtual gluon
 kinematics ($s_0=m_s^2$), yielding
\begin{eqnarray}
\frac{{\rm d}{\cal{B}}}{{\rm d } S_H}=\int_{4 m_l^2}^{(m_B-\sqrt{S_H})^2}
{\rm d} q^2 \frac{{\rm d^2}{\cal{B}}}{{\rm d } x {\rm d } y} 
\frac{1}{2 m_b^3 m_B x (1-x)^2}\, \, .
\label{eq:sudSH}
\end{eqnarray}
The hadronic invariant mass spectrum thus found depends rather
sensitively on $m_b$ (or equivalently $\bar{\Lambda}$),
as can be seen from Fig.~\ref{fig:sH}.
An analogous analysis for the charged current semileptonic $B$ decays
$B \to X_u \ell \nu_\ell$  has been performed in ref.~\cite{FLW},
with similar conclusions.

\section{Power Corrections in the Decay \bxsll \label{sectionhqet}}

The hadronic tensor in eq.~(\ref{eq:hadrontensor}) can be expanded in 
inverse powers of $m_b$ with the help of the 
HQET techniques.
The leading term in this expansion, i.e., ${\cal O}(m_b^0)$ reproduces the 
parton model result. In HQET, the next to leading power corrections are
parameterized in terms of the matrix elements of the kinetic energy and 
the magnetic moment operators $\lambda_1$ and $\lambda_2$, respectively. 
The $B-B^{\ast}$ mass difference yields the value  
$\lambda_2=0.12 ~\mbox{GeV}^2$. In all numerical estimates we shall 
use this value of $\lambda_2$ and, unless otherwise 
stated, we take the value for $\lambda_1$ extracted from an analysis of
data on semileptonic B-decays ($B \to X \ell \nu_{\ell}$), yielding 
$\lambda_1=-0.20 ~\mbox{GeV}^2$ with 
a corresponding value $\bar{\Lambda}=0.39 \, 
\mbox{GeV}$ \cite{gremm}.
For a review on the dispersion in the present values of these 
non-perturbative parameters, see \cite{neubert}.

The contributions of the power corrections
to the structure functions $T_i$ can be decomposed into the sum of
various terms, denoted by $T_{i}^{(j)}$, which can be traced back to well 
defined pieces in the evaluation of the time-ordered product 
in eq.~(\ref{eq:hadtensor}):
\begin{equation}
T_{i}(v.\hat{q},\s) = \sum_{j=0,1,2,s,g,\delta} T_{i}^{(j)}(v.\hat{q}, \s)\,.
\label{Tijhqet}
\end{equation}
The expressions for $T_{i}^{(j)}(v.\hat{q},\s)$, $i=1,2,3$ calculated up to 
$O(m_B/m_b^3)$ are given in \cite{AHHM97}.
After contracting the hadronic and leptonic tensors, one finds
\begin{equation}
        {T^{L/R}}_{\mu \nu} \, {L^{L/R}}^{\mu \nu} = 
                {m_b}^2 \left\{ 2 \, \s \, {T_1}^{L/R} 
                + \left[ (\z)^2 - \frac{1}{4} \u^2 - \s \right] {T_2}^{L/R} 
                \mp \s \, \u \, {T_3}^{L/R} \right\} \, . 
        \label{eqn:tlr}
\end{equation}
With the help of the kinematic identities given in eq.~(\ref{eq:kin}),
we can make the dependence on $x_0$ and $\s_0$ explicit,
\begin{equation}
        {T^{L/R}}_{\mu \nu} \, {L^{L/R}}^{\mu \nu} = 
                {m_b}^2 \left\{ 2  (1-2 x_0+\s_0)  {T_1}^{L/R} 
                + \left[ x_0^2 - \frac{1}{4} \u^2 - \s_0 \right] {T_2}^{L/R} 
                \mp  (1-2 x_0+\s_0) \u \, {T_3}^{L/R} \right\} 
        \label{eqn:tlr2}
\end{equation}
and with this we are able to derive the double differential power corrected 
spectrum $\frac{{\rm d} {\cal{B}}}{{\rm d} x_0 \, {\rm d}\s_0}$ for
\bxsll. 
Integrating eq.~(\ref{eqn:dgds}) over $\u$ first, where 
the variable $\u$ is bounded by 
\begin{eqnarray}
-2 \sqrt{x_0^2-\s_0} \leq \u \leq +2\sqrt{x_0^2-\s_0} \, \, ,
\end{eqnarray}
we arrive at the following expression 
\begin{eqnarray}
\frac{{\rm d}^2 {\cal{B}}}{{\rm d} x_0 \, {\rm d}\s_0}& =&-\frac{8}{\pi} 
{\cal B}_{0}
{\mbox{Im}}\sqrt{x_0^2-\s_0}
\left\{ (1-2 x_0+\s_0)T_1(\s_0,x_0)+\frac{x_0^2-\s_0}{3}T_2(\s_0,x_0) \right\}
+ {\cal{O}}(\lambda_i \alpha_s)
\label{doublediff} \, ,
\end{eqnarray}
where 
\begin{eqnarray}
T_1(\s_0,x_0)&=&\frac{1}{x} \left\{ 
\left( 8 x_0-4  (\frac{\lo}{3}+\lt) \right) \left( |C_9^{\mbox{eff}}(\s)|^2 
+ |C_{10}|^2 \right)
\right. \nonumber \\
&+&
 \left(
32 (-2 \ms^2-2 \s_0-4 \ms^2 \s_0 +x_0+5 \ms^2 x_0+\s_0 x_0+\ms^2 \s_0 x_0) 
+ 16 (\frac{\lo}{3}+\lt) \right. \nonumber \\
&\times& \left. (-5-11 \ms^2+5 \s_0-\ms^2 \s_0+10 x_0 + 22 \ms^2 x_0-10 
x_0^2-10 \ms^2 x_0^2) \right) \frac{|C_7^{\mbox{eff}}|^2}{(\s_0-2 x_0 +1)^2} 
\nonumber \\
&+& \left.
 \left( 
\frac{-32}{\s_0-2 x_0+1} (\ms^2+\s_0-x_0-\ms^2 x_0)-48(\frac{\lo}{3}+\lt) \right) 
Re(C_9^{\mbox{eff}}(\s)) \, C_7^{\mbox{eff}}
\right\} \nonumber \\
&+&
\frac{1}{x^2}  \left\{ 
 \left( \frac{8 \lo}{3} (-2 \s_0-3 x_0+5 x_0^2)+8 \lt (-2 \s_0+x_0+5 x_0^2)
\right)
\left( |C_9^{\mbox{eff}}(\s)|^2 + |C_{10}|^2 \right) \right.
 \nonumber \\
&+&
\left( 
\frac{32 \lo}{3} (6 \ms^2+ 12 \s_0+ 18 \ms^2 \s_0-2 \s_0^2-2 \ms^2 \s_0^2-3 x_0-21 \ms^2 x_0-13 \s_0 x_0-19 \ms^2 \s_0 x_0 
\right.
\nonumber \\
&-& 3 x_0^2+9 \ms^2 x_0^2+ 5 \s_0 x_0^2+ 5 \ms^2 \s_0 x_0^2+ 4 x_0^3+4 \ms^2 x_0^3)
\nonumber \\
&+& 32 \lt(-2 \ms^2-2 \ms^2 \s_0-2 \s_0^2-2 \ms^2 \s_0^2+x_0-\ms^2 x_0-5 \s_0 x_0-11 \ms^2 \s_0 x_0 + x_0^2 
\nonumber \\
&+& \left. 13 \ms^2 x_0^2 + 5 \s_0 x_0^2+ 5 \ms^2 \s_0 x_0^2)\right)
\frac{|C_7^{\mbox{eff}}|^2}{(\s_0-2 x_0 +1)^2} 
 \nonumber \\
&+&
\left( 
\frac{-32 \lo}{3} (-3 \ms^2-5 \s_0+2 \ms^2 \s_0+ 3 x_0+ 6 \ms^2 x_0+ 3 \s_0 x_0-x_0^2-5 \ms^2 x_0^2) \right.
\nonumber \\
&-& \left.\left.
32 \lt(\ms^2+\s_0+ 2 \ms^2 \s_0-x_0+2 \ms^2 x_0 + 3 \s_0 x_0 -3 x_0^2-5 \ms^2 x_0^2) \right)
\frac{Re(C_9^{\mbox{eff}}(\s)) \, C_7^{\mbox{eff}}}{\s_0-2 x_0+1}
\right\} \nonumber \\
&+&\frac{1}{x^3}  \lo (\s_0-x_0^2)\left\{ 
 \frac{32 x_0}{3} \left( |C_9^{\mbox{eff}}(\s)|^2 + |C_{10}|^2 \right) \right.
\nonumber \\
&+&
 \frac{128}{3} (-2 \ms^2 -2 \s_0-4 \ms^2 \s_0+x_0+5 \ms^2 x_0+ \s_0 x_0+\ms^2 \s_0 x_0) \frac{|C_7^{\mbox{eff}}|^2}{(\s_0-2 x_0 +1)^2}  \nonumber \\
&+& \left.
\frac{-128}{3} (\ms^2+\s_0-x_0-\ms^2 x_0)
 \frac{Re(C_9^{\mbox{eff}}(\s)) \, C_7^{\mbox{eff}}}{\s_0-2 x_0+1}
\right\}  \, \, ,\nonumber \\
T_2(\s_0,x_0)&=&\frac{1}{x} \left\{ 
 \left( 16-40 (\frac{\lo}{3}+\lt) \right)
\left( |C_9^{\mbox{eff}}(\s)|^2 + |C_{10}|^2 \right)+
 \left( -64+160(\frac{\lo}{3}+\lt) \right) (1+\ms^2) 
\frac{|C_7^{\mbox{eff}}|^2 }{\s_0-2 x_0+1}
\right\}\nonumber \\
&+&\frac{1}{x^2} \left\{ 
 \left( \frac{112 \lo}{3} (-1+x_0)+ 16 \lt (-3+5 x_0) \right)
\left( |C_9^{\mbox{eff}}(\s)|^2 + |C_{10}|^2 \right) \right.
 \nonumber \\
&+& \left.
 \left( \frac{448 \lo}{3} (1-x_0)+ 64 \lt (5 x_0-1)\right) (1+\ms^2) 
\frac{|C_7^{\mbox{eff}}|^2 }{\s_0-2 x_0+1} 
 -64 \lt Re(C_9^{\mbox{eff}}(\s)) \, C_7^{\mbox{eff}}
\right\} \nonumber \\
&+&\frac{1}{x^3} \lo (\s_0-x_0^2)\left\{ 
 \frac{64}{3} \left( |C_9^{\mbox{eff}}(\s)|^2 + |C_{10}|^2 \right)+
 \frac{-256}{3} (1+\ms^2) \frac{|C_7^{\mbox{eff}}|^2}{\s_0-2 x_0+1}
\right\} \; .
\end{eqnarray}
Here, $x=\s_0 -\ms^2 +i \epsilon$,
$\lo=\lambda_{1}/m_{b}^2$ and $\lt=\lambda_{2}/m_{b}^2$.
As the structure function $T_3$ does not contribute to the branching ratio, 
we did not consider it in our present work.
The Wilson coefficient 
$C_9^{\mbox{eff}}(\s)$ depends both on the variables
$x_0$ and $\s_0$ arising from the matrix element of the Four-Fermi-operators.

The branching ratio for \bxsll is usually expressed in terms
of the measured semileptonic branching ratio ${\cal B}_{sl}$
for the decays $B \to X_c \ell \nu_\ell$. This fixes
the normalization constant ${\cal B}_0$ to be,
\begin{equation}
        {\cal B}_0 \equiv
                {\cal B}_{sl} \frac{3 \, \alpha^2}{16 \pi^2} \frac{
    {\vert V_{ts}^* V_{tb}\vert}^2}{\absvcb^2} \frac{1}{f(\mc) \kappa(\mc)}
                \; ,
\label{eqn:seminorm}
\end{equation}
where
\begin{equation}
        f(\mc) = 1 - 8 \, \mc^2 + 8 \, \mc^6 - \mc^8 - 24 \, \mc^4 \, \ln \mc
        \label{eqn:fr}
\end{equation}
is the phase space factor for $\Gamma (B \rightarrow X_c \ell \nu_{\ell})$
and
the function $\kappa(\mc)$ accounts for both the $O(\alpha_s)$ QCD 
correction to 
the semileptonic decay  width \cite{CM78} and the leading order
$(1/m_b)^2$ power correction \cite{georgi}. 
It reads as:
\begin{equation}
\kappa(\mc) = 1 + \frac{\alpha_s(m_b)}{\pi} g(\mc)
    + \frac{h(\mc)}{2 m_b^2} \; ,
\end{equation}
where 
\begin{eqnarray}
g(\mc)&=& \frac{A_0(\mc)}{f(\mc)} \; , \\
h(\mc) &=& \lambda_1 + \frac{\lambda_2}{f(\mc)} \left[ -9 +24 \mc^2
-72\mc^4 + 72\mc^6 -15\mc^8 -72 \mc^4 \ln \mc \right]\; , 
\label{eqn:ghr}
\end{eqnarray}
and the analytic form of $A_0(\mc)$ can be seen in \cite{FLS}.
Note that the frequently used approximation
$g (z) \approx -\frac{2}{3} ((\pi^2-\frac{31}{4})(1-z)^2 + \frac{3}{2})$
holds within $1.4 \%$ accuracy in the range $0.2 \leq z\leq 0.4$. 
The equation
$g (z) = -1.671+2.04(z-0.3)-2.15(z-0.3)^2$
is accurate for $0.2 \leq z\leq 0.4$ to better than one per mille accuracy
and that is what we have used here.

The double differential ratio given in eq.~(\ref{doublediff}) agrees in the 
$(V-A)$ limit with the corresponding expression derived for the 
semileptonic decay
$B \to X_c \ell \nu_{\ell}$ in \cite{FLS} (their eq.~(3.2)). Taking this limit
amounts to the following transcription:  
\begin{eqnarray}
        C_9^{{\mbox{eff}}} & = & - C_{10} = \frac{1}{2}
                \; , \\
        C_7^{{\mbox{eff}}} & = & 0 
                \; , \\
        \left( \frac{G_F \, \alpha}{\sqrt{2} \, \pi}
                V_{ts}^\ast V_{tb}\right)
                & \rightarrow &
        \left( - \frac{4 \, G_F}{\sqrt{2}} V_{cb} \right)
                \; .
\end{eqnarray}

The hadron energy spectrum can now be obtained by integrating over $\s_0$.
The imaginary part can be obtained using the relation:
\begin{eqnarray}
{\rm Im} \frac{1}{x^n} 
\propto \frac{(-1)^{n-1}}{(n-1) !} \delta^{(n-1)} (\s_0 -\ms^2)
 \, \, .
\end{eqnarray}
The kinematic boundaries are given as:
\begin{eqnarray}
max(\ms^2,-1+2 x_0 +4 \ml^2) \leq &\s_0& \leq x_0^2 \, \, ,\nonumber \\
\ms \leq  & x_0 &  \leq \frac{1}{2} (1+\ms^2-4 \ml^2) \, \, .
\end{eqnarray}
Here we keep $\ml$ as a regulator wherever it is necessary and abbreviate
$C_9^{\mbox{eff}}\equiv C_9^{\mbox{eff}}(\s=1-2 x_0+\ms^2)$.
Including the leading power corrections, the  
hadron energy spectrum in the decay  \bxsll is given below:
\begin{eqnarray}
        \frac{{\rm d}{\cal B}}{{\rm d} x_0} & = & \; {\cal B}_0
 \left\{ 
     \left[
g_0^{(9,10)} + \lo g_1^{(9,10)} +\lt g_2^{(9,10)}
\right]
 \left( |C_9^{\mbox{eff}}|^2 + |C_{10}|^2 \right) \right.  \nonumber \\
&+&\left[
g_0^{(7)} + \lo g_1^{(7)} +\lt g_2^{(7)}
\right]
 \frac{|C_7^{\mbox{eff}}|^2}{x_0-\frac{1}{2}(1+\ms^2)} 
+  \left[
g_0^{(7,9)} + \lo g_1^{(7,9)} +\lt g_2^{(7,9)}
\right]
 Re(C_9^{\mbox{eff}}) \, C_7^{\mbox{eff}}  \nonumber \\
&+&  (\lo h_1^{(9)}+  \lt h_2^{(9)})
\frac{d |C_9^{\mbox{eff}}|^2}{d\s_0}
+  \lo k_1^{(9)}
\frac{d^2 |C_9^{\mbox{eff}}|^2}{d\s_0^2} \nonumber \\
&+&  ( \lo h_1^{(7,9)} + \lt  h_2^{(7,9)} )
\frac{d Re(C_9^{\mbox{eff}})}{d\s_0}  \, C_7^{\mbox{eff}}  
+  \left.  \lo  k_1^{(7,9)}
\frac{d^2 Re(C_9^{\mbox{eff}})}{d\s_0^2}  \, C_7^{\mbox{eff}} 
 \right\} 
 \nonumber \\
&+& \delta(x_0-\frac{1}{2}(1+\ms^2-4 \ml^2)) f_{\delta}(\lo,\lt)
+\delta'(x_0-\frac{1}{2}(1+\ms^2-4 \ml^2)) f_{\delta'}(\lo,\lt) \, \, .
\label{singlediff}
\end{eqnarray}
The functions 
$g_i^{(9,10)},g_i^{(7)},g_i^{(7,9)},h_i^{(9)},h_i^{(7,9)},
k_1^{(9)},k_1^{(7,9)}$  in the above expression are the coefficients of the 
$1/m_b^2$ 
power expansion for different combinations of Wilson coefficients, with
$g_0^{(j,k)}$ being the lowest order (parton model) functions.
They are functions of the variables $x_0$ and $\ms$ and are given 
in appendix \ref{app:auxfunc1}.
The singular functions $ \delta, \delta'$ have support only at the 
lowest order
end point of the spectrum, i.e., at $x_0^{max} \equiv \frac{1}{2}(1+\ms^2-4 
\ml^2)$.
The auxiliary functions $f_{\delta}(\lo,\lt)$ and $f_{\delta'}(\lo,\lt)$ 
vanish in the limit $\lo=\lt=0$.
They are given in appendix {\ref{app:auxfunc}}.
The derivatives of $C_9^{\mbox{eff}}$ are defined as 
$ \frac{d^{n} C_9^{\mbox{eff}}}{d\s_0^n} \equiv 
\frac{d^{n} C_9^{\mbox{eff}}}{d\s^n}(\s=1-2 x_0+\s_0; \s_0=\ms^2)$ $(n=1,2)$.
In the $(V-A)$ limit our eq.~(\ref{singlediff}) for the
hadron energy spectrum in \bxsll agrees with the
corresponding spectrum in $B \to X \ell \nu_\ell$ given in \cite{FLS}
(their eq.~(A1)). Integrating also over $x_0$ the resulting total width
for \bxsll agrees again in the $(V-A)$
limit with the well known result \cite{georgi}.

\begin{figure}[htb]
\vskip -0.0truein
\centerline{\epsfysize=3.5in
{\epsffile{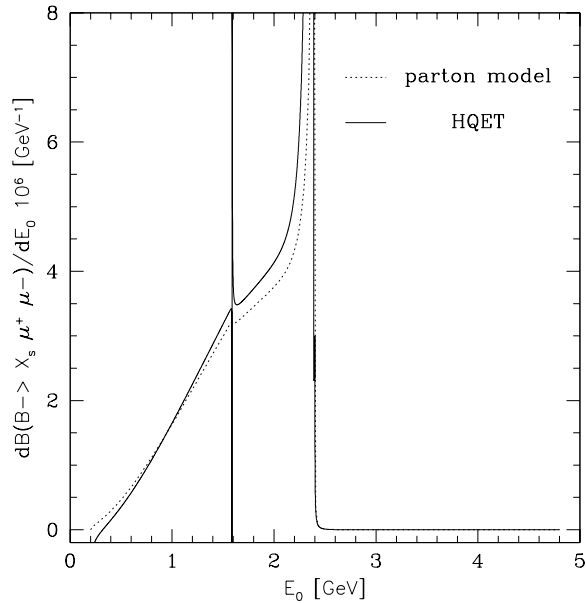}}}
\vskip -0.0truein
\caption[]{ \it Hadron energy spectrum
$ \frac{{\rm d}{\cal B}(B \to X_s \ell^+ \ell^-)}{{\rm d} E_0}$
in the parton model (dotted line) and including leading power corrections 
(solid line). For $m_b/2 < E_0 \leq m_b$ the distributions coincide. 
The parameters used for this plot are the central values given in 
Table~\ref{parameters} and the default values of the HQET parameters
specified in text. }
\label{fig:dbdx0}
\end{figure}
The power-corrected hadron energy spectrum 
$ \frac{{\rm d}{\cal B}(B \to X_s \ell^+ \ell^-)}{{\rm d} E_0}$ 
(with $E_0=m_b x_0$)
is displayed in Fig.~{\ref{fig:dbdx0}} through the solid curve, however, 
without the singular $\delta, \delta^{\prime}$ terms.
Note that before reaching the kinematic lower end point, the 
power-corrected spectrum  becomes negative, as a result of the $\lt$ 
term. This behavior is analogous to what has already been reported for the 
dilepton mass spectrum 
$ \frac{{\rm d}{\cal B}(B \to X_s \ell^+ \ell^-)}{{\rm d} q^2}$
in the high $q^2$ region \cite{AHHM97},
signaling a breakdown of the $\frac{1}{m_b}$ expansion in this region.
The terms with the derivatives of $C_9^{\mbox{eff}}$ in 
eq.~(\ref{singlediff}) give rise to a 
singularity in the hadron energy spectrum at the charm threshold due to 
the cusp in the function $Y(\s)$, when approached from either 
side. The hadron energy spectrum for the parton model is also 
shown in Fig.~{\ref{fig:dbdx0}}, which is finite for all ranges of $E_0$.

What is the region of validity of the hadron energy spectrum derived
in HQET?
It is known that in \bxsll decay there are resonances present,
from which the known six \cite{PDG} populate the $x_0$ (or $E_0$) range 
between the lower end point and the charm threshold. Taking this into
account and what has been remarked earlier, 
one concludes that the HQET spectrum cannot be used near the resonances, 
near the charm threshold and around the lower endpoint. Excluding these
regions, the spectrum calculated in HQET is close to the 
(partonic) perturbative
spectrum as the power corrections are shown to be small.  
The authors of ref.~\cite{buchallaisidorirey},
{\footnote{The ${\cal{O}}(1/m_c^2)$ 
correction to
$\frac{{\rm d}{\cal B}(B \to X_s \ell^+ \ell^-)}{{\rm d} q^2}$ has also been 
calculated in ref.~\cite{chenrupaksavage}, however, the result differs in 
sign from the one in ref.~\cite{buchallaisidorirey}.
It seems that this controversy has been settled in 
favor of ref.~\cite{buchallaisidorirey}.}}
who have performed an $1/m_c$ expansion for the dilepton mass spectrum 
$\frac{{\rm d}{\cal B}(B \to X_s \ell^+ \ell^-)}{{\rm d} q^2}$ 
and who 
also found a charm-threshold singularity, expect a reliable prediction of 
the spectrum for $q^2 \leq 3 m_c^2$ corresponding to
$E_0 \geq \frac{m_b}{2} (1+\ms^2-3 \hat m_c^2) \approx 1.8$ GeV. In this
region, the effect of the $1/m_b$ power corrections on the energy spectrum
is small and various spectra in \bxsll calculated here and in 
ref.~\cite{AHHM97} can be compared with data. 

The leading power corrections to the invariant mass spectrum is found by 
integrating eq.~(\ref{doublediff}) with respect to $x_0$.
We have already discussed the non-trivial hadronic invariant mass spectrum
which results  from the ${\cal O}(\alpha_s)$ bremsstrahlung   
and its Sudakov-improved version.
Since we have consistently dropped everywhere terms of
${\cal O}(\lambda_i \alpha_s)$ (see eq.~(\ref{doublediff})), this is the
only contribution to the invariant mass spectrum also in HQET away from 
$\s_0=\ms^2$, as the result of 
integrating the terms involving power corrections in eq.~(\ref{doublediff}) 
over $x_0$ is a singular function with support only at $\s_0=\ms^2$.
Of course, these corrections contribute to the normalization (i.e., branching
ratio) but leave the perturbative spectrum intact for  $\s_0 \neq \ms^2$.

\section{Hadronic Moments in \bxsll~in HQET}

We start with the derivation of the lowest spectral moments
in the decay \bxsll  at the parton 
level. These moments are worked out by taking into account the
two types of corrections discussed earlier, namely the leading power $1/m_b$ 
and the perturbative ${\cal{O}}(\alpha_s)$ corrections.
To that end, we define:
\begin{equation}
{\cal M}^{(n,m)}_{l^{+} l^{-}} \equiv 
   {1\over {\cal B}_0}\int (\hat s_0-\hat m_s^2)^n  x_0^m\,
   {{\rm d}^2 {\cal B}\over{\rm d}\hat s_0{\rm d} x_0}
   \,{\rm d}\hat s_0{\rm d} x_0\,,
\end{equation}
for integers $n$ and $m$.  These moments are related to the 
corresponding moments $\langle x_0^m(\hat s_0-\hat m_s^2)^n\rangle$  
obtained at the parton level 
by a scaling factor which yields the corrected branching
ratio ${\cal B}={\cal B}_0 {\cal M}_{\ell^+ \ell^-}^{(n,m)}$.
Thus, 
\begin{equation}
   \langle x_0^m(\hat s_0-\hat m_s^2)^n\rangle =
{{\cal B}_0\over {\cal B}}\,
   {\cal M}^{(n,m)}_{l^{+} l^{-}}\,.
\label{momdef}
\end{equation}
The correction factor ${\cal B}_0/{\cal B}$ is given a little later.
We remind that one has to Taylor expand it in terms of the 
${\cal O}(\alpha_s)$ and power corrections. 
The moments can be expressed as double expansion in ${\cal{O}}(\alpha_s)$
and $1/m_b$ and to the accuracy of our calculations can be 
represented in the following form:
 \begin{eqnarray}
 {\cal M}^{(n,m)}_{l^{+} l^{-}}=D_0^{(n,m)}+
\frac{\alpha_s}{\pi} {C_9}^2 A^{(n,m)}+
\lo D_1^{(n,m)} + \lt D_2^{(n,m)} \,\, ,
\end{eqnarray}
with a further decomposition into pieces from different Wilson 
coefficients for $i=0,1,2$:
\begin{eqnarray}
\label{momentexp}
D_i^{(n,m)}=\alpha_i^{(n,m)} {C_7^{{\mbox{eff}}}}^2+
\beta_i^{(n,m)} C_{10}^2+
\gamma_i^{(n,m)} C_7^{{\mbox{eff}}} +\delta_i^{(n,m)}.
\end{eqnarray}
The terms $\gamma_i^{(n,m)}$ and $\delta_i^{(n,m)}$ in              
eq.~(\ref{momentexp}) result from the terms proportional 
to ${\it{Re}}(C_9^{{\mbox{eff}}})C_7^{{\mbox{eff}}}$ and
$|C_9^{{\mbox{eff}}}|^2$  in  eq.~(\ref{doublediff}), respectively.   
The results for   
$\alpha_i^{(n,m)},\beta_i^{(n,m)},  \gamma_i^{(n,m)}, \delta_i^{(n,m)}$
are presented in appendix {\ref{app:moments}}. Out of these, the 
functions $\alpha_i^{(n,m)}$ and $\beta_i^{(n,m)}$
are given analytically, but the other two  
$\gamma_i^{(n,m)}$ and $\delta_i^{(n,m)}$ are given in terms of a 
one-dimensional integral over $x_0$, as these latter functions
involve the coefficient $C_9^{\mbox{eff}}$, which is a complicated
 function of $x_0$.

The leading perturbative contributions for the hadronic invariant mass and 
hadron energy 
moments can be obtained analytically by integrating eq.~(\ref{dbds0}) and 
eq.~(\ref{eq:g1}), respectively, yielding
\begin{eqnarray}
A^{(0,0)}&=&\frac{25-4 \pi^2}{9} \, ,
~~A^{(1,0)}=\frac{91}{675} \, ,
~~A^{(2,0)}=\frac{5}{486} \, ,\nonumber \\
A^{(0,1)}&=&\frac{1381-210 \pi^2}{1350} \, ,
A^{(0,2)}=\frac{2257-320 \pi^2}{5400} \, .
\label{eq:A10}
\end{eqnarray}
The zeroth moment $n=m=0$ is needed for the normalization and we recall 
that the result for $A^{(0,0)}$ was derived by Cabibbo and Maiani 
in the context of the ${\cal O}(\alpha_s)$ correction to the semileptonic
decay rate $B \to X \ell \nu_\ell$ quite 
some time ago \cite{CM78}.
Likewise, the first mixed moment $A^{(1,1)}$ can be extracted from 
the results given  
in \cite{FLS} for the decay $B \to X \ell \nu_{\ell}$ after changing the 
normalization, 
\begin{eqnarray}
\label{eq:A11}
A^{(1,1)}&=&
\frac{3}{50} \, \, .
\end{eqnarray}
For the lowest order parton model contribution 
$D_0^{(n,m)}$, we find, in agreement 
with \cite{FLS}, that the first two hadronic invariant mass moments 
$\langle \s_0-\ms^2 \rangle, \, \langle(\s_0-\ms^2)^2 \rangle$ and the first 
mixed moment $\langle x_0 (\s_0-\ms^2) \rangle$ 
vanish:
\begin{eqnarray}
D_0^{(n,0)}=0 \, \, \,  \mbox{for} \, \, n=1,2 \, \, \, \mbox{and}\, \, \,
D_0^{(1,1)}=0 \, .
\end{eqnarray}
We remark that we have included the $s$-quark mass dependence  in the 
leading term 
and in the power corrections, but omitted it throughout our work in the 
calculation of the explicit $\alpha_s$ term. 
All the expressions derived here for the moments agree in the
$V-A$ limit (and with $\ms=0$ in the perturbative $\alpha_s$ correction 
term) with the corresponding expressions given in \cite{FLS}.
{}From here the full ${\cal{O}}(\alpha_s m_s)$ expressions can be inferred 
after
adjusting the normalization $\Gamma_0 \to {\cal{B}}_0 \frac{2}{3} C_9^2$.
We have checked that a finite $s$-quark mass effects the 
values of the $A^{(n,m)}$  given in eq.~(\ref{eq:A10}-\ref{eq:A11})
by less than $8 \%$ for $m_s=0.2$ GeV.

We can eliminate the hidden dependence on the non-perturbative parameters 
resulting from the $b$-quark mass in the moments 
${\cal M}^{(n,m)}_{l^{+} l^{-}}$ with the help of the HQET mass relation.
As $m_s$ is of order $\Lambda_{QCD}$, to be consistent we keep only 
terms up to order $m_s^2/m_b^2$ \cite{FLSphenom}. An additional 
$m_b$-dependence is in the mass ratios $\ml=\frac{m_l}{m_b}$.
Substituting $m_b$ by the $B$ meson mass using the HQET relation introduces 
additional ${\cal{O}}(1/m_B,1/m_B^2)$ 
terms in the Taylor expansion of eq.~(\ref{momdef}).
We get for the following normalization factor for ${\cal B}/{\cal B}_0 =
{\cal M}_{\ell^+ \ell^-}^{(0,0)}$:
\begin{eqnarray}
{{\cal B}\over {\cal B}_0}&=& \frac{32}{9 m_B^2}(-4 m_B^2-13 m_s^2-
3 (m_B^2-2 m_s^2) \ln(4 \frac{m_l^2}{m_B^2})){C_7^{\mbox{eff}}}^2+
\frac{2}{3 m_B^2} (m_B^2-8 m_s^2)C_{10}^2 \nonumber \\
&+&
\int_{m_s/m_B}^{\frac{1}{2}(1+m_s^2/m_B^2)} d x_0 \frac{64}{m_B^2}
(-m_s^2-4 m_s^2 x_0+2 m_B^2 x_0^2+2 m_s^2 x_0^2)
{\it{Re}}(C_9^{{\mbox{eff}}})C_7^{{\mbox{eff}}} \nonumber \\
&+&
\int_{m_s/m_B}^{\frac{1}{2}(1+m_s^2/m_B^2)} d x_0
\frac{16}{3 m_B^2}(-3 m_s^2+6 m_B^2 x_0^2+6 m_s^2 x_0^2-8 m_B^2 x_0^3)
|C_9^{{\mbox{eff}}}|^2\nonumber \\
&+&\frac{\alpha_s}{\pi} A^{(0,0)} C_9^2+ 
\frac{-64}{3} {C_7^{\mbox{eff}}}^2 \frac{\bar{\Lambda}}{m_B}+
\frac{-32}{3}{C_7^{\mbox{eff}}}^2\frac{\bar{\Lambda}^2}{m_B^2}
+ \left[ \frac{16}{9} (2-3\ln(4 \frac{m_l^2}{m_B^2})){C_7^{\mbox{eff}}}^2+
 \frac{C_{10}^2}{3}  \right. \nonumber \\
&+& \left.
\int_{0}^{\frac{1}{2}} d x_0
(64 x_0^2{\it{Re}}(C_9^{{\mbox{eff}}})C_7^{{\mbox{eff}}}
+\frac{16}{3} (3-4 x_0)x_0^2 |C_9^{{\mbox{eff}}}|^2 ) \right] 
\frac{\lambda_1}{m_B^2}
\nonumber \\
&+& 
 \left[ \frac{16}{3} (4+9\ln(4 \frac{m_l^2}{m_B^2})){C_7^{\mbox{eff}}}^2-
 3 C_{10}^2   \right.\\
&+& \left.
\int_{0}^{\frac{1}{2}} d x_0
(64 (-1-4 x_0+7 x_0^2){\it{Re}}(C_9^{{\mbox{eff}}})C_7^{{\mbox{eff}}}
+16 (-1+15 x_0^2-20 x_0^3)|C_9^{{\mbox{eff}}}|^2 \right] 
\frac{\lambda_2}{m_B^2}
\nonumber  \, .
\end{eqnarray}
Here, the $\frac{\bar{\Lambda}}{m_B}$ and $ \frac{\bar{\Lambda}^2}{m_B^2}$ 
terms result from the expansion of $\ln(4 m_l^2/m_b^2)$.
The first two moments and the first mixed moment,
$\langle x_0 \rangle {\cal B}/{\cal B}_0$, $\langle x_0^2 \rangle {\cal 
B}/{\cal B}_0$, $\langle \hat{s}_0 - \hat{m}_s^2 \rangle {\cal B}/{\cal 
B}_0$, $\langle (\hat{s}_0 - \hat{m}_s^2)^2 \rangle {\cal B}/{\cal B}_0$
and $\langle x_0 (\hat{s}_0 - \hat{m}_s^2) \rangle {\cal B}/{\cal
B}_0$  are presented in appendix {\ref{app:lowmoments}}. 

With this we obtain the moments for the 
physical quantities valid up to ${\cal{O}}(\alpha_s/m_B^2,1/m_B^3)$,
where the second equation corresponds to a further use of
$m_s={\cal{O}}(\Lambda_{QCD})$.
We get for the first two hadronic invariant mass moments
{\footnote{Our first expression for 
$\langle S_H^2\rangle$, eq.~(\ref{sHmoments}), 
does not agree in the coefficient of 
$\langle\hat s_0-\hat m_s^2\rangle $ with the one given in \cite{FLS} 
(their eq.~(4.1)). We point out that $m_B^2$ should have been replaced by 
$m_b^2$ in this expression.
This has been confirmed by Adam Falk (private communication).
Dropping the higher order terms given in their expressions, 
the hadronic moments in HQET derived here and in
 \cite{FLS} agree.}}
\begin{eqnarray}\label{sHmoments}
   \langle S_H\rangle&=&m_s^2+\bar\Lambda^2+(m_B^2-2\bar\Lambda m_B
)\,\langle\hat s_0-\hat m_s^2\rangle
+(2\bar\Lambda m_B-2\bar\Lambda^2-\lambda_1-3\lambda_2)
  \langle  x_0\rangle\,, \nonumber\\
  \langle S_H^2\rangle&=&
m_s^4+2\bar\Lambda^2 m_s^2+
  2 m_s^2 (m_B^2-2\bar\Lambda m_B)
  \langle\hat s_0-\hat m_s^2\rangle
+2m_s^2 (2\bar\Lambda m_B-2\bar\Lambda^2-\lambda_1-3\lambda_2)
\langle  x_0\rangle
  \nonumber\\
  &&\quad\mbox{}+ 
(m_B^4-4\bar\Lambda m_B^3
)\langle (\hat s_0-\hat m_s^2)^2\rangle
+ 4\bar\Lambda^2 m_B^2 \langle x_0^2\rangle+
  4\bar\Lambda m_B^3\langle x_0(\hat s_0-\hat m_s^2)\rangle\,,  \\
&=&
(m_B^4-4\bar\Lambda m_B^3)\langle (\hat s_0-\hat m_s^2)^2\rangle
+ 4\bar\Lambda^2 m_B^2 \langle x_0^2\rangle+
  4\bar\Lambda m_B^3\langle x_0(\hat s_0-\hat m_s^2)\rangle \,,
\nonumber
\end{eqnarray}
and for the hadron energy moments:
\begin{eqnarray}\label{EHmoments}
  \langle E_H\rangle &=& \bar\Lambda-{\lambda_1+3\lambda_2\over2m_B}
  +\left(m_B-\bar\Lambda+{\lambda_1+3\lambda_2\over2m_B}\right)\langle
   x_0\rangle\,,\nonumber\\
  \langle E_H^2\rangle &=& \bar\Lambda^2 + (2\bar\Lambda m_B -
2\bar\Lambda^2
  -\lambda_1-3\lambda_2)\langle  x_0\rangle\\
  &&\quad +(m_B^2-2\bar\Lambda m_B+\bar\Lambda^2+\lambda_1+3\lambda_2)
  \langle x_0^2\rangle\,.\nonumber
\end{eqnarray}
One sees that
there are linear power corrections, ${\cal O}(\bar{\Lambda}/m_B)$,
present in all these hadronic quantities except 
$\langle S_H^2 \rangle$ which starts in
$\frac{\alpha_s}{\pi} \frac{\bar{\Lambda}}{m_B}$.

\subsection{Numerical Estimates of the Hadronic Moments in HQET   
\label{numerics:hqet}}

Using the expressions for the HQET moments given in 
appendix \ref{app:lowmoments},
we  present the numerical results for the hadronic moments in \bxsll,
valid up to ${\cal{O}}(\alpha_s/m_B^2,1/m_B^3)$.
We find: 
\begin{eqnarray}
\langle x_0\rangle&=&0.367 \, (1+0.148 \frac{\alpha_s}{\pi} 
-0.204 \frac{\bar{\Lambda}}{m_B} \frac{\alpha_s}{\pi}
-0.030 \frac{\bar{\Lambda}}{m_B}-0.017\frac{\bar{\Lambda}^2}{m_B^2}
+ 0.884 \frac{\lambda_1}{m_B^2}+3.652\frac{\lambda_2}{m_B^2}) \nonumber \, ,\\
\langle x_0^2\rangle&=&0.147 \, (1+0.324 \frac{\alpha_s}{\pi}
-0.221\frac{\bar{\Lambda}}{m_B}  \frac{\alpha_s}{\pi}
-0.058\frac{\bar{\Lambda}}{m_B}-0.034 \frac{\bar{\Lambda}^2}{m_B^2}
+ 1.206 \frac{\lambda_1}{m_B^2}+4.680\frac{\lambda_2}{m_B^2})\nonumber \, ,\\
\langle x_0(\s_0-\ms^2)\rangle&=&0.041 \frac{\alpha_s}{\pi} 
(1 + 0.083\frac{\bar{\Lambda}}{m_B})
+ 0.124 \frac{\lambda_1}{m_B^2}+0.172 \frac{\lambda_2}{m_B^2}\nonumber \, ,\\
\langle \s_0-\ms^2 \rangle&=&0.093 \frac{\alpha_s}{\pi}
(1+0.083\frac{\bar{\Lambda}}{m_B})
+ 0.641 \frac{\lambda_1}{m_B^2}+0.589\frac{\lambda_2}{m_B^2}\nonumber \, ,\\
\langle (\s_0-\ms^2)^2\rangle&=&0.0071 \frac{\alpha_s}{\pi} 
(1+0.083\frac{\bar{\Lambda}}{m_B})
-0.196 \frac{\lambda_1}{m_B^2} \, .
\end{eqnarray}
As already discussed earlier, the normalizing factor
${\cal B}/ {\cal B}_0$ is also expanded in a Taylor series. 
Thus, in deriving the above results, we have used
 \begin{eqnarray}
{{\cal B}\over {\cal B}_0}&=&25.277 \,\, (1-1.108 \frac{\alpha_s}{\pi} 
-0.083 \frac{\bar{\Lambda}}{m_B}-0.041 \frac{\bar{\Lambda}^2}{m_B^2}
+ 0.546 \frac{\lambda_1}{m_B^2}-3.439\frac{\lambda_2}{m_B^2}) \nonumber \, .
\end{eqnarray}
The parameters used in arriving at the numerical coefficients are given in 
Table~\ref{parameters} and Table~\ref{wilson}.

\begin{figure}[htb]
\vskip -0.0truein
\centerline{\epsfysize=3.5in
{\epsffile{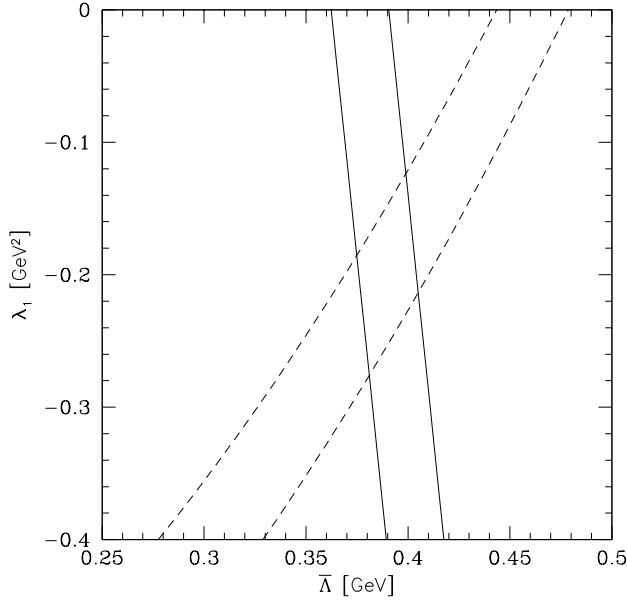}}}
\vskip -0.0truein
\caption[]{ \it $\langle S_H \rangle$ (solid bands) and  $\langle S_H^2 
\rangle$ (dashed bands)
correlation in ($\lambda_1$-$\bar{\Lambda}$) space for fixed values 
$\langle S_H \rangle =1.64$ GeV$^2$ and $\langle S_H^2 \rangle =4.48$ 
GeV$^4$, corresponding to the central values in Table \ref{tab:emoments}.
The curves are forced to meet at the point 
$\lambda_1=-0.2$ GeV$^2$ and $\bar{\Lambda}=0.39$ GeV.}
\label{fig:laml1}
\end{figure}

Inserting the expressions for the moments calculated at the partonic
level into 
eq.~(\ref{sHmoments}) and eq.~(\ref{EHmoments}), 
we find the following expressions for the short-distance hadronic moments, 
valid up to ${\cal{O}}(\alpha_s/m_B^2,1/m_B^3)$:
\begin{eqnarray}
   \langle S_H\rangle&=&m_B^2 (\frac{m_s^2}{m_B^2}
+0.093 \frac{\alpha_s}{\pi} 
-0.069 \frac{\bar{\Lambda}}{m_B} \frac{\alpha_s}{\pi}
+0.735 \frac{\bar{\Lambda}}{m_B}+0.243 \frac{\bar{\Lambda}^2}{m_B^2}
+ 0.273 \frac{\lambda_1}{m_B^2}-0.513\frac{\lambda_2}{m_B^2}) \nonumber \, ,\\
\label{eq:hadmoments}
 \langle S_H^2\rangle&=&m_B^4 (0.0071 \frac{\alpha_s}{\pi} 
+0.138 \frac{\bar{\Lambda}}{m_B} \frac{\alpha_s}{\pi}
+0.587\frac{\bar{\Lambda}^2}{m_B^2}
-0.196 \frac{\lambda_1}{m_B^2}) \, ,\\
   \langle E_H\rangle&=& 0.367 m_B  (1+0.148 \frac{\alpha_s}{\pi} 
-0.352 \frac{\bar{\Lambda}}{m_B} \frac{\alpha_s}{\pi}
+1.691 \frac{\bar{\Lambda}}{m_B}+0.012\frac{\bar{\Lambda}^2}{m_B^2}
+ 0.024 \frac{\lambda_1}{m_B^2}+1.070\frac{\lambda_2}{m_B^2}) \, ,\nonumber \\
 \langle E_H^2\rangle&=&0.147 m_B^2 (1+0.324 \frac{\alpha_s}{\pi} 
-0.128 \frac{\bar{\Lambda}}{m_B} \frac{\alpha_s}{\pi}
+2.954 \frac{\bar{\Lambda}}{m_B}+2.740\frac{\bar{\Lambda}^2}{m_B^2}
-0.299 \frac{\lambda_1}{m_B^2}+0.162\frac{\lambda_2}{m_B^2}) \, .\nonumber
\end{eqnarray}
Setting $m_s=0$ changes the numerical value of the coefficients in the 
expansion given above (in which we already neglected $\alpha_s m_s$) 
by at most $1 \% $.
With the help of the expressions given above,
we have calculated numerically the hadronic moments in HQET for the decay 
$B \to X_s \ell^{+} \ell^{-}$, $\ell=\mu,e$ and have estimated the errors
by varying the parameters within their $\pm 1 \sigma$ ranges given in 
Table~\ref{parameters}. They are presented in Table {\ref{tab:emoments}}
where we have used $\bar{\Lambda}=0.39 \, {\mbox{GeV}}$,  
$\lambda_1=-0.2 \, {\mbox{GeV}}^2$ and $\lambda_2=0.12 \, {\mbox{GeV}}^2$.
Further, using $\alpha_s(m_b)=0.21$,
the explicit dependence of the hadronic moments given in 
eq.~(\ref{eq:hadmoments}) on the HQET parameters
$\lambda_1$ and $\bar{\Lambda}$ can be worked out:
\begin{eqnarray}
   \langle S_H\rangle&=&0.0055 m_B^2(1+
132.61 \frac{\bar{\Lambda}}{m_B}+44.14 \frac{\bar{\Lambda}^2}{m_B^2}
+ 49.66 \frac{\lambda_1}{m_B^2}) \nonumber \, ,\\
 \langle S_H^2\rangle&=& 0.00048 m_B^4(1+
19.41 \frac{\bar{\Lambda}}{m_B} 
+1223.41\frac{\bar{\Lambda}^2}{m_B^2}
-408.39 \frac{\lambda_1}{m_B^2}) \, ,\\
   \langle E_H\rangle&=& 0.372 m_B  (1+
1.64 \frac{\bar{\Lambda}}{m_B}+0.01 \frac{\bar{\Lambda}^2}{m_B^2}
+ 0.02 \frac{\lambda_1}{m_B^2}) \, ,\nonumber \\
 \langle E_H^2\rangle&=&0.150 m_B^2 (1+
2.88 \frac{\bar{\Lambda}}{m_B}+2.68\frac{\bar{\Lambda}^2}{m_B^2}
-0.29 \frac{\lambda_1}{m_B^2}) \, .\nonumber
\end{eqnarray}
While interpreting these numbers, one should bear in mind that there are two
comparable expansion parameters $\bar{\Lambda}/m_B$ and $\alpha_s/\pi$ 
and we have fixed the latter in showing the numbers.
As expected, the
dependence of the energy moments $\langle E_H^n\rangle$ on $\bar{\Lambda}$
and $\lambda_1$ is very weak.
The correlations on the HQET parameters $\lambda_1$ and $\bar{\Lambda}$
which follow from (assumed) fixed
values of the hadronic invariant mass moments  $\langle S_H \rangle$
 and  $\langle S_H^2 \rangle$ are shown in Fig.~\ref{fig:laml1}. We 
have taken the values for the decay $B \to X_s \mu^+ \mu^-$ from Table
\ref{tab:emoments}
for the sake of illustration and have also shown the presently 
irreducible
theoretical errors on these moments following from the input parameters
$m_t$, $\alpha_s$
and the scale $\mu$, given in Table \ref{parameters}. The errors were
calculated
by varying these parameters in the indicated range, one at a time,
and adding the individual errors in quadrature. This exercise
has to be repeated with real data in \bxsll to draw any
quantitative conclusions. 

The theoretical stability of the moments has to be checked against
higher order corrections and the
error estimates presented here will have to be improved.
 The ``BLM-enhanced" two-loop corrections
\cite{BLM} proportional to $\alpha_s^2\beta_0$, where $\beta_0 = 11 -2
n_f/3$ is the first term in the QCD beta function, can be included
at the parton level as has been done in other decays \cite{FLS,GS97}, but
not being crucial to our point we have not
done this. More importantly, higher order corrections in
$\alpha_s$ and $1/m_b^3$ are not included here.
 While we do not think that
the higher orders in $\alpha_s$ will have a significant influence, the  
second moment $\langle S_H^2 \rangle$ is susceptible to the presence of
$1/m_b^3$ corrections as shown for the decay $B \to X
\ell \nu_\ell$ \cite{FL98}. This will considerably enlarge the theoretical
error represented by the dashed band for $\langle S_H^2 \rangle$ in
Fig.~\ref{fig:laml1}. Fortunately, the coefficient of the
$\bar{\Lambda}/m_B$ term in $\langle S_H \rangle$
is large. Hence, a good measurement of  this moment alone constrains
$\bar{\Lambda}$ effectively.
Of course, the utility
of the hadronic moments calculated above is only in conjunction
with the experimental cuts. Since
the optimal experimental cuts in \bxsll remain to be defined, we hope to
return to this and related issue of doing an improved
theoretical error estimate in a future publication.

Related issues in other decays have been studied in literature. The 
classification of the operators 
contributing in ${\cal O}(1/m_b^3)$, estimates of their matrix elements, 
and effects on the decay rates and spectra in the decays 
$B \to X \ell \nu_\ell$ and $B \to (D,D^*) \ell \nu_\ell$ 
have been studied in  refs.~\citer{Shifmanetal94,GK96-2}.
Spectral moments of the photon energy in the decay $B \to X_s \gamma$
have been studied in ref.~\cite{KL95}.
For studies of ${\cal O}(1/m_b^3)$ contributions in this decay and the 
effects of the
experimental cut (on the photon energy) on the photon energy moments, see 
ref.~\cite{Bauer97}.
\begin{table}[h]
        \begin{center}
        \begin{tabular}{|c|l|l|l|l|}
        \hline
        \multicolumn{1}{|c|}{HQET}      &
                \multicolumn{1}{|c|}{$\langle S_H\rangle$  } &
\multicolumn{1}{|c|}{$\langle S_H^2\rangle$ } &
                \multicolumn{1}{|c|}{$\langle E_H\rangle$  } &
\multicolumn{1}{|c|}{$\langle E_H^2\rangle$ } \\
 \hline
\multicolumn{1}{|c|}{{\mbox{}}} &
\multicolumn{1}{|c|}{$({\mbox{GeV}}^2)$ } &
\multicolumn{1}{|c|}{$({\mbox{GeV}}^4)$ } &
\multicolumn{1}{|c|}{$({\mbox{GeV}})$ } &  
\multicolumn{1}{|c|}{$({\mbox{GeV}}^2)$ } \\
 \hline
$\mu^+ \mu^-$&$1.64 \pm 0.06$ &$4.48 \pm 0.29$ & $2.21 \pm 0.04 $&
$5.14 \pm 0.16$ \\
$e^+ e^-$   &$1.79 \pm 0.07 $ &$4.98 \pm 0.29$ & $2.41 \pm 0.06 $&
$6.09 \pm 0.29$\\
        \hline   
        \end{tabular}
        \end{center} 
\caption{\it Hadronic spectral moments for $B \to X_s \mu^{+} \mu^{-}$
and $B \to X_s e^{+} e^{-}$
in HQET with $\bar{\Lambda}=0.39 \, GeV$, $\lambda_1=-0.2 \, GeV^2$,
and $\lambda_2=0.12 \, GeV^2$.
The quoted errors result from varying $\mu, \alpha_s$ and the
top mass within the ranges given in Table \ref{parameters}.}   
\label{tab:emoments}
\end{table}
Finally, concerning the power corrections related to the
$c\bar{c}$ loop in \bxsll, it has been suggested
in \cite{buchallaisidorirey} that an
${\cal{O}}(\Lambda^2_{QCD}/m_c^2)$ expansion in the context of HQET
can be carried out to
take into account such effects  in
the invariant mass spectrum away from the resonances.
Using the expressions (obtained with $m_s=0$)
for the $1/m_c^2$ amplitude, we have calculated the
partonic energy moments
$\triangle \langle x_0^n \rangle$,
which correct the short-distance result at order $\lambda_2/m_c^2$:
\begin{eqnarray}
\triangle \langle x_0^n \rangle {{\cal B}\over {\cal B}_0}
&=&-\frac{256 C_2 \lambda_2}{27 m_c^2}
\int_0^{1/2(1-4 \ml^2)} dx_0 x_0^{n+2} {\rm Re} \left[
F(r) \left( C_9^{\mbox{eff}} (3-2 x_0)+2 C_7^{{\mbox{eff}}}
\frac{-3+4 x_0+2 x_0^2}{2 x_0-1} \right) \right] \; , \nonumber \\
r&=&\frac{1-2 x_0}{4 \mc^2} \; ,
\end{eqnarray}
\begin{equation}\label{frl1}
F(r)=\frac{3}{2r}\left\{ \begin{array}{ll}
\dis\frac{1}{\sqrt{r(1-r)}}\arctan\sqrt{\frac{r}{1-r}}
   -1 &  \qquad\qquad 0< r < 1~, \\
 \dis\frac{1}{2\sqrt{r(r-1)}}\left(
\ln\frac{1-\sqrt{1-1/r}}{1+\sqrt{1-1/r}}+i\pi\right)-1 &
\qquad\qquad r > 1~. \end{array} \right.
\end{equation}
The invariant mass and mixed moments give zero contribution in the order we
are working, with $m_s=0$.
Thus, the correction to the hadronic mass moments are vanishing, if we
further neglect terms proportional to
$\frac{\lambda_2}{m_c^2} \bar{\Lambda}$ and $ \frac{\lambda_2}{m_c^2}
\lambda_i$, with $i=1,2$.
For the hadron energy moments we obtain numerically
\begin{eqnarray}
\triangle \langle E_H \rangle_{1/m_c^2}&=&
m_B \triangle \langle x_0 \rangle= -0.007 \, {\mbox{GeV}} \; ,
\nonumber \\
\triangle \langle E_H^2 \rangle_{1/m_c^2}&=&
m_B^2 \triangle \langle x_0^2 \rangle= -0.013 \, {\mbox{GeV}}^2 \; ,
\end{eqnarray}
leading to a correction of order $-0.3 \%$
 to the short-distance values presented in
Table \ref{tab:moments}.
The power corrections
presented here in the hadron spectrum and
hadronic spectral moments in \bxsll are the first results in this decay.

\section{Hadron Spectra and Moments in the 
Fermi Motion Model \label{FMspectra}}

In this section, we study the non-perturbative effects
associated with the bound state nature of the $B$ hadron 
on the hadronic invariant mass and hadron energy 
distributions in the decay \bxsll. These effects are studied
in the FM model \cite{aliqcd}. The hadronic moments in this model are
compared with the ones calculated in the HQET approach for identical
values of the equivalent parameters. We also define this equivalence and
illustrate this numerically for some values of the FM model parameters 
resulting from fits of data in other $B$ decays. 
With the help of the phenomenological profiles in the FM model, we study the
effects of the experimental cuts used by the CLEO collaboration 
\cite{cleobsll97} on the hadron spectra and spectral moments in the decay 
\bxsll. The resulting  branching ratios and the hadronic invariant 
mass moments are calculated for several values of the FM parameters and 
can be compared directly with data when it becomes available. %
\subsection{Hadron spectra in \bxsll ~in the Fermi motion model 
\protect \cite{aliqcd}}

 The Fermi motion model \cite{aliqcd}  has received a lot
 of phenomenological
attention in $B$ decays, partly boosted by studies in the context of
HQET showing that this model  
can be made to mimic the effects associated with 
the HQET parameters $\bar{\Lambda}$ and
 $\lambda_1$ \cite{Bigietal94,MW}. We further quantify this correspondence
in this paper.
 In the context of rare $B$ decays, this model has 
been employed to calculate the
energy spectra in the decay $B \to X_s+\gamma$ in \cite{ag91}, which
was used subsequently by the CLEO collaboration in their successful search of
this decay \cite{cleobsgamma}.
It has also been used in calculating the dilepton
invariant mass spectrum and FB asymmetry in \bxsll 
in ref.~\cite{AHHM97}.
 
The FM model has two parameters $p_F$ and the spectator quark mass
$m_{q}$. Energy-momentum conservation requires the $b$-quark mass
to be a momentum-dependent parameter determined by the constraint:
\begin {eqnarray}
m_b^2(p) = {m_B}^2 + {m_q}^2 -2m_B \sqrt{p^2 + {m_q}^2} 
\quad ; \quad p = |\vec{p}|\, .
\end{eqnarray}
The $b$-quark momentum $p$ is assumed to have a Gaussian 
distribution, denoted by $\phi(p)$, which is determined by $p_F$
 \begin{equation}
\label{lett13}
 \phi(p)= \frac {4}{\sqrt{\pi}{p_F}^3} \exp (\frac {-p^2}{{p_F}^2}) \; ,
\end{equation}
with the normalization
$ \int_0^\infty \, dp \, p^2 \, \phi(p) = 1 $.
 In this
model, the HQET parameters are calculable in terms of $p_F$ and $m_q$
with 
\begin{eqnarray}
\label{fmtohqet}
\bar{\Lambda} &=& \int_0^\infty dp \, p^2 \phi(p) \sqrt{m_q^2+p^2}, 
\nonumber\\
\lambda_1 &=& -  \int_0^\infty dp \, p^4 \phi(p) = - \frac{3}{2} p_F^2~.
\end{eqnarray}
In addition, for $m_q=0$, one can show that $\bar{\Lambda}=2p_F/\sqrt{\pi}$.
There is, however, no parameter in the FM model analogous to $\lambda_2$ in 
HQET. Curiously, much of the 
HQET {\it malaise} in describing the spectra in the end-point regions is 
related to $\lambda_2$, as also shown in \cite{MW,AHHM97}. For subsequent
use in working out the normalization (decay widths) in the FM model, we
also define an {\it effective} $b$-quark mass by
\begin{equation}
\label{effbmass}
m_b^{{\mbox{eff}}}\equiv(\int_0^\infty  dp \, p^2 \, m_b(p)^5 
\phi(p))^{1/5}~.
\end{equation}
The relation between $m_B$, $m_b$, $\bar{\Lambda}$, $\lambda_1$ and 
$\lambda_2$
in HQET has already been stated. With the 
quantity $m_b^{\mbox{eff}}$ defined in eq.~(\ref{effbmass}) and  
the relations in eqs.~(\ref{fmtohqet}) for $\lambda_1$ and $\bar{\Lambda}$, 
the relation
\begin{equation}
\label{mbfm}
m_B=m_b^{\mbox{eff}}+ \bar{\Lambda} -\lambda_1/(2m_b^{\mbox{eff}})~,
\end{equation}
is found to be satisfied in the FM model to a high accuracy
 (better than $0.7 \%$), which is shown in 
Table \ref{tab:FMhqet} for some representative values of the 
HQET parameters and their FM model equivalents. We shall use the 
HQET parameters $\bar{\Lambda}$ and $\lambda_1$ to characterize also the
FM model parameters, with the relations given in eqs.~(\ref{fmtohqet})
and (\ref{effbmass}) and in Table \ref{tab:FMhqet}.  

\begin{figure}[htb]
\vskip 0.0truein
\centerline{\epsfysize=3.5in
{\epsffile{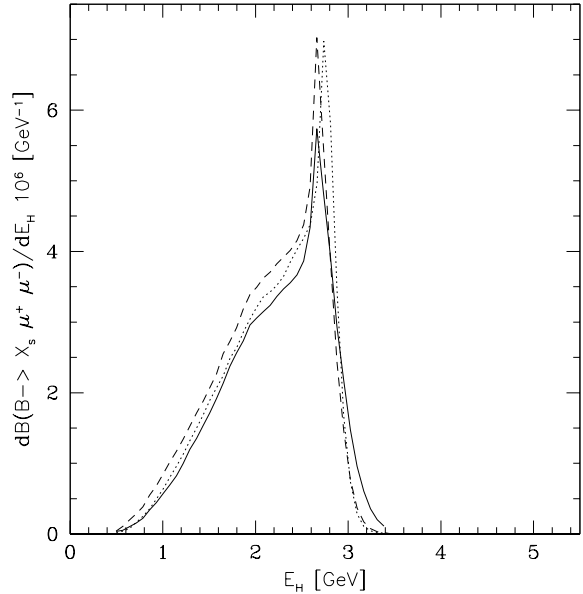}}}
\vskip 0.0truein
\caption[]{ \it Hadron energy spectrum in \bxsll in the Fermi motion 
model based on the perturbative contribution only. The solid, dotted, dashed 
curve corresponds to the parameters
$ (\lambda_1, \bar{\Lambda})=(-0.3,0.5),(-0.1,0.4),(-0.15,0.35)$ in 
(GeV$^2$, GeV), respectively.}
\label{fig:SDEh}   
\end{figure}

With this we turn to discuss the hadron energy spectrum in the 
decay \bxsll in the FM model 
including the ${\cal{O}}(\alpha_s)$ QCD corrections.
The spectrum $\frac{{\rm d}{\cal{B}}}{{\rm d } E_H}(B \to X_s \ell^+ \ell^-)$ 
is composed of a Sudakov improved piece from
$C_9^2$ and the remaining lowest order contribution.
The latter is based on the parton model distribution, which is well 
known and given below for the sake of completeness:
\begin{eqnarray}
        \frac{{\rm d}{\cal B}}{{\rm d} s} & = & 
{\cal{B}}_0 
\frac{\bar{u}}{m_b^6}
    \left\{ 
        \frac{4}{3} (m_b^4-2 m_s^2 m_b^2+m_s^4+m_b^2 s+m_s^2 s-2 s^2)
                \left( |C_9^{\mbox{eff}}(s)|^2 + |C_{10}|^2 \right)
             \right.   \nonumber \\
        & + &
   \frac{16}{3} (2 m_b^6 -2 m_b^4 m_s^2-2 m_b^2 m_s^4+2 m_s^6-m_b^4 s-
14 m_b^2 m_s^2 s-m_s^4 s-m_b^2 s^2-m_s^2 s^2)
                 \frac{|C_7^{\mbox{eff}}|^2}{s}
                \nonumber \\
        &+&     
             \left.  
 16  (m_b^4-2 m_s^2 m_b^2+m_s^4-m_b^2 s -m_s^2 s)
                         Re(C_9^{\mbox{eff}}(s)) \, C_7^{\mbox{eff}} 
                \right\}
\, , \\
\bar{u}  & = &
          \frac{\sqrt{( m_b^2+s-m_s^2)^2 -4 m_b^2 s}}{m_b^2}\, , \nonumber \\
{\cal{B}}_0&=&\frac{{\cal B}_{sl}}{\Gamma_{sl}} 
\frac{G_F^2 |V_{ts}^{\ast} V_{tb}|^2}{192 \pi^3} 
\frac{3 \alpha^2}{16 \pi^2} m_b^5\, , \nonumber \\
\Gamma_{sl}&=&\frac{G_F^2 V_{cb}^2 m_b^5}{192 \pi^3} f(\mc) \kappa(\mc) \, .
\label{eq:Gsl}
\end{eqnarray}  
Note that in the lowest order expression just given, we have    
$|C_9^{\mbox{eff}}(s)|^2=|Y(s)|^2+2 C_9 Re (Y(s))$
with the rest of $C_9^{\mbox{eff}}(s)$  now included in the 
Sudakov-improved piece as can be seen in eq.~(\ref{doubleexpon}).
To be consistent, the total semileptonic width $\Gamma_{sl}$, which enters
via the normalization constant ${\cal{B}}_0$, has also to be calculated
in the FM model with the same set of the model parameters.
We implement the correction in the decay width 
by replacing the $b$-quark mass in $\Gamma_{sl}$ given in 
eq.~(\ref{eq:Gsl}) by $m_b^{\mbox{eff}}$. (See \cite{AHHM97} for further 
quantitative discussions of this point on the branching ratio
for the decay \bxsll.)
The hadronic invariant mass spectrum in the decay \bxsll in this model is 
calculated very much along the same lines.
 The kinematically allowed ranges for the distributions are
$m_X \leq E_H \leq m_B$ and $m_X^2 \leq S_H \leq m_B^2$,
and we recall here that the physical threshold has been implemented by 
demanding that the lowest hadronic invariant mass produced in the decay
\bxsll satisfies $m_X = max(m_K,m_q+m_s)$.
The results for the hadron energy and the hadronic invariant mass spectra
are presented in Figs.~\ref{fig:SDEh} and ~\ref{fig:SDSh}, respectively.  
We do not show the $S_H$ distribution in the entire range, as it tends
monotonically to zero for larger values of $S_H$.

%
%
%
\begin{figure}[htb]
\vskip -1.0truein
\centerline{\epsfysize=8.5in
{\epsffile{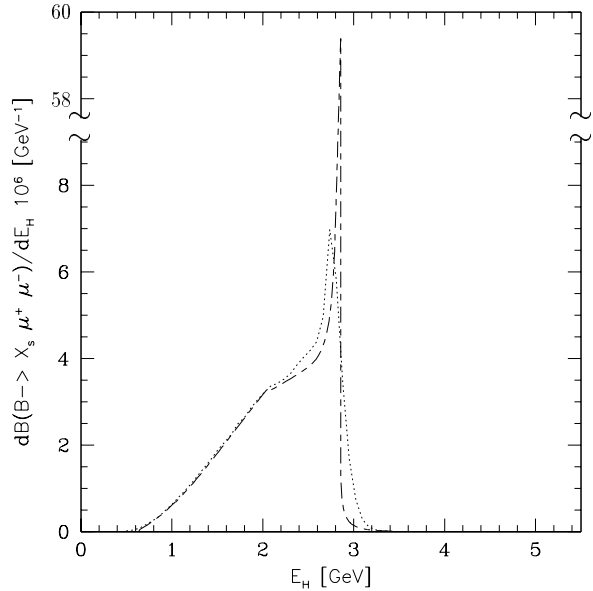}}}
\vskip -4.0truein
\caption[]{ \it Hadron energy spectrum in \bxsll
based on the perturbative contribution only, in the Fermi motion 
model (dotted curve) for $ (p_F, m_q)=(252,300)$ $(MeV,MeV)$, yielding
$m_b^{{\mbox{eff}}}=4.85$ GeV,  and in the parton 
model (long-short dashed curve) for $m_b=4.85$ GeV.}
\label{fig:SDEh485}   
\end{figure}
\begin{figure}[htb]
\vskip -0.0truein
\centerline{\epsfysize=3.5in   
{\epsffile{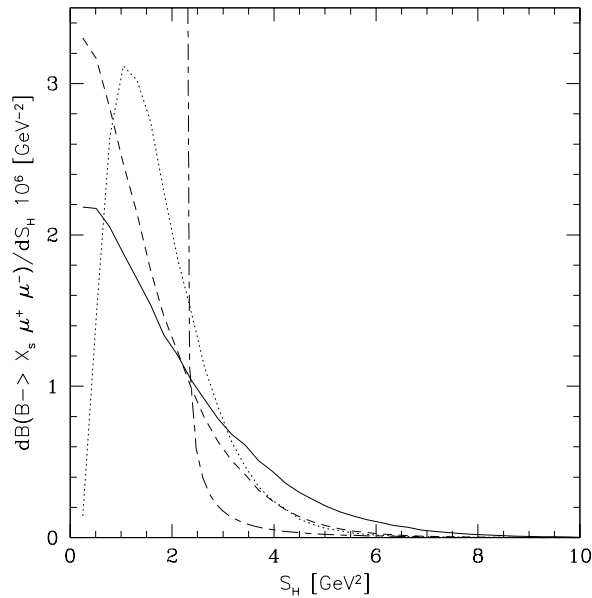}}}
\vskip -0.0truein
\caption[]{ \it Hadronic invariant mass spectrum in the Fermi motion 
model and parton model, based on the perturbative contribution only.
The solid, dotted, dashed
curve corresponds to the parameters
$ (\lambda_1, \bar{\Lambda})=(-0.3,0.5),(-0.1,0.4),(-0.15,0.35)$ in
(GeV$^2$, GeV), respectively.
The parton model (long-short dashed) curve is drawn for $m_b=4.85$ 
GeV.} \label{fig:SDSh}
\end{figure}
\begin{table}[h]
        \begin{center}
        \begin{tabular}{|l|l|l|l|}
        \hline
        \multicolumn{1}{|c|}{$p_F,m_q$ (MeV,MeV)}       & 
                \multicolumn{1}{|c|}{$m_b^{{\mbox{eff}}}$ (GeV)} & 
 \multicolumn{1}{|c|}{$\lambda_1$ $(\mbox{GeV}^2)$} &  
\multicolumn{1}{|c|}{$\bar{\Lambda}$ (GeV) } \\
        \hline \hline
        $(450,0)    $     & $4.76 $  & -0.304 & 0.507 \\
        $(252,300)  $    & $4.85 $  & -0.095 & 0.422  \\
        $(310,0)    $      & $4.92 $  & -0.144 & 0.350  \\
        $(450,150)  $     & $4.73 $  & -0.304 & 0.534 \\
        $(500,150)  $     & $4.68 $  & -0.375 & 0.588  \\
        $(570,150)  $     & $4.60 $  & -0.487 & 0.664  \\
        \hline
        \end{tabular}
        \end{center}
\caption{\it Values of non perturbative parameters
$m_b^{{\mbox{eff}}}$, $\lambda_1$ and $\bar{\Lambda}$   
for different sets of the FM model parameters $(p_F,m_q)$
taken from various fits of the data on $B \to X_s + (J/\psi,\gamma)$
decays discussed in ref.~\cite{AH98-3}.}
\label{tab:FMhqet}
\end{table}

  A number of remarks is in order:
\begin{itemize}
\item The hadron energy spectrum in \bxsll is rather insensitive to the 
model parameters. Also, the difference between the spectra in the FM and
the parton model is rather small as can be seen in Fig.~\ref{fig:SDEh485}. 
Since, away from the lower end-point and
the $c\bar{c}$ threshold, the parton model and HQET have very 
similar spectra (see Fig.~\ref{fig:dbdx0}), the estimates presented in  
Fig.~\ref{fig:SDEh} provide a good phenomenological profile of this
spectrum for the short-distance contribution. Very 
similar conclusions were drawn in
\cite{greubrey} for the corresponding spectrum in the decay $B \to X_u 
\ell \nu_\ell$, where, of course, the added complication of the $c\bar{c}$
threshold is not present.
\item In contrast to the hadron energy spectrum, the hadronic invariant
mass spectrum in \bxsll ~is sensitive to the model parameters, as can be 
seen in Fig.~\ref{fig:SDSh}. Again, one sees a close parallel in the
hadronic invariant mass spectra in \bxsll and   $B 
\to X_u \ell \nu_\ell$, with the latter worked out in \cite{FLW}.
 We think that the present 
theoretical dispersion on the hadron spectra in the decay \bxsll can be 
considerably reduced by the analysis of data in $B \to X_u \ell \nu_\ell$. 
\item The hadronic invariant mass distribution obtained by the 
$O(\alpha_s)$-corrected partonic spectrum and the HQET mass relation 
can only 
be calculated over a limited range of $S_H$, $S_H > m_B \bar{\Lambda}$, 
as shown in Fig.~\ref{fig:sh0}.
The larger is the value of $\bar{\Lambda}$, the smaller is this region. 
Also, in the range where it can be calculated, it depends on the
non-perturbative parameter $m_b$ (or $\bar{\Lambda}$). A comparison of this 
distribution 
and the one in the FM model may be made for the same values of $m_b$ and 
$m_b^{\mbox{eff}}$. This is shown for $m_b=4.85$ GeV in
Fig.~\ref{fig:SDSh} for HQET (long-short dashed curve) to be 
compared 
with the dotted curve in the FM model, which corresponds to
$m_b^{\mbox{eff}}=4.85$ GeV. We see that the two distributions differ though
they are qualitatively similar.

\end{itemize}
\subsection{Numerical Estimates of the Hadronic Moments in FM model and
HQET   \label{numerics}}

To underline the similarity of the HQET and FM descriptions in \bxsll,
and also to make comparison with data when it becomes 
available with the FM model, we have calculated the hadronic moments in 
the FM model 
using the spectra just described.
The  moments are defined as usual:
\begin{eqnarray}
\langle X_H^n\rangle\equiv(\int X_H^n\frac{d{\cal{B}}}{dX_H} dX_H)/{\cal{B}}
\hspace{1cm} {\mbox{for}} \, \, \,\,  X=S,E ~.
\end{eqnarray}
 The values of the moments in both the
HQET approach and the FM for $n=1,2$ are shown in
Table~\ref{tab:moments} for the decay $B \to X_s \mu^+ \mu^-$, with the 
numbers in the parentheses corresponding
to the former. They are based on using the central values of the
parameters given in Table~\ref{parameters} and are
calculated for the same
values of the HQET parameters $\bar{\Lambda}$ and $\lambda_1$, using 
the
transcriptions given in eqs.~(\ref{fmtohqet}).
Both the HQET and the FM model lead to strikingly similar results 
for the hadronic moments shown in this table. With  $\langle 
S_H\rangle \simeq (1.5 - 2.1)$ GeV,
the hadronic invariant mass spectra in \bxsll are expected to be 
dominated by multi-body states.
\begin{table}[h]
        \begin{center}   
        \begin{tabular}{|c|l|l|l|l|}
        \hline
        \multicolumn{1}{|c|}{{\mbox{}}}      &
                \multicolumn{1}{|c|}{$\langle S_H\rangle$  } &
                \multicolumn{1}{|c|}{$\langle S_H^2\rangle$  } &
               \multicolumn{1}{|c|}{$\langle E_H\rangle$  } &
                \multicolumn{1}{|c|}{$\langle E_H^2\rangle$  } \\
 \hline
\multicolumn{1}{|c|}{($\lambda_1,\bar{\Lambda})$ in (GeV$^2$, GeV)} &
\multicolumn{1}{|c|}{$({\mbox{GeV}}^2)$ } &
\multicolumn{1}{|c|}{$({\mbox{GeV}}^4)$ } &
\multicolumn{1}{|c|}{$({\mbox{GeV)}} $ } &
\multicolumn{1}{|c|}{$({\mbox{GeV}}^2)$ } \\
        \hline
    $(-0.3,0.5)$  & 2.03 (2.09)& 6.43 (6.93)&2.23 (2.28)& 5.27 (5.46)   \\
    $(-0.1,0.4)$  & 1.75 (1.80)& 4.04 (4.38) &2.21 (2.22)& 5.19 (5.23)    \\
    $(-0.14,0.35)$  & 1.54 (1.49)& 3.65 (3.64) &2.15 (2.18) &4.94 (5.04)   \\
        \hline
        \end{tabular}
        \end{center}
\caption{\it Hadronic spectral moments for $B \to X_s \mu^{+} \mu^{-}$
in the Fermi motion model (HQET) for the indicated
values of the parameters $(\lambda_1,\bar{\Lambda})$.
 }
\label{tab:moments}
\end{table}
\section{Branching Ratios and Hadron Spectra in \bxsll~with Cuts on 
Invariant Masses}
 The short-distance (SD) contribution (electroweak penguins
and boxes) is expected to be visible away from the resonance regions
dominated by
$B \to X_s (J/\psi,\psi^\prime,...) \to X_s \ell^+ \ell^-$.
So, cuts on the invariant dilepton mass are imposed to get quantitative
control over the long-distance (LD) resonant contribution. 
 For example, the cuts imposed in the recent CLEO analysis
\cite{cleobsll97} given below are typical: 
\begin{eqnarray}
{\mbox{cut A}}&:& 
q^2 \leq  (m_{J/\psi}-0.1 \, {\mbox{GeV}})^2 = 8.98 \, {\mbox{GeV}}^2 \, ,
\nonumber \\
{\mbox{cut B }}&:& 
q^2 \leq  (m_{J/\psi}-0.3 \, {\mbox{GeV}})^2 = 7.82 \, {\mbox{GeV}}^2 \, ,
\nonumber \\
{\mbox{cut C}}&:& 
q^2 \geq  (m_{\psi^{\prime}}+0.1 \, {\mbox{GeV}})^2 = 14.33 \, {\mbox{GeV}}^2
\, . \label{eq:cuts}
\end{eqnarray}
The cuts $A$ and $B$ have been chosen to take into account the
QED radiative corrections as these effects are different in the
$e^+ e^-$ and $\mu^+ \mu^{-}$ modes. In a forthcoming paper 
\cite{AH98-3}, we shall compare the hadron spectra with and without the 
$B \to (J/\psi,\psi^\prime,...) \to X_s \ell^+ \ell^-$ resonant parts after 
imposing these experimental cuts to quantify the theoretical uncertainty due 
to the residual LD-effects. Based on this study, we
argue that the above cuts in $q^2$ greatly reduce the resonant part.
Hence, the resulting distributions and moments with the above 
cuts essentially test (up to the non-perturbative aspects) the SD 
contribution in \bxsll. 

 As mentioned in \cite{cleobsll97}, the dominant $B\bar{B}$ background 
to the decay \bxsll comes from two semileptonic decays of $B$ or $D$
mesons, which produce the lepton pair with two undetected neutrinos.
 To suppress this $B\bar{B}$ background,
it is required that the invariant mass of the final hadronic state is 
less than $t=1.8 \, {\mbox{GeV}}$, which approximately equals $m_D$. 
We define the survival probability of the \bxsll signal 
after the hadronic invariant mass cut:
\begin{eqnarray}
S(t)\equiv (\int_{m_{X}^2}^{t^2} \frac{d{\cal{B}}}{dS_H} dS_H)/{\cal{B}}~,
\label{eq:eff}
\end{eqnarray}
and present $S(t=1.8 ~\mbox{GeV})$ as the
fraction of the branching ratio for \bxsll surviving these cuts  
in Table~\ref{SHoutcome}. We note that the effect
of this cut alone is that between $83\%$ to $92\%$ of the signal for
$B \to X_s \mu^+ \mu^-$ and between $79\%$ to $91\%$ of the signal
in $B \to X_s e^+ e^-$ survives, depending on the FM model parameters.
This shows that while this cut removes a good fraction of
the $B\bar{B}$ background, it allows a very large fraction of the \bxsll
signal to survive. However, this cut does not discriminate between the
SD- and LD- contributions, for which the cuts $A$ - $C$ are effective.

\begin{figure}[t]
\mbox{}\vspace{-2cm}\\
\begin{center}
     \mbox{ }\hspace{-0.7cm}
     \begin{minipage}[t]{7.0cm}
     \mbox{ }\hfill\hspace{1cm}(a)\hfill\mbox{ }
     \epsfig{file=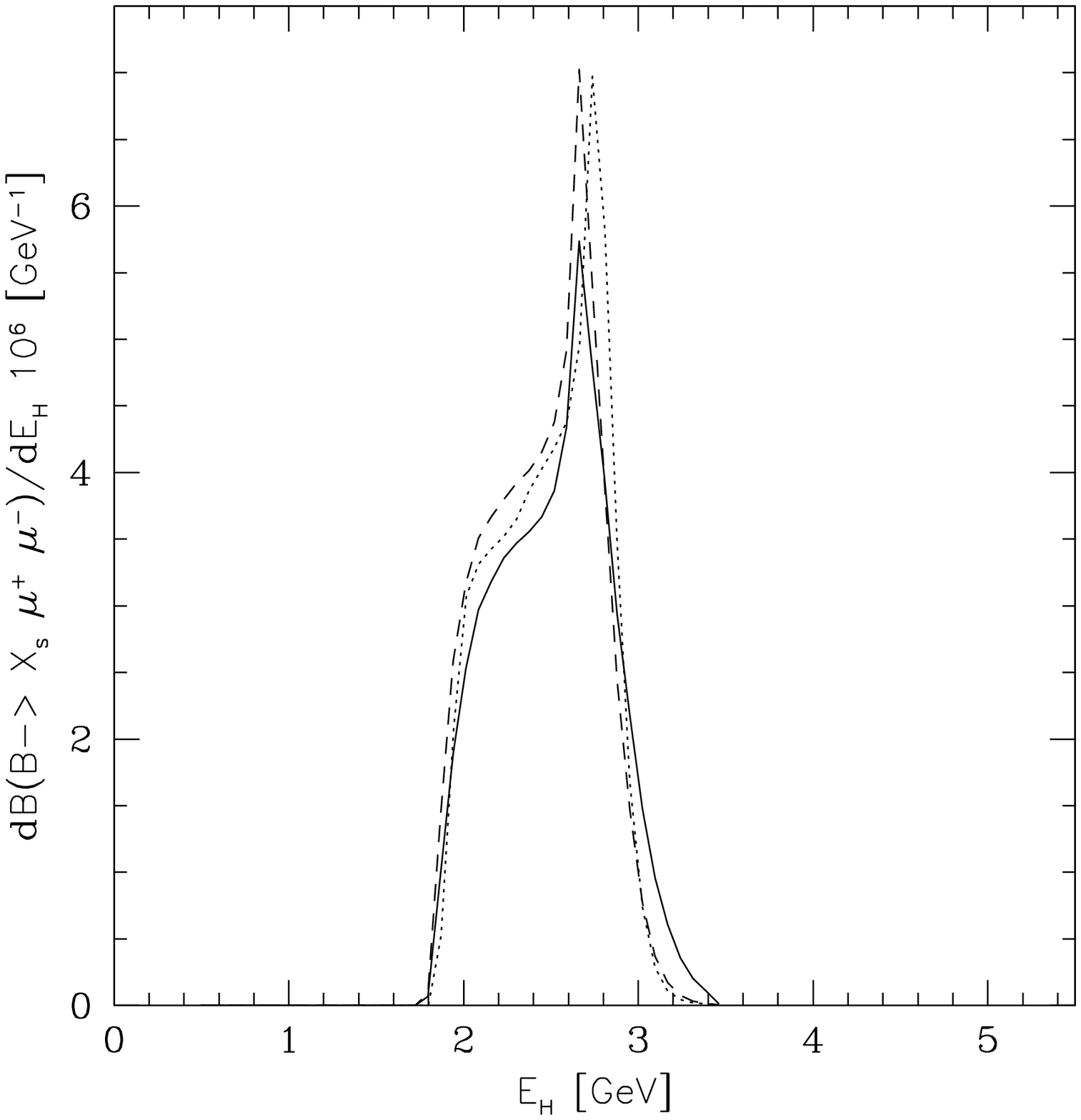,width=7.0cm}
     \end{minipage}
     \hspace{-0.4cm}
     \begin{minipage}[t]{7.0cm}
     \mbox{ }\hfill\hspace{1cm}(b)\hfill\mbox{ }
     \epsfig{file=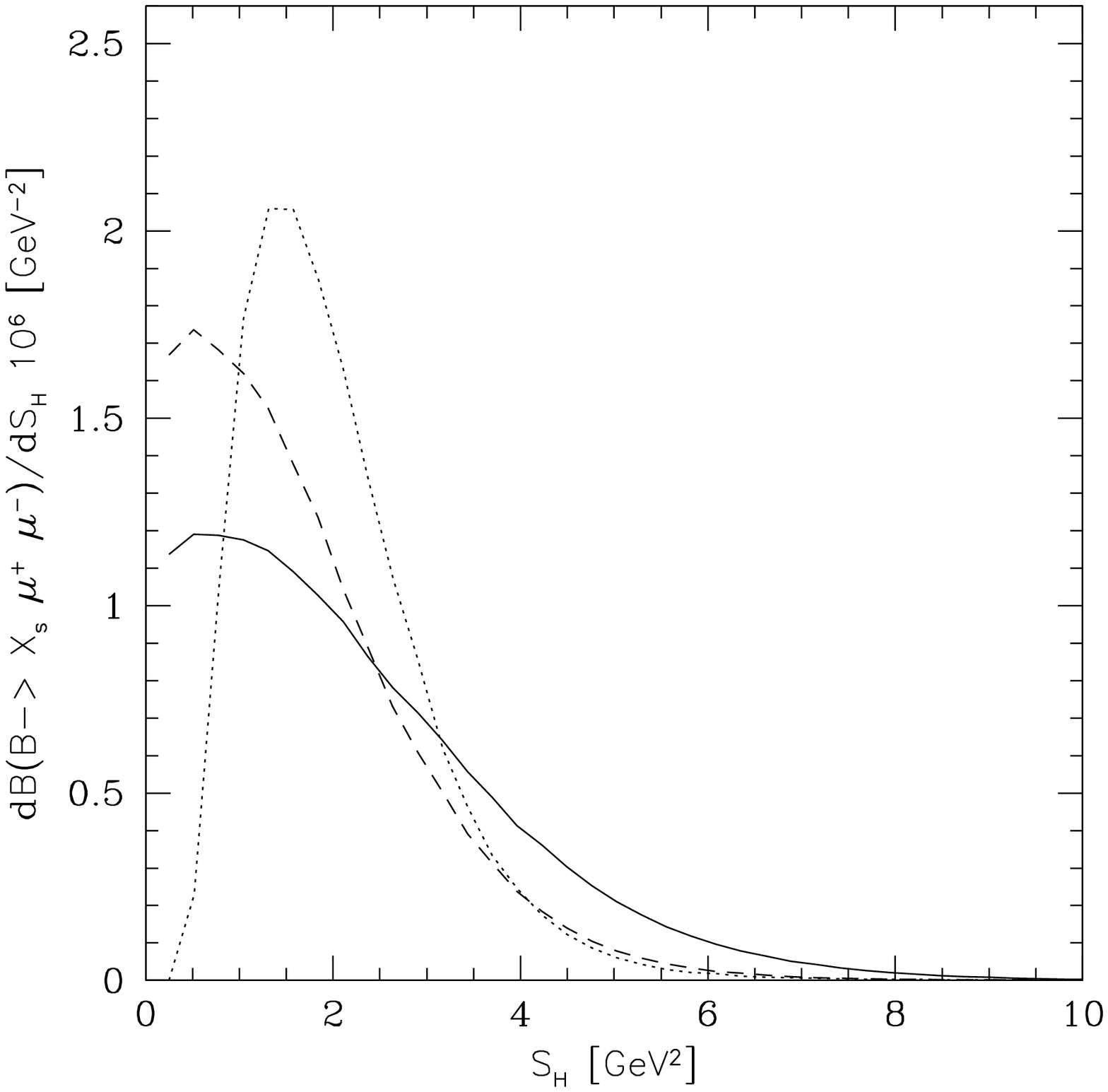,width=7.0cm}
     \end{minipage} \\ 
 \mbox{ }\hspace{-0.7cm}
     \begin{minipage}[t]{7.0cm}
     \mbox{ }\hfill\hspace{1cm}(c)\hfill\mbox{ }
     \epsfig{file=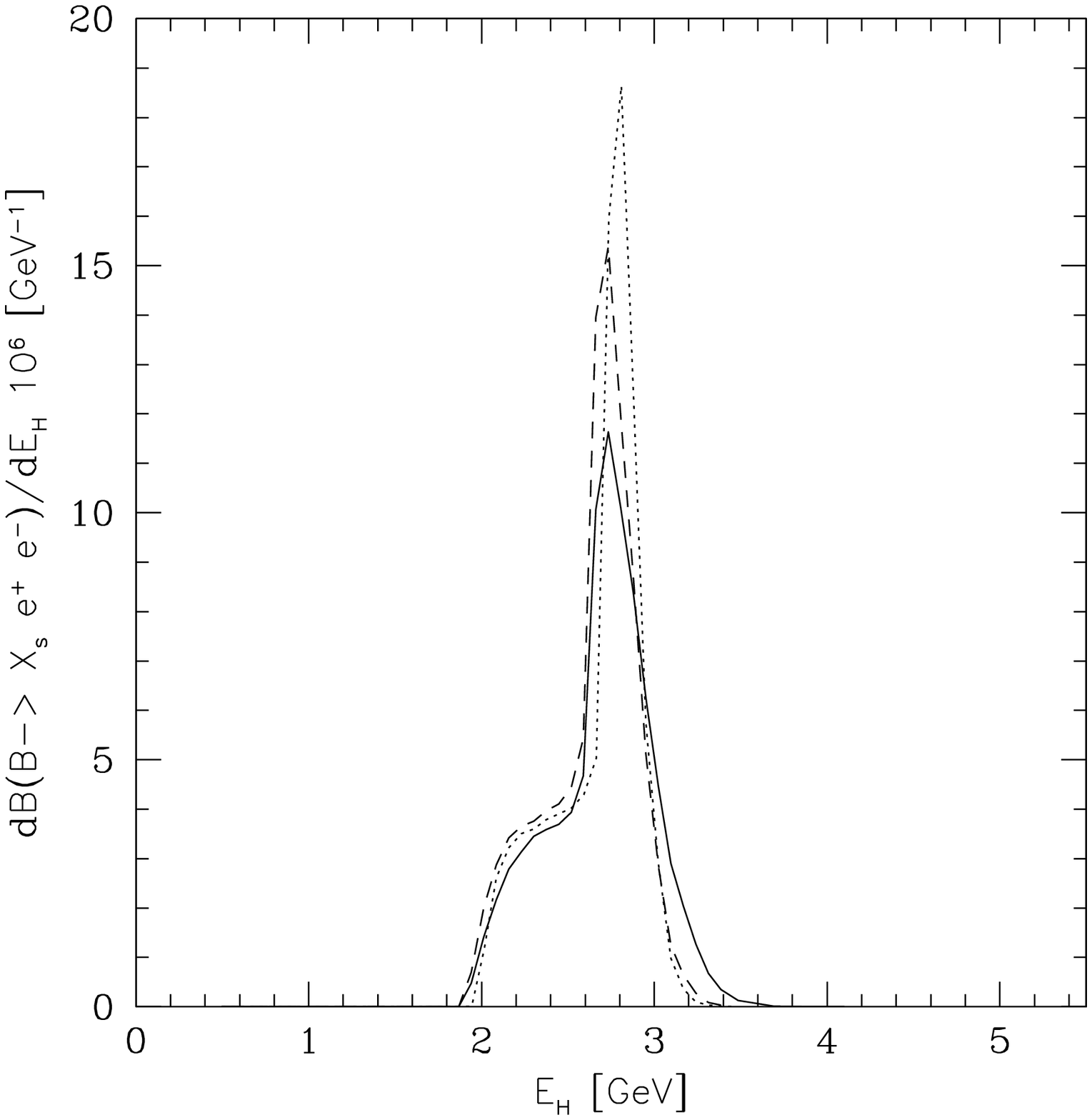,width=7.0cm}
     \end{minipage}
     \hspace{-0.4cm}
     \begin{minipage}[t]{7.0cm}
     \mbox{ }\hfill\hspace{1cm}(d)\hfill\mbox{ }
     \epsfig{file=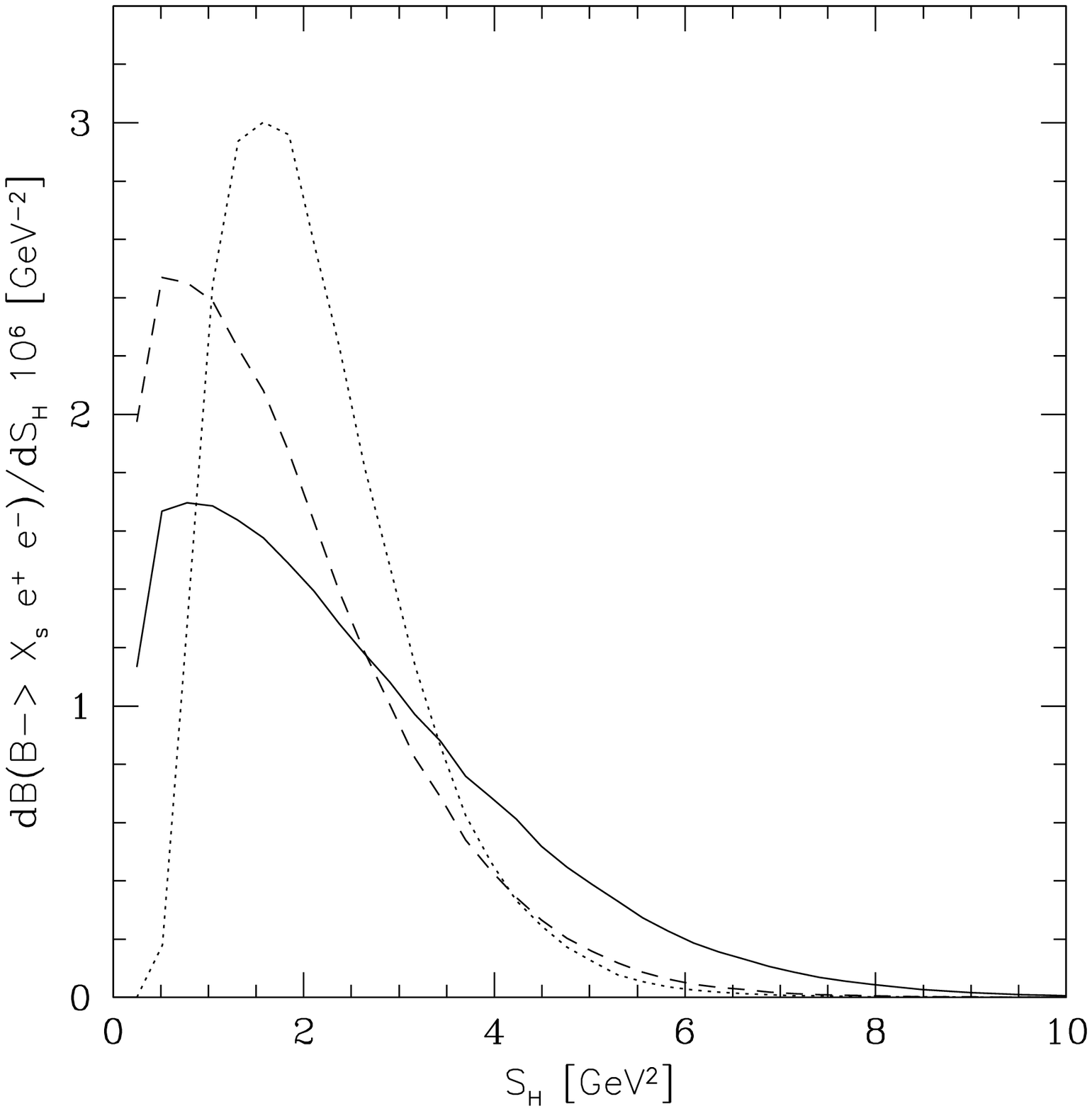,width=7.0cm}
     \end{minipage} \\ 
 \mbox{ }\hspace{-0.7cm}
     \begin{minipage}[t]{7.0cm}
     \mbox{ }\hfill\hspace{1cm}(e)\hfill\mbox{ }
     \epsfig{file=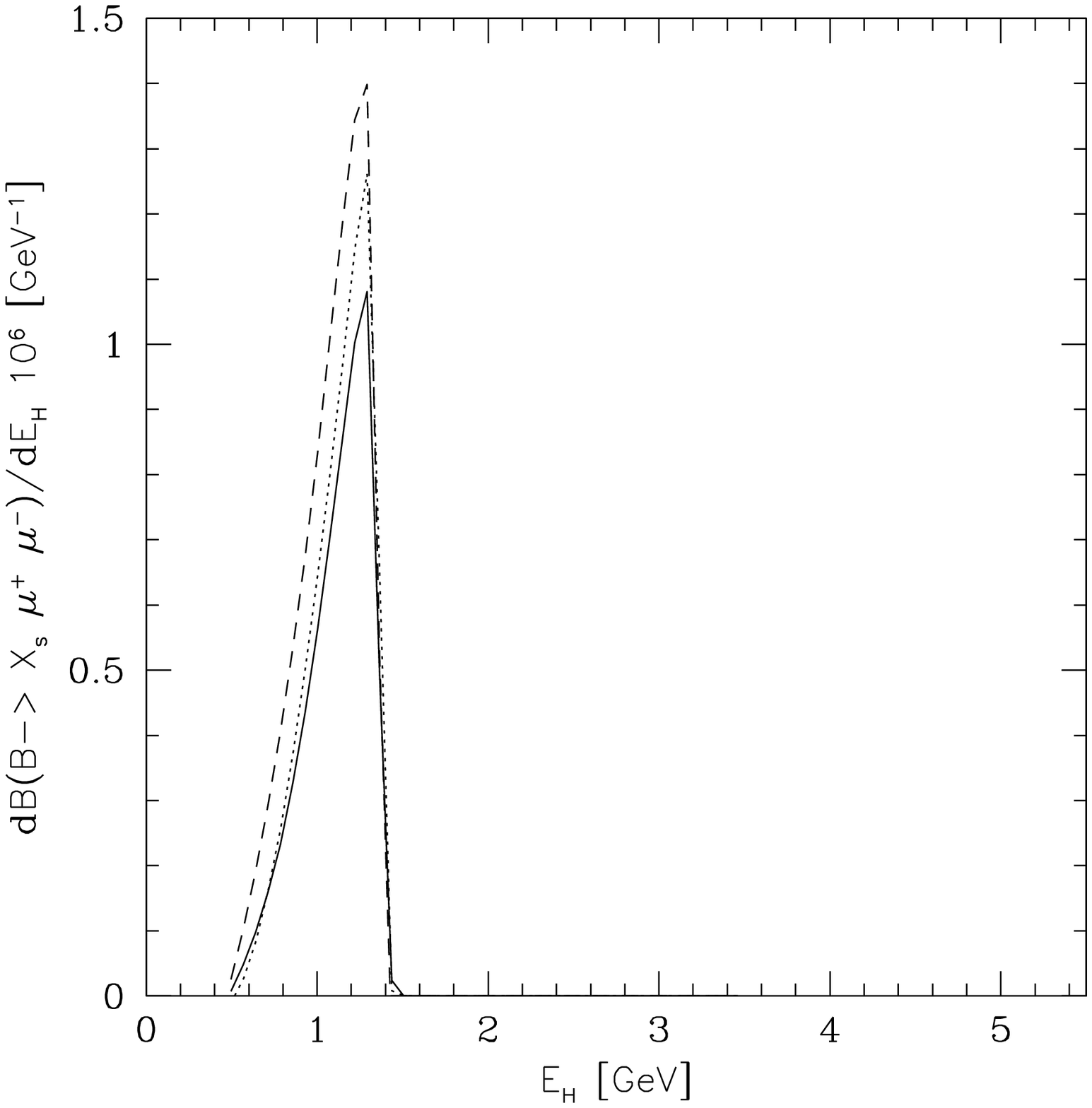,width=7.0cm}
     \end{minipage}
     \hspace{-0.4cm}
     \begin{minipage}[t]{7.0cm}
     \mbox{ }\hfill\hspace{1cm}(f)\hfill\mbox{ }
     \epsfig{file=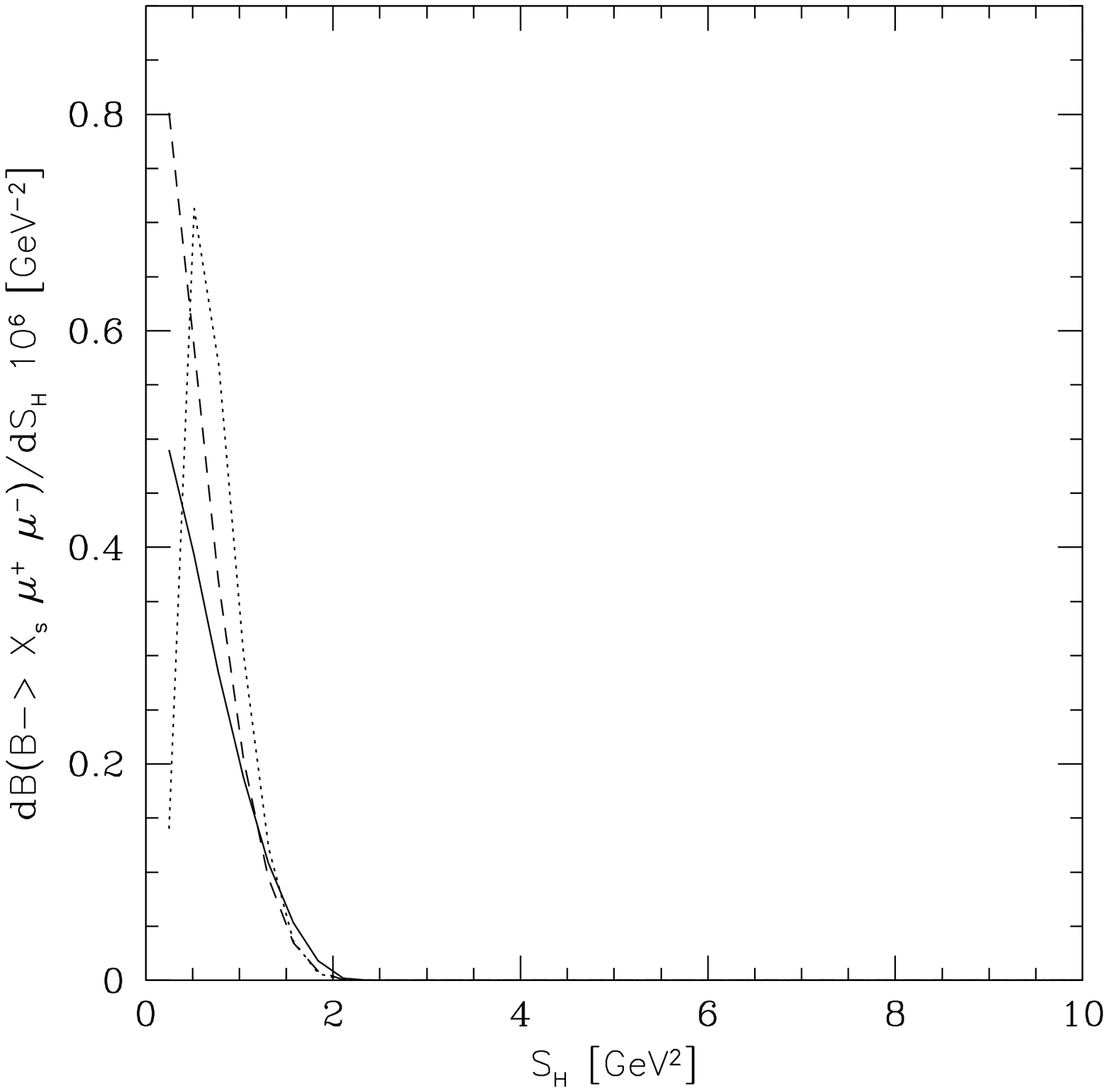,width=7.0cm}
     \end{minipage}
\end{center}  
     \caption{\it 
Hadron spectra in \bxsll in the Fermi motion model with the
cuts on the dilpeton mass defined in eq.~(\ref{eq:cuts}); (a),(c),(e) 
for the hadronic energy 
and (b),(d),(f) for the hadronic invariant mass corresponding to cut
A,B,C, respectively. The  
solid, dotted, dashed curves correspond to the parameters
$ (\lambda_1, \bar{\Lambda})=(-0.3,0.5),(-0.1,0.4),(-0.15,0.35)$ in
(GeV$^2$, GeV), respectively.
}\label{fig:EhLO}
\end{figure}

  With the cut A (B) imposed on the dimuon (dielectron) 
invariant 
mass, we find that between $57\%$ to $65\%$ ($57\%$ to $68\%$) of the \bxsll 
signal survives the additional cut
on the hadronic invariant mass for the SD contribution.
 The theoretical branching ratios for both the dielectron 
and dimuon
cases, calculated using the central values in Table ~\ref{parameters} are 
also given in Table~\ref{SHoutcome}. As estimated in \cite{AHHM97}, the 
uncertainty on the branching ratios resulting from the errors on the
parameters in Table~\ref{parameters} is about $\pm 28\%$ (for the dielectron 
mode) and
$\pm 21 \%$ (for the dimuon case). The wave-function-related uncertainty in
the branching ratios is negligible, as can be seen in Table 
~\ref{SHoutcome}. This reflects that, like in HQET, the corrections to
the decay rates for \bxsll~and $B \to X \ell \nu_\ell$ 
are of order $1/m_b^2$, and a good part of these corrections cancel
in the branching ratio for \bxsll. 
 With the help of the theoretical branching ratios and the survival
probability $S(t=1.8$ GeV), calculated for three sets of the FM parameters,
the branching ratios can be calculated for all six 
cases with the indicated cuts in  Table~\ref{SHoutcome}.
This gives a fair estimate of the theoretical uncertainties on
the partially integrated branching ratios from the $B$-meson wave 
function effects. This table shows that with
$10^7$ $B\bar{B}$ events, ${\cal 
O}(70)$ dimuon and ${\cal O}(100)$ dielectron signal events from 
\bxsll should survive the CLEO cuts A (B) with $m(X_s) <1.8$ GeV.
With cut C, one expects an order of magnitude less events, making 
this region interesting for the LHC experiments.
We show in Fig.~\ref{fig:EhLO} hadron spectra in \bxsll, $\ell^\pm 
=e^\pm,\mu^\pm$, resulting after imposing the CLEO cuts A, B,C, defined in
eq.~(\ref{eq:cuts}). One sees that the general features of the (uncut) 
theoretical distributions remain largely intact: the hadron energy spectra
are relatively insensitive to the FM parameters and the hadronic
invariant mass spectra showing a sensitive dependence on them. Given 
enough data, one can compare the experimental distributions in \bxsll 
directly with the ones presented in  Fig.~\ref{fig:EhLO}.
%
%
%
%
\begin{table}[h]
        \begin{center}
        \begin{tabular}{|l|l|l|l|l|l|l|l|l|}
        \hline
        \multicolumn{1}{|c|}{FM parameters}       &
\multicolumn{1}{|c|}{ $ {\cal{B}}\cdot 10^{-6}$} &
\multicolumn{1}{|c|}{${\cal{B}}\cdot 10^{-6}$} &
\multicolumn{1}{|c|}{No $s$-cut} &
\multicolumn{1}{|c|}{No $s$-cut } &
\multicolumn{1}{|c|}{cut A}  &
\multicolumn{1}{|c|}{cut B}  &
\multicolumn{1}{|c|}{cut C}  &
\multicolumn{1}{|c|}{cut C}  \\
 \multicolumn{1}{|c|}{($\lambda_1,\bar{\Lambda})$ in (GeV$^2$,
GeV)}           &
 \multicolumn{1}{|c|}{$\mu^{+} \mu^{-}$}   &
 \multicolumn{1}{|c|}{$e^+ e^-$}           &
 \multicolumn{1}{|c|}{$\mu^{+} \mu^{-}$}   &
 \multicolumn{1}{|c|}{$e^+ e^-$}           &
 \multicolumn{1}{|c|}{$\mu^{+} \mu^{-}$}   &
 \multicolumn{1}{|c|}{$e^+ e^-$}           &
 \multicolumn{1}{|c|}{$\mu^{+} \mu^{-}$}   &
 \multicolumn{1}{|c|}{$e^+ e^-$}           \\
        \hline \hline
   $(-0.3,0.5)  $  &5.8 &8.6 & $83 \% $  & 79 \%&$57 \% $  & $57 \% $
&$6.4\% $&$4.5\%$\\
   $(-0.1,0.4) $ &5.7 &8.4 & $93 \% $  & 91 \%&$63 \% $  & $68 \% $
&$8.3\%$&$5.8\%$\\
   $(-0.14,0.35)  $  &5.6 &8.3 & $92 \% $  & 90 \%&$65 \% $  & $67 \% $
&$7.9\%$&$5.5\%$\\
        \hline
        \end{tabular}
        \end{center}
\caption{\it Branching ratios for \bxsll, $\ell=\mu,e$ for different FM model
parameters are given in the second and third columns. The values
given in percentage in the fourth to ninth columns represent  the  
survival probability $S(t=1.8 {\mbox{ {\rm {GeV}}}})$,
defined in eq.~(\ref{eq:eff}), with no cut on the dilepton invariant 
mass and with cuts on this variable as defined in eq.~(\ref{eq:cuts}).} 
\label{SHoutcome} \end{table}
%

We have calculated the first two moments of the hadronic invariant mass in 
the FM model by imposing a cut $S_H < t^2$ with $t=1.8 \, \mbox{GeV}$ 
and an optional cut on $q^2$.
\begin{eqnarray}
\label{eq:SHcut}
\langle S_H^n\rangle=
(\int_{m_{X}^2}^{t^2} S_H^n\frac{d^2{\cal{B}}_{cut X}}{dS_Hdq^2} dS_Hdq^2)
/(\int_{m_{X}^2}^{t^2} \frac{d^2{\cal{B}}_{cut X}}{dS_Hdq^2} dS_Hdq^2)
\hspace{1cm} {\mbox{for}} \, \, \,\,  n=1,2 ~.
\end{eqnarray}  
Here the subscript $cut X$ indicates whether we evaluated 
$\langle S_H\rangle$ and $\langle S_H^2 \rangle$ with the cuts 
on the invariant dilepton mass as defined in 
eq.~(\ref{eq:cuts}), or without any cut on the dilepton mass. 
The results are collected in Table \ref{SHcutmoments}.
The moments given in Table \ref{SHcutmoments} can be
compared directly with the data to extract the FM model parameters.
The entries in this table  give a fairly
good idea of what the effects of the experimental cuts on the corresponding
moments in HQET will be, as the FM and HQET yield very similar moments
for equivalent values of the parameters. The 
functional dependence of 
the hadronic moments on the HQET parameters taking into account the 
experimental cuts still remains to be worked out. 

\begin{table}[h]
        \begin{center}
        \begin{tabular}{|l|l|l|l|l|l|l|l|l|l|l|}
        \hline
        \multicolumn{1}{|c|}{FM}       & 
\multicolumn{2}{|c|}{No $s$-cut} & 
\multicolumn{2}{|c|}{No $s$-cut } &  
\multicolumn{2}{|c|}{cut A}  &
\multicolumn{2}{|c|}{cut B}  &
\multicolumn{2}{|c|}{cut C}  \\
 \multicolumn{1}{|c|}{parameters}           & 
 \multicolumn{2}{|c|}{$\mu^{+} \mu^{-}$}   & 
 \multicolumn{2}{|c|}{$e^+ e^-$}           &  
 \multicolumn{2}{|c|}{$\mu^{+} \mu^{-}$}   & 
 \multicolumn{2}{|c|}{$e^+ e^-$}           &  
 \multicolumn{2}{|c|}{$\ell^{+} \ell^{-}$}   \\
 \multicolumn{1}{|c|}{($\lambda_1,\bar{\Lambda})$}           & 
 \multicolumn{1}{|c|}{$\langle S_H \rangle$ } &
\multicolumn{1}{|c|}{$\langle S_H^2\rangle$} &
 \multicolumn{1}{|c|}{$\langle S_H\rangle$  } &
\multicolumn{1}{|c|}{$\langle S_H^2\rangle$ } &
 \multicolumn{1}{|c|}{$\langle S_H\rangle$  } &
\multicolumn{1}{|c|}{$\langle S_H^2\rangle$ } &
 \multicolumn{1}{|c|}{$\langle S_H\rangle$  } &
\multicolumn{1}{|c|}{$\langle S_H^2\rangle$ } &
 \multicolumn{1}{|c|}{$\langle S_H\rangle$  } &
\multicolumn{1}{|c|}{$\langle S_H^2\rangle$ } \\
 \multicolumn{1}{|c|}{GeV$^2$, GeV}           & 
\multicolumn{1}{|c|}{${\mbox{GeV}}^2$ } &
\multicolumn{1}{|c|}{${\mbox{GeV}}^4$ } &
\multicolumn{1}{|c|}{${\mbox{GeV}}^2$ } &
\multicolumn{1}{|c|}{${\mbox{GeV}}^4$ } &
\multicolumn{1}{|c|}{${\mbox{GeV}}^2$ } &
\multicolumn{1}{|c|}{${\mbox{GeV}}^4$ } &
\multicolumn{1}{|c|}{${\mbox{GeV}}^2$ } &
\multicolumn{1}{|c|}{${\mbox{GeV}}^4$ } &
\multicolumn{1}{|c|}{${\mbox{GeV}}^2$ } &
\multicolumn{1}{|c|}{${\mbox{GeV}}^4$ } \\
        \hline \hline
$(-0.3,0.5)  $ &1.47&2.87&1.52&3.05&1.62&3.37&1.66&3.48&0.74&0.69\\
$(-0.1,0.4)  $ &1.57&2.98&1.69&3.37&1.80&3.71&1.88&3.99&0.74&0.63 \\
$(-0.14,0.35)$ &1.31&2.34&1.38&2.55&1.47&2.83&1.52&2.97&0.66&0.54\\
        \hline
        \end{tabular}
        \end{center}
\caption{\it $\langle S_H\rangle$ and $\langle S_H^2\rangle$ for 
\bxsll, $\ell=\mu,e$ for different FM model 
parameters and a hadronic invariant mass cut  $S_H <3.24 \, \mbox{GeV}^2$ 
are given in the second to fifth columns. 
The values in the sixth to eleventh columns have additional cuts on the 
dilepton invariant mass spectrum as defined in eq.~(\ref{eq:cuts}).
The $S_H$-moments with cuts are defined in eq.~(\ref{eq:SHcut}).} 
\label{SHcutmoments}
\end{table}

\section{Summary and Concluding Remarks}

We summarize our results:
\begin{itemize}
\item We have calculated the ${\cal O}(\alpha_s)$ perturbative 
QCD and leading ${\cal O}(1/m_b)$ corrections to the hadron spectra in the 
decay \bxsll, including the Sudakov-improvements in the
perturbative part.
\item We find that the hadronic invariant mass spectrum is calculable
in HQET over a limited range $S_H > m_B \bar{\Lambda}$ and it depends 
sensitively on the parameter $\bar{\Lambda}$ (equivalently $m_b$). 
These features are qualitatively very
similar to the ones found for the hadronic invariant mass spectrum in 
the decay $B \to X_u \ell \nu_\ell$ \cite{FLW}.
\item The $1/m_b$-corrections to the parton model hadron energy spectrum in 
\bxsll are small over most part of this spectrum. However, heavy quark
expansion breaks down near the lower end-point of this spectrum and 
close to the
$c\bar{c}$ threshold. The behavior in the former case 
has a similar origin as the breakdown of HQET near the high end-point
in the dilepton invariant mass spectrum, found in ref.~\cite{AHHM97}.
\item We have calculated the hadronic spectral moments $\langle S_H^n 
\rangle$ and $\langle E_H^n \rangle$ for $n=1,2$ using HQET.
The dependence of these moments on the HQET parameters is worked out 
numerically. In particular, the moments $\langle S_H^n
\rangle$ are sensitive to the parameters $\bar{\Lambda}$ and $\lambda_1$
and they provide complementary constraints on them than the ones
following from the analysis of the decay $B \to X \ell \nu_\ell$.
The simultaneous fit of the data in \bxsll and $B \to X \ell \nu_\ell$ could 
then be used to determine these 
parameters very precisely. This has been illustrated in ref.~\cite{AH98-1} 
based on the present work.
\item The corrections to the hadron energy
moments $\triangle \langle E_H \rangle_{1/m_c^2}$ and
$\triangle \langle E_H^2 \rangle_{1/m_c^2}$ from the leading
${\cal{O}}(\Lambda^2_{QCD}/m_c^2)$ power corrections have been worked out, 
using the results of
ref.~\cite{buchallaisidorirey}. We find that these corrections are very
small. The corresponding corrections in $\triangle \langle S_H^n 
\rangle_{1/m_c^2}$ vanish in the theoretical accuracy we are working.
\item We think that the quantitative knowledge of $\bar{\Lambda}$ and 
$\lambda_1$ from the moments can be used to remove much of the theoretical
uncertainties in the partially integrated decay rates 
in $B \to X_u \ell \nu_\ell$ and \bxsll. Realating the two decay rates would 
enable a precise determination of the CKM matrix element $V_{ub}$.

\item As a phenomenological alternative to HQET, we have worked out the 
hadron spectra and spectral moments in \bxsll~in the Fermi motion model 
\cite{aliqcd}.
This complements the description of the final states in \bxsll
presented in \cite{AHHM97}, where the dilepton invariant mass spectrum
and FB asymmetry were worked out in both the HQET and
FM model approaches. We find that the hadron energy spectrum is stable
against the variation of the FM model parameters. However, the hadronic
invariant mass is sensitive to the input parameters. For equivalent values
of the FM and HQET parameters, the spectral moments are found to be 
remarkably close to each other.

\item We have worked out the hadron spectra and spectral moments in the 
FM model by imposing the CLEO experimental
cuts designed to suppress the resonant $c\bar{c}$ contributions,
as well as the dominant $B\bar{B}$ background
leading to 
the final state $B\bar{B} \to X_s \ell^+ \ell^-$ (+ missing energy). The
parametric dependence of the resulting spectra is studied.  
In particular, the survival probability of the \bxsll~signal is estimated
by imposing a cut on the hadronic invariant mass $S_H < 3.24 
~\mbox{GeV}^2$ and on the dilepton invariant mass as used in the CLEO 
analysis. The spectra and moments can be directly compared with data.
\end{itemize}

 We hope that the work presented here will contribute to precise 
determinations of the HQET parameters and $V_{ub}$ using the inclusive decays
\bxsll~and $B \to X_u \ell \nu_\ell$ in forthcoming B facilities.


\bigskip
\noindent
{\Large \bf Acknowledgements}

 We would like to thank Tomasz Skwarnicki for drawing our attention to
the importance of the hadron spectra in the search for the
decay \bxsll. We thank Christoph Greub for helpful discussions.
Correspondence with Adam Falk and Gino Isidori on
power corrections are thankfully acknowledged.
%
%
\newpage
\centerline{\bf\underline{Appendices}\\}
\begin{appendix}
\renewcommand{\theequation}{\Alph{section}-\arabic{equation}}
\setcounter{equation}{0}

\section{Coefficient Functions 
$g_i^{(9,10)},g_i^{(7)},g_i^{(7,9)},h_i^{(9)},h_i^{(7,9)},
k_1^{(9)},k_1^{(7,9)}$ \label{app:auxfunc1}}
These functions enter in the derivation of the leading $(1/m_b^2)$
corrections to the hadron energy spectrum in \bxsll, given in 
 eq.~(\ref{singlediff}).
\begin{eqnarray}
g_0^{(9,10)}&=&
\sqrt{x_0^2-\ms^2} \frac{32}{3}(-2 \ms^2+3 x_0+3 \ms^2 x_0-4 x_0^2)
 \; , \\
g_1^{(9,10)}&=&
\frac{1}{\sqrt{x_0^2-\ms^2}} \frac{16}{9}(9 \ms^2+23 \ms^4
-18 \ms^2 x_0-18 x_0^2-52 \ms^2 x_0^2+36 x_0^3 + 20 x_0^4)
 \; , \\
g_2^{(9,10)}&=&
\frac{1}{\sqrt{x_0^2-\ms^2}} \frac{16}{3}(3 \ms^2+ 23 \ms^4-3 x_0 
-21 \ms^2 x_0 -6 x_0^2-52 \ms^2 x_0^2+36 x_0^3+20 x_0^4)
 \; , \\
g_0^{(7)}&=&
\sqrt{x_0^2-\ms^2}\frac{64}{3}(10 \ms^2+10 \ms^4-3 x_0-18 \ms^2 x_0
-3 \ms^4 x_0+ 2 x_0^2+2 \ms^2 x_0^2) 
 \; , \\
g_1^{(7)}&=&
\frac{1}{\sqrt{x_0^2-\ms^2}} \frac{1}{(x_0-\frac{1}{2}(1+\ms^2))^2}
\frac{-8}{9}(9 \ms^2+ 34 \ms^4+104 \ms^6+110 \ms^8+31 \ms^{10}-132 \ms^4 x_0
- 312 \ms^6 x_0 
\nonumber \\
&-& 180 \ms^8 x_0-18 x_0^2-170 \ms^2 x_0^2-58 \ms^4 x_0^2+74 \ms^6 x_0^2-20 \ms^8 x_0^2+72 x_0^3+564 \ms^2 x_0^3 + 576 \ms^4 x_0^3
\nonumber \\
&+& 228 \ms^6 x_0^3-116 x_0^4-676 \ms^2 x_0^4-436 \ms^4 x_0^4-20 \ms^6 x_0^4+72 x_0^5+240 \ms^2 x_0^5+ 24 \ms^4 x_0^5)
 \; , \\
g_2^{(7)}&=&
 \frac{1}{\sqrt{x_0^2-\ms^2}} \frac{1}{x_0-\frac{1}{2}(1+\ms^2)}
\frac{16}{3}(27 \ms^2+ 93 \ms^4+97 \ms^6 + 31 \ms^8-3 x_0-63 \ms^2 x_0-189 \ms^4 x_0  
\nonumber \\
&-& 129 \ms^6 x_0-18 x_0^2 -108 \ms^2 x_0^2 -62 \ms^4 x_0^2-20 \ms^6 x_0^2+ 72 x_0^3+ 324 \ms^2 x_0^3+ 180 \ms^4 x_0^3-60 x_0^4
\nonumber \\
&-& 152 \ms^2 x_0^4-20 \ms^4 x_0^4)
 \; , \\
g_0^{(7,9)}&=&
\sqrt{x_0^2-\ms^2} 128 (-2 \ms^2+x_0+\ms^2 x_0)
 \; , \\
g_1^{(7,9)}&=&
\frac{1}{\sqrt{x_0^2-\ms^2}} 64 (\ms^2 +3 \ms^4 +2 \ms^2 x_0-2 x_0^2 -
4 \ms^2 x_0^2)
 \; , \\
g_2^{(7,9)}&=&
\frac{1}{\sqrt{x_0^2-\ms^2}} 64 (5 \ms^2+9 \ms^4-x_0 + 5 \ms^2 x_0-6 x_0^2
-12 \ms^2 x_0^2)
 \; , \\
h_1^{(9,10)}&=&
\frac{32}{9} \sqrt{x_0^2-\ms^2}
(-12 \ms^2-6 \ms^4+9 x_0+19 \ms^2 x_0+3 x_0^2+15 \ms^2 x_0^2-28 x_0^3) 
 \; , \\
h_2^{(9,10)}&=&
\frac{32}{3} \sqrt{x_0^2-\ms^2}
(-6 \ms^4+ 3 x_0+5 \ms^2 x_0+ 3 x_0^2+15 \ms^2 x_0^2-20 x_0^3)
 \; , \\
h_1^{(7,9)}&=&
\frac{128}{3} \sqrt{x_0^2-\ms^2}
(-8 \ms^2-2 \ms^4+3 x_0-3 \ms^2 x_0+5 x_0^2+5 \ms^2 x_0^2)
 \; , \\
h_2^{(7,9)}&=&
\frac{128}{3} \sqrt{x_0^2-\ms^2}
(-4 \ms^2-6 \ms^4+3 x_0-15 \ms^2 x_0+7 x_0^2+15 \ms^2 x_0^2)
 \; , \\
k_1^{(9,10)}&=&
\frac{64}{9} \sqrt{x_0^2-\ms^2}^3
(2 \ms^2-3 x_0-3 \ms^2 x_0+ 4 x_0^2) 
 \; , \\
k_1^{(7,9)}&=&
\frac{-256}{3} \sqrt{x_0^2-\ms^2}^3
(-2 \ms^2+ x_0+ \ms^2 x_0)
 \; .
\label{eq:coefffunc}
\end{eqnarray}

\section{Auxiliary Functions $f_{\delta}(\lo,\lt),f_{\delta'}(\lo,\lt)$
 \label{app:auxfunc}}
The auxiliary functions given below are the coefficients of the singular 
terms in the derivation of the leading 
$(1/m_b^2)$ corrections to the hadron energy spectrum in \bxsll, given in
eq.~(\ref{singlediff}).
 \begin{eqnarray}
 f_{\delta}(\lo,\lt)&=&{\cal B}_0 
 \left\{ 
\left[\frac{2}{9} (1-\ms^2)^3 (5-\ms^2) \lo
\right. \right. 
\nonumber \\
&+& \left. \frac{2}{3} (1-\ms^2)^3 (-1+5 \ms^2) \lt 
\right] \left( |C_9^{\mbox{eff}}|^2 + |C_{10}|^2 \right)
\nonumber \\
&+& \left[\frac{1}{9}(1+ 12 \ml^2-88 \ml^4-4 \ms^2-36 \ml^2 \ms^2-736 \ml^4 \ms^2+ 5 \ms^4+ 24 \ml^2 \ms^4+ 720 \ml^4 \ms^4
\right. \nonumber \\
&+&24 \ml^2 \ms^6+ 160 \ml^4 \ms^6-5 \ms^8 -36 \ml^2 \ms^8 -56 \ml^4 \ms^8 )\frac{\lo}{\ml^2} 
+ \frac{4}{3} (-1+\ms^2) (-3
 \nonumber \\
&+& \left. 14\ml^2-2 \ms^2+166 \ml^2 \ms^2+8 \ms^4+154 \ml^2 \ms^4+2 \ms^6+50 \ml^2 \ms^6-5 \ms^8) \lt
\right] \frac{|C_7^{\mbox{eff}}|^2}{\ml^2}
\nonumber \\
&+&   \left[\frac{8}{3} (1-\ms^2)^3 (7+\ms^2) \lo 
+ \frac{8}{3} (1-\ms^2)^3 (13+15 \ms^2 ) \lt
\right]
 Re(C_9^{\mbox{eff}}) \, C_7^{\mbox{eff}} \nonumber \\
&+& \left. \lo (-1+\ms^2)^5 (\frac{2}{9} \frac{d |C_9^{\mbox{eff}}|^2}{d\s_0} 
+ \frac{8}{3} \frac{d Re(C_9^{\mbox{eff}})}{d\s_0}  \, C_7^{\mbox{eff}})
 \right\}.
\end{eqnarray}
\begin{eqnarray}
 f_{\delta'}(\lo,\lt)&=&{\cal B}_0 \lo 
 \left\{ 
    \frac{1}{9} (1-\ms^2)^5 \left( |C_9^{\mbox{eff}}|^2 + |C_{10}|^2 \right)
\right. \nonumber \\
&+& \frac{2}{9} (-1+\ms^2)^3 (-1+14 \ml^2+\ms^2+52 \ml^2 \ms^2+\ms^4
+14 \ml^2 \ms^4 -\ms^6)
 \frac{|C_7^{\mbox{eff}}|^2}{\ml^2} \nonumber \\
&+& \left.  \frac{4}{3} (1-\ms^2)^5
 Re(C_9^{\mbox{eff}}) \, C_7^{\mbox{eff}}  \right\}.
\end{eqnarray}

\section{The Functions $\alpha_i, \beta_i, \gamma_i, \delta_i$ 
\label{app:moments}}
The functions entering in the definition of the hadron moments 
in eq.~(\ref{momentexp}) are given in this appendix. Note that the 
functions $\alpha_i^{(n,m)}$ and $\beta_i^{(n,m)}$ multiply the
Wilson coefficients $|C_7^{\mbox{eff}}|^2$ and $C_{10}^2$, respectively.
Their results are given in a closed form.  The 
functions $\gamma_i^{(n,m)}$ multiply the Wilson coefficients 
$C_7^{\mbox{eff}} Re (C_9^{\mbox{eff}})$, of which 
$Re (C_9^{{\mbox{eff}}})$ is an implicit function of
$x_0$. Likewise, the functions $\delta_i^{(n,m)}$ multiply the Wilson
coefficient $|C_9^{{\mbox{eff}}}|^2$. 
The expressions for $\gamma_i^{(n,m)}$ and
$\delta_i^{(n,m)}$ are given in the form of one-dimensional integrals over 
$x_0$.

{\bf \underline{ The functions $\mathbf \alpha_i^{(n,m)}$}}
\begin{eqnarray}
\alpha_0^{(0,0)}&=&\frac{16}{9}(-8-26\ms^2+18 \ms^4+22 \ms^6-11 \ms^8)+ 
   \frac{32}{3}(-1+\ms^2)^3(1+\ms^2) \ln(4 {\hat m_l}^2)  \nonumber \\
&+& 
   \frac{64}{3} \ms^4 (-9-2 \ms^2+\ms^4) \ln(\ms)  \; ,\\
\alpha_1^{(0,0)}&=&\frac{1}{2} \alpha_0^{(0,0)} \; , \\
\alpha_2^{(0,0)}&=&\frac{8}{3}(-4+38 \ms^2-42 \ms^4-26 \ms^6-15 \ms^8) + 
   16 (-1+\ms^2)^2(3+8 \ms^2+5 \ms^4) \ln(4 {\hat m_l}^2)\nonumber \\
&+& 
   32 \ms^2(-8-17\ms^2-2\ms^4+5\ms^6) \ln (\ms)  \; ,\\
\alpha_0^{(0,1)}&=&\frac{2}{9}(-41-49 \ms^2+256 \ms^4-128 \ms^6-43 \ms^8) + 
   \frac{16}{3}(-1+\ms^2)^3 (1+\ms^2)^2 \ln(4 {\hat m_l}^2) \nonumber \\
&+& 
   \frac{16}{3} \ms^4 (3-\ms^2-2 \ms^4) \ln(\ms)  \; , \\
\alpha_1^{(0,1)}&=&\alpha_1^{(0,0)}  \; ,\\
\alpha_2^{(0,1)}&=&\frac{4}{9}(21+167 \ms^2+128 \ms^4-276 \ms^6-319 \ms^8)
+ \frac{16}{3}(-1+\ms^2)^2 (3+14 \ms^2+21 \ms^4+10 \ms^6) 
\ln(4 {\hat m_l}^2)\nonumber \\
&+& \frac{32}{3} \ms^2
      (3-24 \ms^2-18 \ms^4+\ms^6) \ln(\ms) \; , \\
\alpha_0^{(0,2)}&=&\frac{2}{45}(-119-144 \ms^2+45 \ms^4+320 \ms^6+45 \ms^8) + 
   \frac{8}{3}(-1+\ms^4)^3 \ln(4 {\hat m_l}^2)  \nonumber \\
& -& 
   \frac{16}{3} \ms^6 (8+3 \ms^2) \ln(\ms) \; , \\
\alpha_1^{(0,2)}&=&\frac{1}{27}(-127-278\ms^2+1075\ms^4-800 \ms^6+49\ms^8)
 +\frac{4}{9}(1-\ms^2)^3 (-7-17\ms^2-5\ms^4+5\ms^6)
\ln(4 {\hat m_l}^2)\nonumber \\
&+& \frac{8}{9} \ms^4 (18-38 \ms^2-13 \ms^4) \ln(\ms)  \; , \\
\alpha_2^{(0,2)}&=&\frac{1}{9}(27-46 \ms^2+1681 \ms^4-688 \ms^6-1189 \ms^8)
+ \frac{4}{3}(-1+\ms^4)^2 (3+20 \ms^2+25 \ms^4)
\ln(4 {\hat m_l}^2)\nonumber \\
&-& \frac{8}{3} \ms^4 (18+54 \ms^2+47 \ms^4) \ln(\ms) \; , \\
\alpha_0^{(1,0)}&=&0 \;  ,\\
\alpha_1^{(1,0)}&=&\frac{2}{9}(-23-159 \ms^2-112 \ms^4+304 \ms^6-45 \ms^8)
- \frac{16}{3} (-1+\ms^2)^4 (1+\ms^2)\ln(4 {\hat m_l}^2)\nonumber \\
& +& 
  \frac{16}{3} \ms^4 (-39-7 \ms^2+6 \ms^4)\ln(\ms) \; , \\
\alpha_2^{(1,0)}&=&\frac{2}{9}(-93-469 \ms^2+704 \ms^4-127 \ms^8) + 
\frac{16}{3}(-1+\ms^2)^3 (3+8 \ms^2+5 \ms^4) \ln(4 {\hat m_l}^2)\nonumber \\ 
&-& 
   \frac{112}{3} \ms^4 (3+3 \ms^2+2 \ms^4)\ln(\ms) \; ,   \\
\alpha_0^{(1,1)}&=&0 \; ,  \\
\alpha_1^{(1,1)}&=&\frac{2}{27}(-4-131 \ms^2+307 \ms^4-416 \ms^6+178 \ms^8)
-\frac{8}{9}(-1+\ms^2)^4 (1+6 \ms^2+5 \ms^4)\ln(4 {\hat m_l}^2)\nonumber \\ 
&+& \frac{16}{9} \ms^4 (9-35 \ms^2-7 \ms^4) \ln(\ms) \; ,  \\
\alpha_2^{(1,1)}&=&\frac{2}{9}(-60-185 \ms^2+173 \ms^4+160 \ms^6+70 \ms^8)
+\frac{8}{3}(-1+\ms^2)^3 (1+\ms^2)^2 (3+5 \ms^2)\ln(4 {\hat m_l}^2)\nonumber \\
&+& \frac{16}{3} \ms^4 (3-21 \ms^2-13 \ms^4)\ln(\ms)  \; , \\
\alpha_0^{(2,0)}&=&0  \; , \\
\alpha_1^{(2,0)}&=&\frac{8}{135}(119-176 \ms^2-1085 \ms^4+400 \ms^6+835 \ms^8)
+\frac{32}{9}(1-\ms^2)^5 (1+\ms^2) \ln(4 {\hat m_l}^2)\nonumber \\
&-& 
  \frac{64}{9} \ms^6 (28 + 5 \ms^2)\ln(\ms)  \; , \\
\alpha_2^{(2,0)}&=&0 \, .
\end{eqnarray}
{\bf \underline{ The functions $\mathbf \beta_i^{(n,m)}$}}
\begin{eqnarray}
\beta_0^{(0,0)}&=&
 \frac{2}{3} (1-8 \ms^2+8 \ms^6-\ms^8-24 \ms^4 \ln(\ms)) \; ,\\
\beta_1^{(0,0)}&=&\frac{1}{2}\beta_0^{(0,0)} \; , \\ 
\beta_2^{(0,0)}&=&-3+8 \ms^2-24 \ms^4+24 \ms^6-5 \ms^8- 
       24 \ms^4 \ln(\ms) \; , \\
\beta_0^{(0,1)}&=&\frac{1}{30}(7-25 \ms^2+160 \ms^4-160 \ms^6+25 \ms^8 - 
       7 \ms^{10} + 120 \ms^4 \ln(\ms) + 120 \ms^6 \ln(\ms)) \; ,\\
\beta_1^{(0,1)}&=&\beta_1^{(0,0)} \; , \\
\beta_2^{(0,1)}&=&\frac{1}{3} \ms^2 (7-20 \ms^2+20 \ms^6-7 \ms^8+24
 \ln(\ms) - 
       48 \ms^2 \ln(\ms)) \; ,  \\
\beta_0^{(0,2)}&=&\frac{2}{45}(2-3 \ms^2-30 \ms^4+30 \ms^8+3\ms^{10}-2\ms^{12} 
-        120 \ms^6\ln(\ms) ) \; , \\
\beta_1^{(0,2)}&=&\frac{1}{270}(43-135 \ms^2+1260 \ms^4-1440 \ms^6+405 \ms^8 - 
       153 \ms^{10} + 20 \ms^{12}+1080 \ms^4 \ln(\ms)  \nonumber \\
&+& 840 \ms^6 \ln(\ms)) \; , \\
\beta_2^{(0,2)}&=&\frac{1}{90}(13-315 \ms^2+1500 \ms^4-1560 \ms^6+315 \ms^8 + 
147 \ms^{10}-100 \ms^{12}+360 \ms^4\ln(\ms) 
{\mbox{\hspace{2.4cm}}} \nonumber \\
 &+& 840 \ms^6 \ln(\ms) ) \; , \\
\beta_0^{(1,0)}&=&0 \; , \\
\beta_1^{(1,0)}&=&\frac{1}{30}(13-135 \ms^2-160 \ms^4+320 \ms^6-45 \ms^8+
7 \ms^{10}-600 \ms^4 \ln(\ms)-120 \ms^6 \ln(\ms)) \; , \\
\beta_2^{(1,0)}&=&\frac{1}{6}(3-9 \ms^2+16 \ms^4-48 \ms^6+45 \ms^8-7 \ms^{10}+
24 \ms^4  \ln(\ms)-72 \ms^6  \ln(\ms)) \; , \\
\beta_0^{(1,1)}&=&0 \; , \\
\beta_1^{(1,1)}&=&\frac{1}{270}(23-45\ms^2+1080\ms^4-1440 \ms^6+585\ms^8-243
\ms^{10}+40 \ms^{12}+1080 \ms^4 \ln(\ms) \nonumber \\
&+&600 \ms^6 \ln(\ms) ) \; , \\
\beta_2^{(1,1)}&=&\frac{1}{90}(13+45 \ms^2-120 \ms^4-45 \ms^8+147 \ms^{10}-40 \ms^{12}+360 \ms^4\ln(\ms)-600 \ms^6 \ln(\ms) ) \; , \\
\beta_0^{(2,0)}&=&0 \; , \\
\beta_1^{(2,0)}&=&\frac{16}{135}(-1+9\ms^2-45 \ms^4+45 \ms^8-9 \ms^{10}+\ms^{12}-120 \ms^6\ln(\ms)) \; , \\
\beta_2^{(2,0)}&=&0 \;  .
\end{eqnarray}
{\bf \underline{ The functions $\mathbf \gamma_i^{(n,m)}$}}

Note that in the expressions given below 
$C_9^{\mbox{eff}}\equiv C_9^{\mbox{eff}}(\s=1-2 x_0+\ms^2)$.
The lower and upper limits of the $x_0$-integrals are:
$x_0^{min}=\ms$ and $x_0^{max}=\frac{1}{2}(1+\ms^2-4 \ml^2)$.
\begin{eqnarray}
\gamma_0^{(0,0)}&=&128 \int_{x_0^{min}}^{x_0^{max}} d x_0
\sqrt{x_0^2-\ms^2}(-2 \ms^2+x_0+\ms^2 x_0)
{\it{Re}}(C_9^{{\mbox{eff}}}) \; , \\
\gamma_1^{(0,0)}&=&\frac{1}{2} \gamma_0^{(0,0)} \; , \\ 
\gamma_2^{(0,0)}&=&\int_{x_0^{min}}^{x_0^{max}} d x_0
\frac{64}{\sqrt{x_0^2-\ms^2}} (4 \ms^2+14 \ms^4-x_0-\ms^2 x_0-12 \ms^4 x_0
-4 x_0^2-22 \ms^2 x_0^2+7 x_0^3 
{\mbox{\hspace{2.6cm}}}
\nonumber \\
&+&15 \ms^2 x_0^3)
{\it{Re}}(C_9^{{\mbox{eff}}}) \; , \\
\gamma_0^{(0,1)}&=&128 \int_{x_0^{min}}^{x_0^{max}} d x_0
 x_0 \sqrt{x_0^2-\ms^2} (-2 \ms^2+x_0+\ms^2 x_0)
{\it{Re}}(C_9^{{\mbox{eff}}}) \; , \\
\gamma_1^{(0,1)}&=& \gamma_1^{(0,0)} \; , \\
\gamma_2^{(0,1)}&=&\frac{64}{3} \int_{x_0^{min}}^{x_0^{max}} d x_0
\frac{1}{\sqrt{x_0^2-\ms^2}}(4 \ms^4+6 \ms^6+9  \ms^2 x_0 +57 \ms^4 x_0-
3 x_0^2-14 \ms^2 x_0^2-57 \ms^4 x_0^2\nonumber \\
&-&9 x_0^3-81 \ms^2 x_0^3+28 x_0^4+60 \ms^2 x_0^4)
{\it{Re}}(C_9^{{\mbox{eff}}}) \; , \\
\gamma_0^{(0,2)}&=&128 \int_{x_0^{min}}^{x_0^{max}} d x_0
x_0^2 \sqrt{x_0^2-\ms^2} (-2 \ms^2+x_0+\ms^2 x_0)
{\it{Re}}(C_9^{{\mbox{eff}}}) \; , \\
\gamma_1^{(0,2)}&=&\frac{64}{3} \int_{x_0^{min}}^{x_0^{max}} d x_0
\sqrt{x_0^2-\ms^2} (-4 \ms^4-10 \ms^2 x_0+2 \ms^4 x_0+6 x_0^2 +16 \ms^2 x_0^2
-5 x_0^3 \nonumber \\
&-&5 \ms^2 x_0^3)
{\it{Re}}(C_9^{{\mbox{eff}}}) \; , \\
\gamma_2^{(0,2)}&=&\frac{64}{3} \int_{x_0^{min}}^{x_0^{max}} d x_0
\frac{x_0}{\sqrt{x_0^2-\ms^2}} (8 \ms^4+12 \ms^6+6 \ms^2 x_0+72 \ms^4 x_0-3 x_0^2-25 \ms^2 x_0^2-78 \ms^4 x_0^2 \nonumber \\
&-&6 x_0^3-96 \ms^2 x_0^3+35 x_0^4+75 \ms^2 x_0^4)
{\it{Re}}(C_9^{{\mbox{eff}}}) \; , \\
\gamma_0^{(1,0)}&=&0 \; , \\
\gamma_1^{(1,0)}&=&128 \int_{x_0^{min}}^{x_0^{max}} d x_0
\frac{(x_0-1)}{\sqrt{x_0^2-\ms^2}} (-2 \ms^4+ \ms^2 x_0+\ms^4 x_0+2 \ms^2 x_0^2-x_0^3-\ms^2 x_0^3){\it{Re}}(C_9^{{\mbox{eff}}}) \; , \\
\gamma_2^{(1,0)}&=&\frac{128}{3} 
\int_{x_0^{min}}^{x_0^{max}}  d x_0 \sqrt{x_0^2-\ms^2}
(-4 \ms^2 -6 \ms^4+3 x_0-15 \ms^2 x_0+7 x_0^2+15 \ms^2 x_0^2)
{\it{Re}}(C_9^{{\mbox{eff}}}) \; , \\
\gamma_0^{(1,1)}&=&0 \; , \\
\gamma_1^{(1,1)}&=&\frac{128}{3} 
\int_{x_0^{min}}^{x_0^{max}}  d x_0 \frac{1}{\sqrt{x_0^2-\ms^2}}
(4 \ms^6+4 \ms^4 x_0-2 \ms^6 x_0 -3 \ms^2 x_0^2 -17 \ms^4 x_0^2+\ms^2 x_0^3+7 \ms^4 x_0^3 \nonumber \\
&+& 3 x_0^4+13 \ms^2 x_0^4-5 x_0^5-5 \ms^2 x_0^5)
{\it{Re}}(C_9^{{\mbox{eff}}}) \; , \\
\gamma_2^{(1,1)}&=&\frac{128}{3} 
\int_{x_0^{min}}^{x_0^{max}}  d x_0 x_0 \sqrt{x_0^2-\ms^2}
(-4 \ms^2-6 \ms^4+3 x_0-15 \ms^2 x_0+7 x_0^2+15 \ms^2 x_0^2)
{\it{Re}}(C_9^{{\mbox{eff}}}) \; , \\
\gamma_0^{(2,0)}&=&0 \; , \\
\gamma_1^{(2,0)}&=& \frac{512}{3} \int_{x_0^{min}}^{x_0^{max}} 
d x_0 \sqrt{x_0^2-\ms^2}
(-2 \ms^4 + \ms^2 x_0 + \ms^4 x_0+2 \ms^2 x_0^2-x_0^3 
- \ms^2 x_0^3) {\it{Re}}(C_9^{{\mbox{eff}}}) \; , \\
\gamma_2^{(2,0)}&=&0 \; .
\end{eqnarray}
{\bf \underline{ The functions $\mathbf \delta_i^{(n,m)}$}}
\begin{eqnarray}
\delta_0^{(0,0)}&=&\frac{32}{3} \int_{x_0^{min}}^{x_0^{max}} d x_0
\sqrt{x_0^2-\ms^2} (-2 \ms^2+3 x_0+3 \ms^2 x_0-4 x_0^2)
|C_9^{{\mbox{eff}}}|^2 \; , \\
\delta_1^{(0,0)}&=&\frac{1}{2}\delta_0^{(0,0)} \; , \\ 
\delta_2^{(0,0)}&=& \int_{x_0^{min}}^{x_0^{max}} d x_0
\frac{16}{\sqrt{x_0^2-\ms^2}}(6 \ms^4-x_0-9 \ms^2 x_0-12 \ms^4 x_0+6 \ms^2 x_0^2+15 x_0^3+15 \ms^2 x_0^3 
{\mbox{\hspace{3.0cm}}} \nonumber \\
&-& 20 x_0^4)
|C_9^{{\mbox{eff}}}|^2 \; , \\
\delta_0^{(0,1)}&=&\frac{32}{3} \int_{x_0^{min}}^{x_0^{max}} d x_0
x_0 \sqrt{x_0^2-\ms^2}  (-2 \ms^2+3 x_0+3 \ms^2 x_0-4 x_0^2)
|C_9^{{\mbox{eff}}}|^2 \; , \\
\delta_1^{(0,1)}&=&\delta_1^{(0,0)}  \; , \\
\delta_2^{(0,1)}&=&\frac{16}{3} \int_{x_0^{min}}^{x_0^{max}} d x_0
\frac{1}{\sqrt{x_0^2-\ms^2}} (6 \ms^6-3 \ms^2 x_0+13 \ms^4 x_0-3 x_0^2-30 \ms^2 x_0^2-57 \ms^4 x_0^2 +3 x_0^3 \nonumber \\
&+&43 \ms^2 x_0^3+48 x_0^4+60 \ms^2 x_0^4-80 x_0^5)
|C_9^{{\mbox{eff}}}|^2 \; , \\
\delta_0^{(0,2)}&=&\frac{32}{3} \int_{x_0^{min}}^{x_0^{max}} d x_0
 x_0^2 \sqrt{x_0^2-\ms^2} (-2 \ms^2+3 x_0+3 \ms^2 x_0-4 x_0^2)
|C_9^{{\mbox{eff}}}|^2 \; , \\
\delta_1^{(0,2)}&=&\frac{16}{9} \int_{x_0^{min}}^{x_0^{max}} d x_0
\sqrt{x_0^2-\ms^2} (-4 \ms^4-6 \ms^2 x_0+6 \ms^4 x_0+18 x_0^2+20 \ms^2 x_0^2-39 x_0^3-15 \ms^2 x_0^3 \nonumber \\
&+& 20 x_0^4)
|C_9^{{\mbox{eff}}}|^2 \; , \\
\delta_2^{(0,2)}&=&\frac{16}{3} \int_{x_0^{min}}^{x_0^{max}} d x_0
\frac{x_0}{\sqrt{x_0^2-\ms^2}}(12 \ms^6-6 \ms^2 x_0+8 \ms^4 x_0-3 x_0^2-33 \ms^2 x_0^2-78 \ms^4 x_0^2+6 x_0^3 \nonumber \\
&+&68 \ms^2 x_0^3+51 x_0^4+75 \ms^2 x_0^4-
100 x_0^5)
|C_9^{{\mbox{eff}}}|^2 \; , \\
\delta_0^{(1,0)}&=&0 \; , \\
\delta_1^{(1,0)}&=&\frac{32}{3} \int_{x_0^{min}}^{x_0^{max}} d x_0
\frac{(x_0-1)}{\sqrt{x_0^2-\ms^2}} (-2 \ms^4+ 3\ms^2 x_0+3\ms^4 x_0-2 \ms^2 x_0^2-3 x_0^3-3\ms^2 x_0^3+4 x_0^4)
|C_9^{{\mbox{eff}}}|^2 \, , \\
\delta_2^{(1,0)}&=&\frac{32}{3} \int_{x_0^{min}}^{x_0^{max}} d x_0
\sqrt{x_0^2-\ms^2} (-6 \ms^4+3 x_0+5 \ms^2 x_0+3 x_0^2+15 \ms^2 x_0^2-20 x_0^3)
|C_9^{{\mbox{eff}}}|^2 \; , \\
\delta_0^{(1,1)}&=& 0 \; , \\
\delta_1^{(1,1)}&=&\frac{32}{9} \int_{x_0^{min}}^{x_0^{max}} d x_0
\frac{1}{\sqrt{x_0^2-\ms^2}} (4 \ms^6-6 \ms^6 x_0-9 \ms^2 x_0^2-15 \ms^4 x_0^2+27 \ms^2 x_0^3+ 21 \ms^4 x_0^3 \nonumber \\
&+&9 x_0^4-9 \ms^2 x_0^4-27 x_0^5-15 \ms^2 x_0^5+20 x_0^6)
|C_9^{{\mbox{eff}}}|^2 \; , \\
\delta_2^{(1,1)}&=&\frac{32}{3} \int_{x_0^{min}}^{x_0^{max}} d x_0 x_0
\sqrt{x_0^2-\ms^2} (-6 \ms^4+3 x_0+5 \ms^2 x_0+3 x_0^2+15 \ms^2 x_0^2-20 x_0^3)
|C_9^{{\mbox{eff}}}|^2 \; , \\
\delta_0^{(2,0)}&=&0 \; , \\
\delta_1^{(2,0)}&=&\frac{128}{9} \int_{x_0^{min}}^{x_0^{max}} 
d x_0 \sqrt{x_0^2-\ms^2}
(-2 \ms^4 +3 \ms^2 x_0 +3 \ms^4 x_0-2 \ms^2 x_0^2-3 x_0^3 \nonumber \\
&-& 3 \ms^2 x_0^3+
4 x_0^4) |C_9^{{\mbox{eff}}}|^2 \; , \\
\delta_2^{(2,0)}&=&0 \; .
\end{eqnarray}
\section{Lowest Hadronic Moments (Parton Level) \label{app:lowmoments}}

\begin{eqnarray}
\langle x_0\rangle {{\cal B}\over {\cal B}_0} &=&
\frac{2}{9 m_B^2}(-41 m_B^2-49 m_s^2-
24 (m_B^2- m_s^2) \ln(4 \frac{m_l^2}{m_B^2})){C_7^{\mbox{eff}}}^2+
\frac{1}{30 m_B^2} (7 m_B^2-25 m_s^2)C_{10}^2 \nonumber \\
&+&
\int_{m_s/m_B}^{\frac{1}{2}(1+m_s^2/m_B^2)} d x_0 \frac{64}{m_B^2} x_0
(-m_s^2-4 m_s^2 x_0+2 m_B^2 x_0^2+2 m_s^2 x_0^2)
{\it{Re}}(C_9^{{\mbox{eff}}})C_7^{{\mbox{eff}}} \nonumber \\
&+&
\int_{m_s/m_B}^{\frac{1}{2}(1+m_s^2/m_B^2)} d x_0
\frac{16}{3 m_B^2} x_0(-3 m_s^2+6 m_B^2 x_0^2+6 m_s^2 x_0^2-8 m_B^2 x_0^3)
|C_9^{{\mbox{eff}}}|^2\nonumber \\
&+&\frac{\alpha_s}{\pi} A^{(0,1)} C_9^2+ 
\frac{-32}{3} {C_7^{\mbox{eff}}}^2 \frac{\bar{\Lambda}}{m_B}+
\frac{-16}{3}{C_7^{\mbox{eff}}}^2\frac{\bar{\Lambda}^2}{m_B^2}
+ \left[ \frac{-16}{9} (1+3\ln(4 \frac{m_l^2}{m_B^2})){C_7^{\mbox{eff}}}^2+
 \frac{C_{10}^2}{3}  \right. \nonumber \\
&+& \left.
\int_{0}^{\frac{1}{2}} d x_0 
(64 x_0^2{\it{Re}}(C_9^{{\mbox{eff}}})C_7^{{\mbox{eff}}}
+\frac{16}{3} (3-4 x_0)x_0^2 |C_9^{{\mbox{eff}}}|^2 ) \right] 
\frac{\lambda_1}{m_B^2}
\nonumber \\
&+& 
 \left[ \frac{4}{3} (19+12 \ln(4 \frac{m_l^2}{m_B^2})){C_7^{\mbox{eff}}}^2
+\int_{0}^{\frac{1}{2}} d x_0
(\frac{64}{3} x_0(-3-9 x_0+28 x_0^2){\it{Re}}
(C_9^{{\mbox{eff}}})C_7^{{\mbox{eff}}}
 \right. \nonumber \\
&+& \left.
\frac{16}{3} x_0 (-3+3 x_0+48 x_0^2-80 x_0^3)|C_9^{{\mbox{eff}}}|^2 \right] 
\frac{\lambda_2}{m_B^2} \, ,
\end{eqnarray}
\begin{eqnarray}
\langle x_0^2\rangle {{\cal B}\over {\cal B}_0} &=&
\frac{2}{45 m_B^{12}}(-119 m_B^{12}-144 m_B^{10} m_s^2-
60 (m_B^{12}- m_s^{12}) \ln(4 \frac{m_l^2}{m_B^2})){C_7^{\mbox{eff}}}^2+
\frac{2}{45 m_B^2} (2 m_B^2-3 m_s^2)C_{10}^2 \nonumber \\
&+&
\int_{m_s/m_B}^{\frac{1}{2}(1+m_s^2/m_B^2)} d x_0 \frac{64}{m_B^2} x_0^2
(-m_s^2-4 m_s^2 x_0+2 m_B^2 x_0^2+2 m_s^2 x_0^2)
{\it{Re}}(C_9^{{\mbox{eff}}})C_7^{{\mbox{eff}}} \nonumber \\
&+&
\int_{m_s/m_B}^{\frac{1}{2}(1+m_s^2/m_B^2)} d x_0
\frac{16}{3 m_B^2} x_0^2(-3 m_s^2+6 m_B^2 x_0^2+6 m_s^2 x_0^2-8 m_B^2 x_0^3)
|C_9^{{\mbox{eff}}}|^2\nonumber \\
&+&\frac{\alpha_s}{\pi} A^{(0,2)} C_9^2+ 
\frac{-16}{3} {C_7^{\mbox{eff}}}^2 \frac{\bar{\Lambda}}{m_B}+
\frac{-8}{3}{C_7^{\mbox{eff}}}^2\frac{\bar{\Lambda}^2}{m_B^2}
+ \left[ \frac{-1}{27} (55+84\ln(4 \frac{m_l^2}{m_B^2})){C_7^{\mbox{eff}}}^2+
 43 \frac{C_{10}^2}{270}  \right. \nonumber \\
&+& \left.
\int_{0}^{\frac{1}{2}} d x_0 
(\frac{64}{3} (6-5 x_0) x_0^3{\it{Re}}(C_9^{{\mbox{eff}}})C_7^{{\mbox{eff}}}
+\frac{16}{9} (18-39 x_0+20 x_0^2)x_0^3 |C_9^{{\mbox{eff}}}|^2 ) \right] 
\frac{\lambda_1}{m_B^2}
\nonumber \\
&+& 
 \left[ (11+4 \ln(4 \frac{m_l^2}{m_B^2})){C_7^{\mbox{eff}}}^2
  +13 \frac{C_{10}^2}{90} 
+ \int_{0}^{\frac{1}{2}} d x_0
(\frac{64}{3} x_0^2(-3-6 x_0+35 x_0^2){\it{Re}}(C_9^{{\mbox{eff}}})C_7^{{\mbox{eff}}} \right. \nonumber \\
&+& \left. 
\frac{16}{3} x_0^2 (-3+6 x_0+51 x_0^2-100 x_0^3)|C_9^{{\mbox{eff}}}|^2 \right] 
\frac{\lambda_2}{m_B^2} \, ,
\end{eqnarray}
\begin{eqnarray}
\langle x_0 (\s_0-\ms^2) \rangle {{\cal B}\over {\cal B}_0} &=&
\frac{\alpha_s}{\pi} A^{(1,1)} C_9^2
+ \left[ \frac{-8}{27} (1+3\ln(4 \frac{m_l^2}{m_B^2})){C_7^{\mbox{eff}}}^2+
 23 \frac{C_{10}^2}{270}  \right. \nonumber \\
&+& \left.
\int_{0}^{\frac{1}{2}} d x_0 
(\frac{128}{3} (3-5 x_0) x_0^3{\it{Re}}(C_9^{{\mbox{eff}}})C_7^{{\mbox{eff}}}
+\frac{32}{9} (9-27 x_0+20 x_0^2)x_0^3 |C_9^{{\mbox{eff}}}|^2 ) \right] 
\frac{\lambda_1}{m_B^2}
\nonumber \\
&+& 
 \left[ \frac{-8}{3}(5+3 \ln(4 \frac{m_l^2}{m_B^2})){C_7^{\mbox{eff}}}^2
  +13 \frac{C_{10}^2}{90} \right.\\
&+& \left.
\int_{0}^{\frac{1}{2}} d x_0
(\frac{128}{3} x_0^3(3+7 x_0){\it{Re}}(C_9^{{\mbox{eff}}})C_7^{{\mbox{eff}}}
+\frac{32}{3} x_0^3 (3+3 x_0-20 x_0^2)|C_9^{{\mbox{eff}}}|^2 \right] 
\frac{\lambda_2}{m_B^2}
\nonumber   \, ,
\end{eqnarray}
\begin{eqnarray}
\langle \s_0-\ms^2 \rangle {{\cal B}\over {\cal B}_0} &=&
\frac{\alpha_s}{\pi} A^{(1,0)} C_9^2
+ \left[ \frac{-2}{9} (23+24 \ln(4 \frac{m_l^2}{m_B^2})){C_7^{\mbox{eff}}}^2+
 13 \frac{C_{10}^2}{30}  \right. \nonumber \\
&+& \left.
\int_{0}^{\frac{1}{2}} d x_0 
(128 (1- x_0) x_0^2{\it{Re}}(C_9^{{\mbox{eff}}})C_7^{{\mbox{eff}}}
+\frac{32}{3} (3-7 x_0+4 x_0^2)x_0^2 |C_9^{{\mbox{eff}}}|^2 ) \right] 
\frac{\lambda_1}{m_B^2}
\nonumber \\
&+& 
 \left[ \frac{-2}{3}(31+24 \ln(4 \frac{m_l^2}{m_B^2})){C_7^{\mbox{eff}}}^2
  +\frac{C_{10}^2}{2} \right.\\
&+& \left.
\int_{0}^{\frac{1}{2}} d x_0
(\frac{128}{3} x_0^2 (3+7 x_0){\it{Re}}(C_9^{{\mbox{eff}}})C_7^{{\mbox{eff}}}
+\frac{32}{3} x_0^2 (3+3 x_0-20 x_0^2)|C_9^{{\mbox{eff}}}|^2 \right] 
\frac{\lambda_2}{m_B^2}
\nonumber  \, ,
\end{eqnarray}
\begin{eqnarray}
\langle (\s_0-\ms^2)^2 \rangle {{\cal B}\over {\cal B}_0} &=&
\frac{\alpha_s}{\pi} A^{(2,0)} C_9^2
+ \left[ \frac{8}{135} (119+60 \ln(4 \frac{m_l^2}{m_B^2})){C_7^{\mbox{eff}}}^2
 -16 \frac{C_{10}^2}{135}  \right.  \\
&+& \left.
\int_{0}^{\frac{1}{2}} d x_0 
(\frac{-512}{3} x_0^4{\it{Re}}(C_9^{{\mbox{eff}}})C_7^{{\mbox{eff}}}
+\frac{128}{9} (-3+4 x_0)x_0^4 |C_9^{{\mbox{eff}}}|^2 ) \right] 
\frac{\lambda_1}{m_B^2}
\nonumber \, .
\end{eqnarray}

\end{appendix}

\end{document}